\documentclass[english]{article}
\usepackage[T1]{fontenc}
\usepackage[latin9]{inputenc}
\usepackage{babel}
\usepackage{verbatim}
\usepackage{amsmath}
\usepackage{amssymb}
\usepackage{graphicx}
\PassOptionsToPackage{version=3}{mhchem}
\usepackage{mhchem}
\usepackage[unicode=true,pdfusetitle,
 bookmarks=true,bookmarksnumbered=false,bookmarksopen=false,
 breaklinks=false,pdfborder={0 0 0},pdfborderstyle={},backref=false,colorlinks=false]
 {hyperref}

\makeatletter

\let\SF@@footnote\footnote
\def\footnote{\ifx\protect\@typeset@protect
    \expandafter\SF@@footnote
  \else
    \expandafter\SF@gobble@opt
  \fi
}
\expandafter\def\csname SF@gobble@opt \endcsname{\@ifnextchar[
  \SF@gobble@twobracket
  \@gobble
}
\edef\SF@gobble@opt{\noexpand\protect
  \expandafter\noexpand\csname SF@gobble@opt \endcsname}
\def\SF@gobble@twobracket[#1]#2{}

\makeatletter
\def\@xfootnote[#1]{\protected@xdef\@thefnmark{#1}\@footnotemark\@footnotetext}
\makeatother
\makeatletter
\def\titlefootnote{\ifx\protect\@typeset@protect\expandafter\footnote\else\expandafter\@gobble\fi}
\makeatother

\usepackage{fullpage}
\usepackage{xr}
\usepackage[normalem]{ulem}
\usepackage{authblk}

\title{Cumulant-based formulation of higher-order fluorescence correlation spectroscopy}
\author{Farshad Abdollah-Nia\thanks{fabdo@rams.colostate.edu}} 
\affil{\emph{Department of Physics, Colorado State University, Fort Collins, Colorado 80523}} 

\date{}

\makeatother

\begin{document}
\maketitle

\global\long\def\var{\mathrm{var}}%

\global\long\def\E{\mathrm{E}}%

\begin{abstract}
Extended derivations regarding the cumulant-based formulation of higher-order
fluorescence correlation spectroscopy (FCS) are presented. First,
we review multivariate cumulants and their relation to photon counting
distributions in single and multi-detector experiments. Then we derive
the factorized correlation functions which describe system of diffusing
and reacting molecules using two independent approaches. Finally, we 
calculate the variance of these correlation functions up to the fourth 
order.
\end{abstract}
\tableofcontents{}\newpage{}

\section{Introduction}

Fluorescence correlation spectroscopy (FCS) is a powerful technique
for a time-resolved analysis of reaction, diffusion, and flow properties
of individual fluorescent molecules moving through a laser-illuminated
probe region\cite{elson74,krichevsky02,vanorden04}. Conventional
FCS is based on second-order correlation functions only, hence insufficient
for measuring the parameters which describe molecular reactions or
mixtures\cite{abdollahnia16a,Wu16} such as species populations and
brightness values and the kinetic rate constants. Higher-order correlation
functions can provide the necessary information for a complete measurement
of the underlying reaction or mixture parameters. Previous work to
define higher-order correlations based on higher moments of the fluorescence
signal lead to complex expressions for mixtures of diffusing molecules\cite{palmer87,palmer89intensity,palmer89mixture},
thus no extension of such approach to include reactions has been proposed.
More recently, a formulation based on the cumulants of the fluorescence
intensity or the factorial cumulants of the detected photon counts
has been published, which results in correlation functions incorporating
reaction and diffusion properties in a simple factorized form\cite{melnykov09}.
To experimentally apply the technique to the study of fast molecular
kinetics, difficulties due to shot noise and detector artifacts have
been recently overcome\cite{abdollahnia16a,abdollahnia16b}.

Theoretically, the formulation of correlation functions based on multi-variate
cumulants utilizes a variety of concepts not commonly covered in the
available literature on FCS. A detailed description of the theoretical
basis and derivations can make the underlying work for some earlier
publications\cite{abdollahnia16a,abdollahnia16b, melnykov09} more accessible.
This document has been produced with such purpose in mind, and presents
no further results and findings. 

The material in this document appears in developmental order. That
is, some introductory content is presented first and a continuous
forward flow of reading is assumed. However, informed reader may skip
to the desired section. Effort has been made to facilitate independent
reading of this document by including the necessary preliminary information,
derivations, and relations.

\section{Mathematical introduction and notation \label{sec:Mathematical-introduction}}

In this section we review the definitions, notation, and relations
between multivariate moments, central moments, cumulants, and their
factorial counterparts. For a more thorough treatment, the reader
may refer to \cite{balak98,kendall94}. 

In what follows, we use $\mu_{i_{1},i_{2},\ldots,i_{l}}$ to denote
$\mu_{i_{1},i_{2},\ldots,i_{l},0,\ldots,0}$ and use $\kappa_{i_{1},i_{2},\ldots,i_{l}}$
to denote $\kappa_{i_{1},i_{2},\ldots,i_{l},0,\ldots,0}$. Also, we
will use the differential operators 
\begin{align*}
\mathrm{D}_{i} & =\frac{\partial}{\partial t_{i}}\\
\mathrm{D}_{i}^{r} & =\frac{\partial^{r}}{\partial t_{i}^{r}}
\end{align*}
We will occasionally exchange the order of product and summation in
the following way:
\begin{align*}
\prod_{i=1}^{k}\sum_{j_{i}=0}^{r_{i}}f(r_{i},j_{i}) & =[f(r_{1},0)+\ldots+f(r_{1},r_{1})]\ldots[f(r_{k},0)+\ldots+f(r_{k},r_{k})]\\
 & =\sum_{j_{1}=0}^{r_{1}}\ldots\sum_{j_{k}=0}^{r_{k}}f(r_{1},j_{1})f(r_{2},j_{2})\ldots f(r_{k},j_{k})\\
 & =\sum_{j_{1}=0}^{r_{1}}\ldots\sum_{j_{k}=0}^{r_{k}}\prod_{i=1}^{k}f(r_{i},j_{i})
\end{align*}
For simplicity, we assume the moments, etc. exist and the expansions
converge.

\subsection{(Central) moments and cumulants\label{subsec:(Central)-moments-and}}

Let $\vec{X}=(X_{1},X_{2},\ldots,X_{k})$ be a multivariate random
vector. For $\vec{r}=(r_{1},r_{2},\ldots,r_{k})$ the $\vec{r}$th
\emph{moment} of $\vec{X}$ is defined as 
\[
\mu_{\vec{r}}^{\prime}=\mathrm{E}[\prod_{i=1}^{k}X_{i}^{r_{i}}]
\]
where $\mathrm{E}$ denotes the expectation operator. In particular,
\[
\mu_{\vec{u}_{i}}^{\prime}=\mathrm{E}[X_{i}]
\]
where $\vec{u}_{i}=(0,\ldots,0,1,0,\ldots,0)$ has only the $i$th
element equal to 1.

The $\vec{r}$th \emph{central moment} of $\vec{X}$ is defined as
\begin{equation}
\mu_{\vec{r}}=\mathrm{E}[\prod_{i=1}^{k}(X_{i}-\mathrm{E}[X_{i}])^{r_{i}}]\label{eq:mc-1-3}
\end{equation}
We denote the \emph{moment generating function }of $\vec{X}$ by 
\begin{align}
M^{\prime}(\vec{t}) & =\mathrm{E}[\mathrm{e}^{\vec{t}\cdot\vec{X}}]\label{eq:mc-3-1-1}\\
 & =\sum_{r_{1}=0}^{\infty}\ldots\sum_{r_{k}=0}^{\infty}\mu_{\vec{r}}^{\prime}\prod_{i=1}^{k}t_{i}^{r_{i}}/r_{i}!\nonumber 
\end{align}
and the \emph{central moment generating function} of $\vec{X}$ by
\begin{align}
M(\vec{t}) & =\mathrm{E}[\mathrm{e}^{\vec{t}\cdot(\vec{X}-\vec{\theta})}]\label{eq:mc-3-15-1}\\
 & =\sum_{r_{1}=0}^{\infty}\ldots\sum_{r_{k}=0}^{\infty}\mu_{\vec{r}}\prod_{i=1}^{k}t_{i}^{r_{i}}/r_{i}!\label{eq:mc-3-15-1-1}
\end{align}
where 
\begin{align*}
\vec{\theta} & =\left(\mathrm{E}[X_{1}],\mathrm{E}[X_{2}],\ldots,\mathrm{E}[X_{k}]\right)
\end{align*}
 and $\mu_{\vec{r}}$ is the $\vec{r}$th \emph{central moment} (or
moment about the mean) of $\vec{X}$.

The \emph{cumulant generating function} of $\vec{X}$ is defined as
\begin{align}
K(\vec{t}) & =\ln M^{\prime}(\vec{t})\label{eq:mc-3-14-1}\\
 & =\sum_{r_{1}=0}^{\infty}\ldots\sum_{r_{k}=0}^{\infty}\kappa_{\vec{r}}\prod_{i=1}^{k}t_{i}^{r_{i}}/r_{i}!\label{eq:mc-3-14-2}
\end{align}
which also defines $\kappa_{\vec{r}}$, the $\vec{r}$th \emph{cumulant}
of $\vec{X}$, through multiple differentiation of $K(\vec{t})$
\[
\kappa_{\vec{r}}=\left[\mathrm{D}_{1}^{r_{1}}\ldots\mathrm{D}_{k}^{r_{k}}K(\vec{t})\right]_{\vec{t}=\vec{0}}
\]

One can proceed to substitute the expansions (\ref{eq:mc-3-14-2})
and (\ref{eq:mc-3-15-1-1}) into relation (\ref{eq:mc-3-14-1}), or
into $M^{\prime}(\vec{t})=\exp(K(\vec{t}))$, then expand the $\ln$
or the $\exp$ function and set the terms of equal order in $\prod t_{i}^{r_{i}}$
equal to each other to obtain the relationship between the cumulants
$\kappa_{\vec{r}}$ and the moments $\mu_{\vec{r}}^{\prime}$. Then
setting all the first order moments $\mu_{\vec{u}_{i}}^{\prime}$
equal to zero in the resultant expressions and dropping the primes
will yield the relations between the cumulants $\kappa_{\vec{r}}$
and the central moments $\mu_{\vec{r}}$. Alternatively, one can write,
using (\ref{eq:mc-3-15-1}), 
\begin{align}
M(\vec{t}) & =\mathrm{e}^{-\vec{t}\cdot\vec{\theta}}\mathrm{E}[\mathrm{e}^{\vec{t}\cdot\vec{X}}]\nonumber \\
 & =\mathrm{e}^{-\vec{t}\cdot\vec{\theta}}M^{\prime}(\vec{t})\nonumber \\
 & =\mathrm{e}^{-\vec{t}\cdot\vec{\theta}}e^{K(\vec{t})}\nonumber \\
 & =\exp[K(\vec{t})-\vec{t}\cdot\vec{\theta}]\label{eq:mc-3-15-2}
\end{align}
and expand to obtain the relations between $\kappa_{\vec{r}}$ and
$\mu_{\vec{r}}$. The central moments, and therefore $M(\vec{t})$,
do not depend on the mean values $\mu_{\vec{u}_{i}}^{\prime}=\theta_{i}$.
In other words, $\mu_{\vec{u}_{i}}=0$, thus terms of individual $t_{i}$
do not appear in the expansion of $M(\vec{t})$. Thus (\ref{eq:mc-3-15-2})
immediately shows that the first cumulant of each $X_{i}$ is equal
to its mean, 
\[
\kappa_{\vec{u}_{i}}=\mu_{\vec{u}_{i}}^{\prime}=\mathrm{E}[X_{i}]
\]
Also, all higher order central moments are independent of mean in
the sense that they are invariant under translation of origin. Therefore
(\ref{eq:mc-3-15-2}) shows that all higher order cumulants are also
independent of mean, i.e. invariant under change of origin. 

Cumulants share the described invariance property with central moments.
However, cumulants also have another property which makes them unique:
Each cumulant of a sum of independent random variables is equal to
the sum of the corresponding cumulants of those random variables.
To see this, consider two independent random vectors $\vec{X}$ and
$\vec{Y}$. Directly from definition (\ref{eq:mc-3-1-1}) we have
for the moment generating function of their sum, $\vec{X}+\vec{Y},$
\begin{align*}
M_{\vec{X}+\vec{Y}}(\vec{t}) & =\mathrm{E}[\mathrm{e}^{\vec{t}\cdot(\vec{X}+\vec{Y})}]\\
 & =\mathrm{E}[\mathrm{e}^{\vec{t}\cdot\vec{X}}\mathrm{e}^{\vec{t}\cdot\vec{Y}}]\\
 & =\mathrm{E}[\mathrm{e}^{\vec{t}\cdot\vec{X}}]\mathrm{E}[\mathrm{e}^{\vec{t}\cdot\vec{Y}}]\tag{indep. \ensuremath{\vec{X}}, \ensuremath{\vec{Y}}}\\
 & =M_{\vec{X}}(\vec{t})M_{\vec{Y}}(\vec{t})
\end{align*}
Now from (\ref{eq:mc-3-14-1})
\begin{align*}
K_{\vec{X}+\vec{Y}}(\vec{t}) & =\ln[M_{\vec{X}+\vec{Y}}(\vec{t})]\\
 & =\ln[M_{\vec{X}}(\vec{t})M_{\vec{Y}}(\vec{t})]\\
 & =\ln[M_{\vec{X}}(\vec{t})]+\ln[M_{\vec{Y}}(\vec{t})]\\
 & =K_{\vec{X}}(\vec{t})+K_{\vec{Y}}(\vec{t})
\end{align*}
Expanding and setting the coefficients equal we get: 
\begin{equation}
\kappa_{\vec{r}}(\vec{X}+\vec{Y})=\kappa_{\vec{r}}(\vec{X})+\kappa_{\vec{r}}(\vec{Y})\label{eq:cum_of_sum}
\end{equation}
This property makes cumulants particularly useful in the study of
single molecules in solution, where each molecule can be considered
as an independent emitter of photons. Fluorescence Cumulant Analysis
(FCA) \cite{muller04,wu05} and Higher Order Fluorescence Correlation
Spectroscopy \cite{melnykov09} are examples of such applications. 

To obtain relations between multivariate cumulants and central moments,
it will be more convenient to use recursive formulas rather than the
expansion method mentioned above. Such formulas are obtained through
successive differentiation of the relevant moment generating function,
as presented by Balakrishnan \emph{et al} \cite{balak98}. For $1\leq l\leq k$
let us find the central moment $\mu_{r_{1},r_{2,},\ldots,r_{l}}$
in terms of the cumulants $\kappa_{i_{1},i_{2,},\ldots,i_{l}}$ of
order $(r_{1},r_{2},\ldots,r_{l})$ and less (i.e. $i_{m}\leq r_{m}$
for all $m$), and the central moments $\mu_{i_{1},i_{2,},\ldots,i_{l}}$
of order $(r_{1},r_{2},\ldots,r_{l}-1)$ and less:
\begin{align}
\mu_{r_{1},r_{2,},\ldots,r_{l}}= & \left[\mathrm{D}_{1}^{r_{1}}\mathrm{D}_{2}^{r_{2}}\ldots\mathrm{D}_{l}^{r_{l}}\mathrm{e}^{K(\vec{t})-\vec{t}\cdot\vec{\theta}}\right]_{\vec{t}=\vec{0}}\nonumber \\
= & \left[\mathrm{D}_{1}^{r_{1}}\mathrm{D}_{2}^{r_{2}}\ldots\mathrm{D}_{l}^{r_{l}-1}\mathrm{\{D}_{l}[K(\vec{t})-\vec{t}\cdot\vec{\theta}]\}\mathrm{e}^{K(\vec{t})-\vec{t}\cdot\vec{\theta}}\right]_{\vec{t}=\vec{0}}\nonumber \\
= & \left[\mathrm{D}_{1}^{r_{1}}\mathrm{D}_{2}^{r_{2}}\ldots\mathrm{D}_{l}^{r_{l}-1}\mathrm{\{D}_{l}K(\vec{t})-\theta_{l}\}\mathrm{e}^{K(\vec{t})-\vec{t}\cdot\vec{\theta}}\right]_{\vec{t}=\vec{0}}\nonumber \\
= & \left[\mathrm{D}_{1}^{r_{1}}\mathrm{D}_{2}^{r_{2}}\ldots\mathrm{D}_{l}^{r_{l}-1}\mathrm{\{D}_{l}K(\vec{t})\}\mathrm{e}^{K(\vec{t})-\vec{t}\cdot\vec{\theta}}\right]_{\vec{t}=\vec{0}}\nonumber \\
 & -\mathrm{E}[X_{l}]\left[\mathrm{D}_{1}^{r_{1}}\mathrm{D}_{2}^{r_{2}}\ldots\mathrm{D}_{l}^{r_{l}-1}\mathrm{e}^{K(\vec{t})-\vec{t}\cdot\vec{\theta}}\right]_{\vec{t}=\vec{0}}\label{eq:mc-4-1}
\end{align}
It is easy to see for multiple derivatives of a product 
\[
\mathrm{D}_{1}^{r_{1}}\left[f(\vec{t})g(\vec{t})\right]=\sum_{i=0}^{r_{1}}\binom{r_{1}}{i}\left[\mathrm{D}_{1}^{i}f(\vec{t})\right]\left[\mathrm{D}_{1}^{r_{1}-i}g(\vec{t})\right]
\]
From there,
\begin{align*}
\mathrm{D}_{1}^{r_{1}}\ldots\mathrm{D}_{l}^{r_{l}}\left[f(\vec{t})g(\vec{t})\right]= & \sum_{i_{1}=0}^{r_{1}}\ldots\sum_{i_{l}=0}^{r_{l}}\binom{r_{1}}{i_{1}}\ldots\binom{r_{l}}{i_{l}}\\
 & \times\left[\mathrm{D}_{1}^{i_{1}}\ldots\mathrm{D}_{l}^{i_{l}}f(\vec{t})\right]\left[\mathrm{D}_{1}^{r_{1}-i_{1}}\ldots\mathrm{D}_{l}^{r_{l}-i_{l}}g(\vec{t})\right]
\end{align*}
Applying this to (\ref{eq:mc-4-1}) with $f(\vec{t})=e^{K(\vec{t})-\vec{t}\cdot\vec{\theta}}=M(\vec{t})$
and $g(\vec{t})=\mathrm{D}_{l}K(\vec{t})$ we obtain
\begin{align}
\mu_{r_{1},r_{2,},\ldots,r_{l}}= & \sum_{i_{1}=0}^{r_{1}}\ldots\sum_{i_{l}=0}^{r_{l}-1}\binom{r_{1}}{i_{1}}\ldots\binom{r_{l}-1}{i_{l}}\nonumber \\
 & \times\left\{ \left[\mathrm{D}_{1}^{i_{1}}\ldots\mathrm{D}_{l}^{i_{l}}M(\vec{t})\right]\left[\mathrm{D}_{1}^{r_{1}-i_{1}}\ldots\mathrm{D}_{l}^{r_{l}-i_{l}}K(\vec{t})\right]\right\} _{\vec{t}=\vec{0}}\nonumber \\
 & -\mathrm{E}[X_{l}]\left[\mathrm{D}_{1}^{r_{1}}\mathrm{D}_{2}^{r_{2}}\ldots\mathrm{D}_{l}^{r_{l}-1}\mathrm{e}^{K(\vec{t})-\vec{t}\cdot\vec{\theta}}\right]_{\vec{t}=\vec{0}}\nonumber \\
= & \sum_{i_{1}=0}^{r_{1}}\ldots\sum_{i_{l}=0}^{r_{l}-1}\binom{r_{1}}{i_{1}}\ldots\binom{r_{l}-1}{i_{l}}\kappa_{r_{1}-i_{1},\ldots,r_{l}-i_{l}}\mu_{i_{1},\ldots,i_{l}}\nonumber \\
 & -\mathrm{E}[X_{l}]\mu_{r_{1},r_{2},\ldots,r_{l}-1}\label{eq:mc-6-1}
\end{align}

We write the first few relations up to a total order of four. For
a univariate distribution, we have directly from definition (\ref{eq:mc-1-3}):
\begin{align*}
\mu_{0} & =1\\
\mu_{1} & =0
\end{align*}
Then from (\ref{eq:mc-3-14-1}) 
\[
\kappa_{0}=\ln\mu_{0}=0
\]
The recursive formula (\ref{eq:mc-6-1}) now generates the rest:
\[
\mu_{1}=\kappa_{1}-\mathrm{E}[X]
\]
which equals zero, thus
\[
\kappa_{1}=\mathrm{E}[X]
\]
Again from (\ref{eq:mc-6-1}), 
\begin{align*}
\mu_{2} & =\kappa_{2}\mu_{0}+\kappa_{1}\mu_{1}-\mathrm{E}[X]\mu_{1}\\
 & =\kappa_{2}
\end{align*}
\begin{align*}
\mu_{3} & =\kappa_{3}\mu_{0}+\kappa_{2}\mu_{1}+\kappa_{1}\mu_{2}-\mathrm{E}[X]\mu_{2}\\
 & =\kappa_{3}
\end{align*}
And for $m\geq4$:
\[
\mu_{m}=\kappa_{m}+\sum_{i=2}^{m-2}\binom{m-1}{i}\kappa_{m-1}\mu_{i}\tag{m\ensuremath{\geq}4}
\]
For example,
\begin{align*}
\mu_{4} & =\kappa_{4}+3\kappa_{2}\mu_{2}\\
 & =\kappa_{4}+3\kappa_{2}^{2}
\end{align*}
conversely,
\[
\kappa_{4}=\mu_{4}-3\mu_{2}^{2}
\]
and so forth.

For a bivariate distribution, we have directly from the definitions,
(\ref{eq:mc-1-3}):
\begin{align*}
\mu_{0,0} & =1\\
\mu_{1,0} & =\mu_{0,1}=0
\end{align*}
and from (\ref{eq:mc-3-14-1}):
\[
\kappa_{0,0}=\ln\mu_{0,0}=0
\]
The recursive formula (\ref{eq:mc-6-1}) now yields
\[
\mu_{0,1}=\kappa_{0,1}\mu_{0,0}-\mathrm{E}[X_{2}]\mu_{0,0}
\]
which vanishes, thus
\[
\kappa_{0,1}=\mathrm{E}[X_{2}]
\]
$\kappa_{1,0}$ also follows the univariate relation:
\[
\kappa_{1,0}=\mathrm{E}[X_{1}]
\]

As evident from basic definitions, $\kappa_{m,0}$, $\mu_{m,0}$,
$\kappa_{0,m}$ and $\mu_{0,m}$ are identical to univariate cases
and follow their relations:
\begin{align*}
\kappa_{2,0} & =\mu_{2,0} & \kappa_{0,2} & =\mu_{0,2}\\
\kappa_{3,0} & =\mu_{3,0} & \kappa_{0,3} & =\mu_{0,3}\\
\kappa_{4,0} & =\mu_{4,0}-3\mu_{2,0}^{2} & \kappa_{0,4} & =\mu_{0,4}-3\mu_{0,4}^{2}
\end{align*}
and so forth.

A useful symmetry property also follows from the definitions: exchanging
the subscripts in any valid relation between $\kappa$s and $\mu$s
will produce a valid relation. Thus we can find the expression for
$\kappa_{n,m}$ (or $\mu_{n,m}$) from the that of $\kappa_{m,n}$
(or $\mu_{m,n}$). 

Once again our recursive formula (\ref{eq:mc-6-1}) can be applied:
\begin{align*}
\mu_{1,1} & =\kappa_{1,1}\mu_{0,0}+\kappa_{0,1}\mu_{1,1}-\mathrm{E}[X_{2}]\mu_{1,0}\\
 & =\kappa_{1,1}
\end{align*}
\begin{align*}
\mu_{2,1} & =\kappa_{2,1}\mu_{0,0}+\kappa_{1,1}\mu_{1,0}+\kappa_{0,1}\mu_{2,0}-\mathrm{E}[X_{2}]\mu_{2,0}\\
 & =\kappa_{2,1}
\end{align*}
By symmetry,
\[
\mu_{1,2}=\kappa_{1,2}
\]
In general, for $m+n\geq4$,
\[
\mu_{m,n}=\kappa_{m,n}+\mathop{\sum_{i=0}^{m}\sum_{j=0}^{n}}_{2\leqslant i+j\leqslant m+n-2}\binom{m}{i}\binom{n-1}{j}\kappa_{m-i,n-j}\mu_{i,j}\tag{m+n\ensuremath{\geq}4}
\]
For example,
\begin{align*}
\mu_{2,2} & =\kappa_{2,2}+2\kappa_{1,1}\mu_{1,1}+\kappa_{0,2}\mu_{2,0}\\
 & =\kappa_{2,2}+2\kappa_{1,1}^{2}+\kappa_{0,2}\kappa_{2,0}
\end{align*}
Conversely,
\[
\kappa_{2,2}=\mu_{2,2}-2\mu_{1,1}^{2}-\mu_{0,2}\mu_{2,0}
\]
As other examples,
\begin{align*}
\mu_{3,1} & =\kappa_{3,1}+3\kappa_{1,1}\mu_{2,0}\\
 & =\kappa_{3,1}+3\kappa_{1,1}\kappa_{2,0}
\end{align*}
\begin{align*}
\mu_{1,3} & =\kappa_{1,3}+\kappa_{1,1}\mu_{0,2}+2\kappa_{0,2}\mu_{1,1}\\
 & =\kappa_{1,3}+3\kappa_{1,1}\kappa_{0,2}
\end{align*}
which could also be obtained by swapping the subscripts. Conversely,
\begin{align*}
\kappa_{3,1} & =\mu_{3,1}-3\mu_{1,1}\mu_{2,0}\\
\kappa_{1,3} & =\mu_{1,3}-3\mu_{1,1}\mu_{0,2}
\end{align*}
and so forth.

\subsection{Factorial moments and factorial cumulants}

The tools developed in this section are particularly useful for discreet
random variables. In what follows, suppose $X_{i}$ can take only
non-negative integer values $\{0,1,\ldots\}$.

The $\vec{r}$th (\emph{descending}) \emph{factorial moment} of $\vec{X}$
is defined as
\[
\mu_{[\vec{r]}}^{\prime}=\mathrm{E}[\prod_{i=1}^{k}X_{i}^{[r_{i}]}]
\]
where 
\begin{align*}
x^{[r]} & =x(x-1)(x-2)\ldots(x-r+1)\\
 & =\frac{x!}{(x-r)!}\\
 & =\sum_{j=1}^{r}s(r,j)x^{j}
\end{align*}
denotes the $r$th \emph{(falling) factorial power} of $x$. Because
$x\in\{0,1,\ldots\}$, $x^{[r]}=0$ for $r>x$. The constants $s(r,j)$
are called the Stirling numbers of the first kind. Conversely,
\[
x^{r}=\sum_{j=1}^{r}S(r,j)x^{[j]}
\]
defines the Stirling numbers of the second kind, $S(r,j)$. 

Thus we have
\begin{align}
\mu_{[r_{1},\ldots,r_{k}]}^{\prime} & =\mathrm{E}[\prod_{i=1}^{k}\sum_{j_{i}=1}^{r_{i}}s(r_{i},j_{i})X_{i}^{j_{i}}]\nonumber \\
 & =\mathrm{E}[\sum_{j_{1}=1}^{r_{1}}\ldots\sum_{j_{k}=1}^{r_{k}}\prod_{i=1}^{k}s(r_{i},j_{i})X_{i}^{j_{i}}]\nonumber \\
 & =\sum_{j_{1}=1}^{r_{1}}\ldots\sum_{j_{k}=1}^{r_{k}}s(r_{1},j_{1})\ldots s(r_{k},j_{k})\mathrm{E}[\prod_{i=1}^{k}X_{i}^{j_{i}}]\nonumber \\
 & =\sum_{j_{1}=1}^{r_{1}}\ldots\sum_{j_{k}=1}^{r_{k}}s(r_{1},j_{1})\ldots s(r_{k},j_{k})\mu_{j_{1},\ldots,j_{k}}^{\prime}\label{eq:mc-2-6}
\end{align}
which gives factorial moments in terms of moments. Similarly,
\begin{equation}
\mu_{r_{1},\ldots,r_{k}}^{\prime}=\sum_{j_{1}=1}^{r_{1}}\ldots\sum_{j_{k}=1}^{r_{k}}S(r_{1},j_{1})\ldots S(r_{k},j_{k})\mu_{[j_{1},\ldots,j_{k}]}^{\prime}\label{eq:mc-3-8}
\end{equation}
giving moments in terms of factorial moments.

The \emph{probability generating function} of $\vec{X}$ is defined
as 
\begin{align}
\Omega(\vec{t}) & =\mathrm{E}[\prod_{i=1}^{k}t_{i}^{X_{i}}]\label{eq:pgf-def}\\
 & =\sum_{x_{1}=0}^{\infty}\ldots\sum_{x_{k}=0}^{\infty}p_{\vec{x}}\prod_{i=1}^{k}t_{i}^{x_{i}}\nonumber 
\end{align}
where 
\begin{align*}
p_{\vec{x}} & =\Pr(\vec{X}=\vec{x})\\
 & =\left[\frac{\mathrm{D}_{1}^{x_{1}}\ldots\mathrm{D}_{k}^{x_{k}}}{x_{1}!\ldots x_{k}!}\Omega(\vec{t})\right]_{\vec{t}=\vec{0}}
\end{align*}
 is the probability that $\vec{X}$ takes the value $\vec{x}=(x_{1},x_{2},\ldots,x_{k})$.

Now we define $\Phi(\vec{t})$ through a change of variable in the
probability generating function, $\Omega(\vec{t})$: 
\begin{align}
\Phi(\vec{t}) & =\Omega(\vec{1}+\vec{t})\label{eq:fmgf-def}\\
 & =\sum_{x_{1}=0}^{\infty}\ldots\sum_{x_{k}=0}^{\infty}p_{\vec{x}}\prod_{i=1}^{k}(1+t_{i})^{x_{i}}\nonumber \\
 & =\sum_{x_{1}=0}^{\infty}\ldots\sum_{x_{k}=0}^{\infty}p_{\vec{x}}\prod_{i=1}^{k}\sum_{r_{i}=0}^{x_{i}}\binom{x_{i}}{r_{i}}t_{i}^{r_{i}}\nonumber \\
 & =\sum_{x_{1}=0}^{\infty}\ldots\sum_{x_{k}=0}^{\infty}p_{\vec{x}}\prod_{i=1}^{k}\sum_{r_{i}=0}^{x_{i}}\frac{x_{i}^{[r_{i}]}}{r_{i}!}t_{i}^{r_{i}}\nonumber \\
 & =\sum_{x_{1}=0}^{\infty}\ldots\sum_{x_{k}=0}^{\infty}p_{\vec{x}}\sum_{r_{1}=0}^{x_{1}}\ldots\sum_{r_{k}=0}^{x_{k}}\prod_{i=1}^{k}x_{i}^{[r_{i}]}\frac{t_{i}^{r_{i}}}{r_{i}!}\nonumber \\
\intertext{\text{Since }\ensuremath{x_{i}^{[r_{i}]}=0}\text{ for }\ensuremath{r_{i}>x_{i}}} & =\sum_{x_{1}=0}^{\infty}\ldots\sum_{x_{k}=0}^{\infty}p_{\vec{x}}\sum_{r_{1}=0}^{\infty}\ldots\sum_{r_{k}=0}^{\infty}\prod_{i=1}^{k}x_{i}^{[r_{i}]}\frac{t_{i}^{r_{i}}}{r_{i}!}\nonumber \\
\intertext{\text{Now we can switch the summation order}} & =\sum_{r_{1}=0}^{\infty}\ldots\sum_{r_{k}=0}^{\infty}\left\{ \sum_{x_{1}=0}^{\infty}\ldots\sum_{x_{k}=0}^{\infty}p_{\vec{x}}\prod_{j=1}^{k}x_{j}^{[r_{j}]}\right\} \prod_{i=1}^{k}\frac{t_{i}^{r_{i}}}{r_{i}!}\nonumber \\
 & =\sum_{r_{1}=0}^{\infty}\ldots\sum_{r_{k}=0}^{\infty}\mu_{[\vec{r}]}^{\prime}\prod_{i=1}^{k}\frac{t_{i}^{r_{i}}}{r_{i}!}\nonumber 
\end{align}
Therefore, $\Phi(\vec{t})=\Omega(\vec{1}+\vec{t})$ is in fact the
\emph{factorial moment generating function} of $\vec{X}$. 

The \emph{factorial cumulant generating function} of $\vec{X}$ is
then defined as
\begin{align}
\Psi(\vec{t}) & =\ln\Phi(\vec{t})\label{eq:fcgf-def}\\
 & =\sum_{r_{1}=0}^{\infty}\ldots\sum_{r_{k}=0}^{\infty}\kappa_{[\vec{r}]}\prod_{i=1}^{k}t_{i}^{r_{i}}/r_{i}!\nonumber 
\end{align}
which also defines $\kappa_{[\vec{r}]}$, the $\vec{r}$th \emph{factorial
cumulant} of $\vec{X}$, through multiple differentiation of $\Psi(\vec{t})$:
\[
\kappa_{[\vec{r}]}=\left[\mathrm{D}_{1}^{r_{1}}\ldots\mathrm{D}_{k}^{r_{k}}\Psi(\vec{t})\right]_{\vec{t}=\vec{0}}
\]

As with ordinary moments and cumulants, one can proceed to expand
$\Psi(\vec{t})$ and $\ln\Phi(\vec{t})$ in (\ref{eq:fcgf-def}) and
compare term by term to obtain the relations between $\kappa_{[\vec{r}]}$
and $\mu_{[\vec{r}]}^{\prime}$. This procedure is identical to that
of finding the relations between $\kappa_{\vec{r}}$ and $\mu_{\vec{r}}^{\prime}$,
noting the similarity between (\ref{eq:fcgf-def}) and $K(\vec{t})=\ln M^{\prime}(\vec{t})$.
Therefore the relations between factorial moments and factorial cumulants
are formally similar to those between moments and cumulants, obtained
in section~\ref{subsec:(Central)-moments-and}.

It remains to determine the relations between cumulants and factorial
cumulants. To this purpose, we first examine the relation between
the moment generating function, $M^{\prime}(\vec{t})$, and the factorial
moment generating function, $\Phi(\vec{t})$, of $\vec{X}$. Using
the change of variables $\vec{s}(\vec{t})=(\mathrm{e}^{t_{1}},\ldots,\mathrm{e}^{t_{k}})$
in (\ref{eq:pgf-def}) and comparing with (\ref{eq:mc-3-1-1}) we
can write
\begin{align*}
\Omega\left(\vec{s}(\vec{t})\right) & =\mathrm{E}[\prod_{i=1}^{k}\mathrm{e}^{t_{i}X_{i}}]\\
 & \mathrm{=E}[\mathrm{e}^{\vec{t}\cdot\vec{X}}]\\
 & =M^{\prime}(\vec{t})
\end{align*}
On the other hand, (\ref{eq:fmgf-def}) gives $\Omega\left(\vec{s}(\vec{t})\right)=\Phi\left(\vec{s}(\vec{t})-\vec{1}\right)$,
thus we have
\begin{equation}
M^{\prime}(\vec{t})=\Phi\left(\vec{s}(\vec{t})-\vec{1}\right)\label{eq:mgf-fmgf-rel}
\end{equation}
Taking the natural logarithm of both sides we also have 
\begin{equation}
K(\vec{t})=\Psi\left(\vec{s}(\vec{t})-\vec{1}\right)\label{eq:cgf-fcgf-rel}
\end{equation}
as the relation between the cumulant generating function, $K(\vec{t})$,
and the factorial cumulant generating function, $\Psi(\vec{t})$,
of $\vec{X}$. Expanding both sides of (\ref{eq:mgf-fmgf-rel}) and
equating terms of equal order in $\prod t_{i}^{r_{i}}$ yields the
relations between moments and factorial moments, which were previously
found in (\ref{eq:mc-2-6}) and (\ref{eq:mc-3-8}) through direct
use of definitions. Noting the similarity between (\ref{eq:mgf-fmgf-rel})
and (\ref{eq:cgf-fcgf-rel}), formally similar relations are then
obtained through expansion of (\ref{eq:cgf-fcgf-rel}) between cumulants
and factorial cumulants:
\begin{align*}
\kappa_{[\vec{r}]} & =\sum_{j_{1}=1}^{r_{1}}\ldots\sum_{j_{k}=1}^{r_{k}}s(r_{1},j_{1})\ldots s(r_{k},j_{k})\kappa_{\vec{j}}\\
\kappa_{\vec{r}} & =\sum_{j_{1}=1}^{r_{1}}\ldots\sum_{j_{k}=1}^{r_{k}}S(r_{1},j_{1})\ldots S(r_{k},j_{k})\kappa_{[\vec{j}]}
\end{align*}

It will be instructive to actually expand (\ref{eq:mgf-fmgf-rel})
and equate the corresponding coefficients of $\prod t_{i}^{r_{i}}$
on the two sides. Expanding the right hand side gives: 
\begin{align*}
M^{\prime}(\vec{t}) & =\Phi\left(\vec{s}(\vec{t})-\vec{1}\right)\\
 & =\sum_{j_{1}=0}^{\infty}\ldots\sum_{j_{k}=0}^{\infty}\mu_{[\vec{j}]}^{\prime}\prod_{i=1}^{k}\frac{(\mathrm{e}^{t_{i}}-1){}^{j_{i}}}{j_{i}!}\\
 & =\sum_{j_{1}=0}^{\infty}\ldots\sum_{j_{k}=0}^{\infty}\mu_{[\vec{j}]}^{\prime}\prod_{i=1}^{k}\sum_{n_{i}=0}^{j_{i}}\binom{j_{i}}{n_{i}}\frac{(-1)^{j_{i}-n_{i}}\mathrm{e}^{n_{i}t_{i}}}{j_{i}!}\\
 & =\sum_{j_{1}=0}^{\infty}\ldots\sum_{j_{k}=0}^{\infty}\mu_{[\vec{j}]}^{\prime}\prod_{i=1}^{k}\sum_{n_{i}=0}^{j_{i}}\frac{(-1)^{j_{i}-n_{i}}}{n_{i}!(j_{i}-n_{i})!}\sum_{r_{i}=0}^{\infty}\frac{n_{i}^{r_{i}}t_{i}^{r_{i}}}{r_{i}!}\\
 & =\sum_{r_{1}=0}^{\infty}\ldots\sum_{r_{k}=0}^{\infty}\left\{ \sum_{j_{1}=0}^{\infty}\ldots\sum_{j_{k}=0}^{\infty}\mu_{[\vec{j}]}^{\prime}\prod_{l=1}^{k}\sum_{n_{l}=0}^{j_{l}}\frac{(-1)^{j_{l}-n_{l}}n_{l}^{r_{l}}}{n_{l}!(j_{l}-n_{l})!}\right\} \prod_{i=1}^{k}\frac{t_{i}^{r_{i}}}{r_{i}!}\\
\intertext{\text{while the left hand side is equal to}} & =\sum_{r_{1}=0}^{\infty}\ldots\sum_{r_{k}=0}^{\infty}\mu_{\vec{r}}^{\prime}\prod_{i=1}^{k}\frac{t_{i}^{r_{i}}}{r_{i}!}
\end{align*}
Therefore, 
\begin{align}
\mu_{\vec{r}}^{\prime} & =\sum_{j_{1}=0}^{\infty}\ldots\sum_{j_{k}=0}^{\infty}\mu_{[\vec{j}]}^{\prime}\prod_{l=1}^{k}\sum_{n_{l}=0}^{j_{l}}\frac{(-1)^{j_{l}-n_{l}}n_{l}^{r_{l}}}{n_{l}!(j_{l}-n_{l})!}\label{eq:m_in_fm}
\end{align}
Let us rewrite (\ref{eq:mc-3-8}) as:
\begin{align*}
\mu_{\vec{r}}^{\prime} & =\sum_{j_{1}=1}^{r_{1}}\ldots\sum_{j_{k}=1}^{r_{k}}\mu_{[\vec{j}]}^{\prime}\prod_{l=1}^{k}S(r_{l},j_{l})\\
\intertext{\text{Since }\ensuremath{S(r,j)=0}\text{ for }\ensuremath{j>r}\text{ and for }\ensuremath{j=0}} & =\sum_{j_{1}=0}^{\infty}\ldots\sum_{j_{k}=0}^{\infty}\mu_{[\vec{j}]}^{\prime}\prod_{l=1}^{k}S(r_{l},j_{l})
\end{align*}
Comparing with (\ref{eq:m_in_fm}) we get 
\[
S(r,j)=\sum_{n=0}^{j}\frac{(-1)^{j-n}n^{r}}{n!(j-n)!}
\]
an explicit expression for the Stirling numbers of the second kind.

We finish this section by tabulating the relations between cumulants
and factorial cumulants up to 4th order for future reference.
\begin{align}
 & \left.\begin{array}{l}
\kappa_{[1]}=\kappa_{1}\\
\kappa_{[2]}=\kappa_{2}-\kappa_{1}\\
\kappa_{[3]}=\kappa_{3}-3\kappa_{2}+2\kappa_{1}\\
\kappa_{[4]}=\kappa_{4}-6\kappa_{3}+11\kappa_{2}-6\kappa_{1}
\end{array}\right\} \nonumber \\
 & \left.\begin{array}{l}
\kappa_{[1,1]}=\kappa_{1,1}\\
\kappa_{[1,2]}=\kappa_{1,2}-\kappa_{1,1}\\
\kappa_{[1,3]}=\kappa_{1,3}-3\kappa_{1,2}+2\kappa_{1,1}\\
\kappa_{[2,2]}=\kappa_{2,2}-\kappa_{2,1}-\kappa_{1,2}+\kappa_{1,1}
\end{array}\right\} \nonumber \\
\intertext{\text{Conversely}} & \left.\begin{array}{l}
\kappa_{2}=\kappa_{[2]}+\kappa_{[1]}\\
\kappa_{3}=\kappa_{[3]}+3\kappa_{[2]}+\kappa_{[1]}\\
\kappa_{4}=\kappa_{[4]}+6\kappa_{[3]}+7\kappa_{[2]}+\kappa_{[1]}
\end{array}\right\} \nonumber \\
 & \left.\begin{array}{l}
\kappa_{1,2}=\kappa_{[1,2]}+\kappa_{[1,1]}\\
\kappa_{1,3}=\kappa_{[1,3]}+3\kappa_{[1,2]}+\kappa_{[1,1]}\\
\kappa_{2,2}=\kappa_{[2,2]}+\kappa_{[2,1]}+\kappa_{[1,2]}+\kappa_{[1,1]}
\end{array}\right\} \label{eq:c_fc_list}
\end{align}

\section{The relation between fluorescence intensity and photon counting distributions}

Consider a fluorescence detection experiment involving $k$ channels
and denote the fluorescence light intensity in channel $i$ at time
$t$ by $I_{i}(t)$. Typically, different channels consist of different
detectors and/or signals at different lag times. 

Let us for the time being limit our discussion to a single channel
$i$ and drop the subscript $i$ in the relevant quantities. Consider
a time interval (bin) of size $T$ and set the origin at its beginning:
$[0,T)$. Absorbing all efficiency parameters into $I$, or assuming
perfect efficiency, the probability of detecting a photon in the infinitesimal
interval $[t_{j},t_{j}+\mathrm{d}t_{j})$ is given by $I_{i}(t_{j})\mathrm{d}t_{j}$.
The probability of detecting no photons in the interval $[t_{j-1},t_{j})$
is given by the product of the probabilities of detecting no photons
in each infinitesimal interval $\delta t^{\prime}$ in that interval:
\begin{align*}
\lim_{\delta t^{\prime}\rightarrow0}\prod_{\delta t^{\prime}\epsilon[t_{j-1},t_{j})}(1-I(t^{\prime})\mathrm{\delta}t^{\prime}) & =\lim_{\delta t^{\prime}\rightarrow0}\prod_{\delta t^{\prime}\epsilon[t_{j-1},t_{j})}\exp[-I(t^{\prime})\mathrm{\delta}t^{\prime}]\\
 & =\exp\left[-\int_{t_{j-1}}^{t_{j}}I(t^{\prime})\mathrm{d}t^{\prime}\right]
\end{align*}
Let us also define $t_{0}=0$ and $t_{n+1}=T$. For $j=1,2,\ldots,n$
the probability of detecting exactly $n$ photons at $n$ infinitesimal
intervals $[t_{j},t_{j}+\mathrm{d}t_{j})$ where $0\leqslant t_{1}<\dots<t_{n}<T$
is given by 
\begin{align*}
\left\{ \prod_{j=1}^{n+1}\exp\left[-\int_{t_{j-1}}^{t_{j}}I(t_{j}^{\prime})\mathrm{d}t_{j}^{\prime}\right]\right\} \left\{ \prod_{j=1}^{n}I(t_{j})\mathrm{d}t_{j}\right\} \\
=\exp\left[-\int_{0}^{T}I(t^{\prime})\mathrm{d}t^{\prime}\right]\prod_{j=1}^{n}I(t_{j})\mathrm{d}t_{j}
\end{align*}
Now considering various placements of $t_{j}$ subject to the condition
$0\leqslant t_{1}<\dots<t_{n}<T$, the probability of detecting exactly
$n$ photons in the whole bin $[0,T)$ given a particular $I(t)$
becomes 
\begin{align*}
P(n;T|I) & =\exp\left[-\int_{0}^{T}I(t^{\prime})\mathrm{d}t^{\prime}\right]\int_{0}^{T}\mathrm{d}t_{1}\int_{t_{1}}^{T}\mathrm{d}t_{2}\ldots\int_{t_{n-1}}^{T}\mathrm{d}t_{n}I(t_{1})I(t_{2})\ldots I(t_{n})\\
 & =\exp\left[-\int_{0}^{T}I(t^{\prime})\mathrm{d}t^{\prime}\right]\int_{0}^{T}\mathrm{d}t_{n}\int_{0}^{t_{n}}\mathrm{d}t_{n-1}\ldots\int_{0}^{t_{2}}\mathrm{d}t_{1}I(t_{1})I(t_{2})\ldots I(t_{n})
\end{align*}
The integrand $I(t_{1})I(t_{2})\ldots I(t_{n})$ is symmetric: it
has the same value at any permutation of the $t_{j}$s. Any permutation
of $t_{j}$s in $0\leqslant t_{1}<\dots<t_{n}<T$ covers a separate
block of the parameter space, and the integration yields the same
value over each block. The union of all such blocks, or permutations
of the $t_{j}$s, covers the entire span of $0<t_{j}<T$ for all $j$.
There are $n!$ blocks, therefore we have
\begin{align}
P(n;T|I) & =\exp\left[-\int_{0}^{T}I(t^{\prime})\mathrm{d}t^{\prime}\right]\frac{1}{n!}\int_{0}^{T}\mathrm{d}t_{1}\int_{0}^{T}\mathrm{d}t_{2}\ldots\int_{0}^{T}\mathrm{d}t_{n}I(t_{1})I(t_{2})\ldots I(t_{n})\nonumber \\
 & =\frac{1}{n!}\left[\int_{0}^{T}I(t)\mathrm{d}t\right]^{n}\exp\left[-\int_{0}^{T}I(t^{\prime})\mathrm{d}t^{\prime}\right]\label{eq:Pn_given_I}
\end{align}

Now we take the fluctuations of $I(t)$ into account. Let $P(I)\mathrm{d}I$
denote the probability that the fluorescence intensity takes a value
within an infinitesimal variation $\mathrm{d}I$ around $I(t)$ defined
over $[0,T)$. Assuming a stationary process, binning a long experimental
time and averaging a quantity over all such bins, as done in an FCS
experiment, is equivalent to averaging that quantity over an ensemble
of $I(t)$ (ergodicity). The probability of detecting $n$ photons
in each bin is therefore
\begin{equation}
P(n;T)=\int P(n;T|I)P(I)\mathrm{d}I\label{eq:Pn_all_I}
\end{equation}
where $P(n;T|I)$ is given by (\ref{eq:Pn_given_I}). A more useful
random variable is the \emph{integrated intensity} within the bin
time $T$, defined as
\begin{equation}
W=\int_{0}^{T}I(t)\mathrm{d}t\label{eq:W_def}
\end{equation}
The probability of detecting $n$ photons in the bin, (\ref{eq:Pn_all_I}),
can be written as
\[
P(n;T)=\int P(n;T|W)P(W)\mathrm{d}W
\]
where, by (\ref{eq:Pn_given_I}), 
\begin{align*}
P(n;T|W) & =\frac{W^{n}}{n!}\mathrm{e}^{-W}\\
 & =\mathrm{Poi}(n,W)
\end{align*}
Thus we obtain Mandel's formula\cite{mandel58}:
\[
P(n;T)=\int\mathrm{Poi}(n,W)P(W)\mathrm{d}W
\]

We can now consider the case of multiple channels, and a common bin
size $T$. The fluorescence intensities $I_{i}(t)$ at different channels
are not necessarily independent. Neither are the integrated intensities,
$W_{i}=\int_{0}^{T}I_{i}(t)\mathrm{d}t$. However, given particular
$W_{1},W_{2},\ldots,W_{k}$ the occurrence probabilities of photons
in distinct channels are, by definition, independent:
\begin{align}
P(n_{1},n_{2},\ldots,n_{k};T|W_{1},W_{2},\ldots,W_{k}) & =\prod_{i=1}^{k}P(n_{i};T|W_{i})\label{eq:Pn_given_W_mult}\\
 & =\prod_{i=1}^{k}\mathrm{Pio}(n_{i},W_{i})\nonumber 
\end{align}
''Distinct'' channels in practice can be bins (overlapping with
no cross-talk, or non-overlapping) in independent detector signals,
or non-overlapping bins in a single detector signal. Denoting the
joint probability of integrated intensities taking the values $W_{1},W_{2},\allowbreak\ldots,W_{k}$
by $P(W_{1},W_{2},\ldots,W_{k})$ and integrating (\ref{eq:Pn_given_W_mult})
we obtain for the joint probability of detecting $n_{i}$ photons
in channel $i$:
\begin{equation}
P(n_{1},n_{2},\ldots,n_{k};T)=\int\int\ldots\int P(W_{1},W_{2},\ldots,W_{k})\prod_{i=1}^{k}\mathrm{Poi}(n_{i},W_{i})\mathrm{d}W_{i}\label{eq:Mandels_mult}
\end{equation}
which is the multivariate form of Mandel's formula.

To abbreviate notation, we define $\vec{\boldsymbol{W}}=(W_{1},W_{2},\allowbreak\ldots,W_{k})$
and $\vec{\boldsymbol{n}}=(n_{1},n_{2},\allowbreak\ldots,n_{k})$
using boldface symbols in our general treatment of multiple channels,
to be distinguished from vectors introduced later specific to two-time
correlation experiments.

Let us now examine the relations between moments (or cumulants) of
$\vec{\boldsymbol{W}}$ and moments (or cumulants) of $\vec{\boldsymbol{n}}$.
The probability generating function of $\vec{\boldsymbol{n}}$ is
\begin{align*}
\Omega_{\vec{\boldsymbol{n}}}(\vec{t}) & =\sum_{\vec{\boldsymbol{n}}}P(\vec{\boldsymbol{n}};T)\prod_{i=1}^{k}t_{i}^{n_{i}}
\end{align*}
where the shorthand notation 
\[
\sum_{\vec{\boldsymbol{n}}}=\sum_{n_{1}=0}^{\infty}\ldots\sum_{n_{k}=0}^{\infty}
\]
has been used. The factorial moment generating function of $\vec{\boldsymbol{n}}$
is 
\begin{align}
\Phi_{\vec{\boldsymbol{n}}}(\vec{t}) & =\Omega_{\vec{\boldsymbol{n}}}(\vec{t}+\vec{1})\nonumber \\
 & =\sum_{\vec{\boldsymbol{n}}}P(\vec{\boldsymbol{n}};T)\prod_{i=1}^{k}(t_{i}+1)^{n_{i}}\nonumber \\
\intertext{\text{using Mandel's formula, \eqref{eq:Mandels_mult},}} & =\sum_{\vec{\boldsymbol{n}}}\left\{ \int_{\vec{\boldsymbol{W}}}P(\vec{\boldsymbol{W}})\mathrm{d}^{k}W\prod_{l=1}^{k}\frac{W_{l}^{n_{l}}}{n_{l}!}\mathrm{e}^{-W_{l}}\right\} \prod_{i=1}^{k}(t_{i}+1)^{n_{i}}\nonumber \\
 & =\int_{\vec{\boldsymbol{W}}}P(\vec{\boldsymbol{W}})\mathrm{d}^{k}W\left\{ \prod_{l=1}^{k}\mathrm{e}^{-W_{l}}\right\} \sum_{\vec{\boldsymbol{n}}}\prod_{i=1}^{k}\frac{W_{i}^{n_{i}}}{n_{i}!}(t_{i}+1)^{n_{i}}\nonumber \\
 & =\int_{\vec{\boldsymbol{W}}}P(\vec{\boldsymbol{W}})\mathrm{d}^{k}W\left\{ \prod_{l=1}^{k}\mathrm{e}^{-W_{l}}\right\} \prod_{i=1}^{k}\sum_{n_{i}}\frac{(W_{i}t_{i}+W_{i})^{n_{i}}}{n_{i}!}\nonumber \\
 & =\int_{\vec{\boldsymbol{W}}}P(\vec{\boldsymbol{W}})\mathrm{d}^{k}W\left\{ \prod_{l=1}^{k}\mathrm{e}^{-W_{l}}\right\} \prod_{i=1}^{k}\mathrm{e}^{W_{i}t_{i}+W_{i}}\nonumber \\
 & =\int_{\vec{\boldsymbol{W}}}P(\vec{\boldsymbol{W}})\mathrm{d}^{k}W\prod_{i=1}^{k}\mathrm{e}^{W_{i}t_{i}}\nonumber \\
 & =\mathrm{E}[\mathrm{e}^{\vec{\boldsymbol{W}}\cdot\vec{t}}]\nonumber \\
 & =M_{\vec{\boldsymbol{W}}}^{\prime}(\vec{t})\label{eq:fmgf_n_mgf_W}
\end{align}
the moment generating function of $\vec{\boldsymbol{W}}$. This shows
that the factorial moments of $\vec{\boldsymbol{n}}$ are equal to
the (ordinary) moments of $\vec{\boldsymbol{W}}$. 

Taking the natural logarithm of (\ref{eq:fmgf_n_mgf_W}) we also obtain:
\begin{equation}
\Psi_{\vec{\boldsymbol{n}}}(\vec{t})=K_{\vec{\boldsymbol{W}}}(\vec{t})\label{eq:fcgf_n_cgf_W}
\end{equation}
That is, the factorial cumulants of $\vec{\boldsymbol{n}}$ are equal
to the (ordinary) cumulants of $\vec{\boldsymbol{W}}$.

Alternatively, we can calculate the factorial moments of $\vec{\boldsymbol{n}}$
directly:
\begin{align*}
\mu_{[\vec{r}]}(\vec{\boldsymbol{n}}) & =\sum_{\vec{\boldsymbol{n}}}P(\vec{\boldsymbol{n}})\prod_{i=1}^{k}n_{i}^{[r_{i}]}\\
\intertext{\text{and since }\ensuremath{n^{[r]}=0}\text{ for \ensuremath{n<r}}} & =\sum_{\vec{\boldsymbol{n}}\geqslant\vec{r}}\left\{ \int_{\vec{\boldsymbol{W}}}P(\vec{\boldsymbol{W}})\mathrm{d}^{k}W\prod_{l=1}^{k}\frac{W_{l}^{n_{l}}}{n_{l}!}\mathrm{e}^{-W_{l}}\right\} \prod_{i=1}^{k}\frac{n_{i}!}{(n_{i}-r_{i})!}\\
 & =\int_{\vec{\boldsymbol{W}}}P(\vec{\boldsymbol{W}})\mathrm{d}^{k}W\left\{ \prod_{l=1}^{k}\mathrm{e}^{-W_{l}}\right\} \prod_{i=1}^{k}\sum_{n_{i}=r_{i}}^{\infty}\frac{W_{i}^{n_{i}}}{(n_{i}-r_{i})!}\\
 & =\int_{\vec{\boldsymbol{W}}}P(\vec{\boldsymbol{W}})\mathrm{d}^{k}W\left\{ \prod_{l=1}^{k}\mathrm{e}^{-W_{l}}\right\} \prod_{i=1}^{k}W_{i}^{r_{i}}\mathrm{e}^{W_{i}}\\
 & =\int_{\vec{\boldsymbol{W}}}P(\vec{\boldsymbol{W}})\mathrm{d}^{k}W\prod_{i=1}^{k}W_{i}^{r_{i}}\\
 & =\mu_{\vec{r}}^{\prime}(\vec{\boldsymbol{W}})
\end{align*}
Then we can argue that the f.m.g.f. of $\vec{\boldsymbol{n}}$ is
equal to m.g.f. of $\vec{\boldsymbol{W}}$. From there, (\ref{eq:fcgf_n_cgf_W})
and the equality of the factorial cumulants of $\vec{\boldsymbol{n}}$
with the corresponding cumulants of $\vec{\boldsymbol{W}}$ follow:
\begin{equation}
\kappa_{[\vec{r}]}(\vec{\boldsymbol{n}})=\kappa_{\vec{r}}(\vec{\boldsymbol{W}})\label{eq:fc_n_c_W}
\end{equation}

\subsection[Two-time correlations]{Two-time correlations\protect\titlefootnote[*]{The content of this section is available in reference \cite{abdollahnia16a}
and is brought here for the sake of continuity.}}

We can now use Equation~(\ref{eq:fc_n_c_W}) to describe the correlation
functions between two lag times, $0$ and $t$. A number of methods
have been proposed due to the experimental limitations which arise
from detector artifacts, namely dead-time, after-pulsing, and cross-talk.
The effects of these artifacts in higher-order correlations extend
to all time scales, beyond the better-known effects in second-order
FCS which occur at short lag times only. Multi-detector and/or ``sub-binning''
approaches have been proposed to overcome these issues in higher-order
FCS. These methods have been described in another article\cite{abdollahnia16a}
along with the advantages and disadvantages of each method. 

In brief, a single-detector method with no modification usually suffers
most severely from detector artifacts. Earlier work tried to estimate
and account for these artifacts\cite{melnykov09,hillesheim05,wu06},
however, the extra modeling, approximations, and calibrations can
make such analysis more complicated and less versatile. Using two
or more detectors may avoid dead-time and after-pulsing artifacts,
however, cross-talk between the detectors may become an issue if not
experimentally prevented. Sub-binning refers to using smaller, non-overlapping
intervals (sub-bins) within the larger sampling intervals (bins).
The independent sub-bins virtually convert a single-detector experiment
to a multi-detector one with no cross-talk issue. In a true multi-detector
experiment, sub-binning also helps avoid the cross-talk artifact.
Therefore, sub-binning yields artifact-free results in both single-detector
and multi-detector experiments.\cite{abdollahnia16a} 

The discussion below assumes no sub-binning. For sub-binning we can
employ the multi-detector formulation without sub-binning; no separate
formulation is needed.

\subsubsection{Single-Detector Experiment}

When only one detector is used in the experiment (without sub-binning),
we can denote the fluorescence intensity at the detection point at
zero lag time by $I(0)$ and at lag time $t$ by $I(t)$. Take $W(0)$
and $W(t)$ to denote the corresponding integrated intensities over
some binning time $T$, that is
\begin{align*}
W(t) & =\int_{t}^{t+T}I(t^{\prime})\mathrm{d}t^{\prime}
\end{align*}
Also, $n(0)$ and $n(t)$ denote the number of photons detected in
the corresponding bins. The random vectors $\vec{I}=\left[I(0),I(t)\right]$,
$\vec{W}=\left[W(0),W(t)\right]$, and $\vec{n}_{\mathrm{1d}}=\left[n(0),n(t)\right]$
are then defined accordingly. Relation (\ref{eq:fc_n_c_W}) then directly
yields
\begin{equation}
\kappa_{[p,q]}\left[n(0),n(t)\right]=\kappa_{p,q}\left[W(0),W(t)\right]\label{eq:kappa_n1d_kappa_W}
\end{equation}

\subsubsection{Multi-Detector Experiment}

We can use $n_{\mathrm{d}}$ independent detectors (without sub-binning)
to obtain correlations of order $(p,q)$ with $p,q\leq n_{\mathrm{d}}$.
We assume that the beamsplitters and the detection efficiencies are
adjusted such that the light intensity is equal for all detectors
at any moment. Then take $I(0)$ to denote the fluorescence light
intensity arriving at any detector at lag time zero, and $I(t)$ to
denote that intensity at lag time $t$. Correspondingly, $W(0)$ and
$W(t)$ are defined by integration over a sampling interval (bin)
of size $T$, as in the single-detector case. The random vectors $\vec{I}=\left[I(0),I(t)\right]$
and $\vec{W}=\left[W(0),W(t)\right]$ are also defined accordingly.
The photon count $n_{i}(t)$ in the $i$th detector during a bin is
a distinct random variable for each detector, $i$. Therefore we define
the vector 
\begin{equation}
\vec{n}=\left[n_{1}(0),n_{2}(0),\ldots,n_{p}(0),n_{1}(t),n_{2}(t),\ldots,n_{q}(t)\right]\label{eq:vec_n_multi}
\end{equation}
Relation (\ref{eq:fc_n_c_W}) directly yields
\begin{align}
\kappa_{\left[\vec{1}_{p+q}\right]}(\vec{n}) & =\kappa_{\vec{1}_{p+q}}\left[W(0),\ldots,W(0),W(t),\ldots,W(t)\right]\nonumber \\
 & =\kappa_{p,q}\left[W(0),W(t)\right]\nonumber \\
 & =\kappa_{p,q}(\vec{W})\label{eq:kappa_n_kappa_W}
\end{align}
where $\vec{1}_{p+q}=(1,1,\ldots,1)$ has $p+q$ elements. 

In particular, for the case of two detectors named $\mathrm{A}$ and
$\mathrm{B}$ we have:
\[
\kappa_{[1,1,1]}\left[n_{\mathrm{A}}(0),n_{\mathrm{B}}(0),n_{\mathrm{A}}(t)\right]=\kappa_{2,1}\left[W(0),W(t)\right]
\]
\[
\kappa_{[1,1,1]}\left[n_{\mathrm{A}}(0),n_{\mathrm{A}}(t),n_{\mathrm{B}}(t)\right]=\kappa_{1,2}\left[W(0),W(t)\right]
\]
\[
\kappa_{[1,1,1,1]}\left[n_{\mathrm{A}}(0),n_{\mathrm{B}}(0),n_{\mathrm{A}}(t),n_{\mathrm{B}}(t)\right]=\kappa_{2,2}\left[W(0),W(t)\right]
\]

\section{Modeling correlations for diffusing and reacting molecules\label{sec:Molecules-in-Solution}}

In this section we derive the relations describing higher order fluorescence
correlations for molecules of a single-species which simultaneously
diffuse and undergo reaction between multiple states. We assume the
molecules have the same diffusion constant in all the ration states.
This multi-state system reduces to a multi-species non-interacting
system when the reaction rates are set to zero, with identical diffusion
constant for all species assumed. A mixture of reacting and non-reacting
species can also be described by setting only a subset of the reaction
rates equal to zero. 

Palmer and Thompson \cite{palmer87} defined higher-order correlations
using higher-order moments of intensity. For mixtures of diffusing
molecules, the moment-based definition of correlation functions leads
to complex expressions that depend on lower-order correlation functions.
No such expression has been proposed to include reactions of diffusing
molecules due to the increased complexity of formulation. Later, Melnykov
and Hall\cite{melnykov09}, following the approach developed by Müller
in the study of Fluorescence Cumulant Analysis \cite{muller04}, presented
a definition of higher-order correlations based on higher-order cumulants.
In their derivation, Melnykov and Hall used the additive property
of cumulants to arrive at simple factorized expressions for the cumulant-based
higher-order correlation functions describing systems of diffusing
molecules with reactions. 

In comparison to moments, the computation of cumulants from the experimental
photon stream is only slightly more complicated: moments are first
computed, then converted to cumulants using tabulated relations. However,
with the cumulant-based formulation, the theoretical relations which
describe systems of diffusing and reacting molecules are much simpler
than with the moment-based formulation. Most importantly, with the
cumulant-based formulation, the expressions factorize into pure reaction
and diffusion parts for systems with independent reaction and diffusion
processes. This allows for the experimental removal of any dependence
on the molecular detection function (MDF, defined as the combination
of laser intensity distribution, collection point-spread function,
and pinhole aperture\cite{schiro07}) and on the diffusion constant
by using a reference measurement\cite{abdollahnia16b}. 

In this section, before presenting the derivation reported by Melnykov
and Hall, we present an alternative derivation starting from simpler
premises and use a reverse reasoning process: we start from the explicit
integrals following the definition of higher-order moments, (\ref{eq:mup_I}),
and find conversion relations by only demanding simple final expressions
which are factorized into reaction and diffusion parts, (\ref{eq:F_def}),
without assuming any knowledge about cumulants, their properties,
and their relation to moments. Only then, we show that such conversion
relations are in general equivalent to the well-known conversion relations
between moments and cumulants. We label this approach the Palmer-Thompson
approach because the explicit expression of  integrals using Dirac
and Kronecker delta functions was inspired by the work of those authors.
On the other hand, Melnykov and Hall used the well-known additive
property of cumulants to directly derive the simple factorized expressions
for a multi-particle system based on the moments for a single particle.
While the approach by Melnykov and Hall is more concise and elegant,
the Palmer-Thompson approach is more elaborate and instructive therefore
it is discussed first. The two derivations are formally independent
and the reader may skip to the second approach if desired (after the
introductory discussion below).

Some preliminary discussion comes first. Take the random vector $\vec{W}=\left[W(0),W(t)\right]$,
where, following (\ref{eq:W_def}),
\begin{align*}
W(t) & =\int_{t}^{t+T}I(t^{\prime})\mathrm{d}t^{\prime}
\end{align*}
The general dependence of non-correlated higher moments of $W(t)$
on the bin size, $T$, has been studied by Wu and Müller \cite{wu05}
for non-interacting diffusing molecules through the introduction of
``binning functions''. In this report, however, we limit our attention
to the case of small bin sizes. For a short bin size $T$ over which
the variations of intensity can be neglected, we have
\[
W(t)\approx TI(t)
\]
\begin{equation}
\mu_{m,n}^{\prime}\left[W(0),W(t)\right]\approx T^{m+n}\mu_{m,n}^{\prime}\left[I(0),I(t)\right]\label{eq:mup_W}
\end{equation}
which becomes exact in the limit $T\to0$. 

Absorbing any detection efficiency factors into $I$, we can write
\begin{equation}
I(t)=\sum_{s}^{J}Q_{s}\int_{V}L(\vec{r})C_{s}(\vec{r},t)\mathrm{d}^{3}r\label{eq:I}
\end{equation}
where $L(\vec{r})$ is the laser illumination profile normalized to
its peak value, $J$ is the number of molecular states, $Q_{s}$ is
the brightness of state $s$ at illumination peak in counts per unit
time per molecule, $C_{s}$ is the concentration of the particles
in state $s$ at position $\vec{r}$ and time $t$, and $V$ is an
integration volume that includes the illuminated region. $V$ can
be taken to be the entire sample volume containing a fixed number
of molecules, $M$.

Therefore we have, in the limit $T\to0$,

\begin{align}
\lefteqn{\mu_{m,n}^{\prime}\left[I(0),I(t)\right]}\nonumber \\
 & \hphantom{\mu_{m,n}^{\prime}}=\left\langle I^{m}(0)I^{n}(t)\right\rangle \nonumber \\
 & \hphantom{\mu_{m,n}^{\prime}}=\sum_{s_{1}=1}^{J}\ldots\sum_{s_{m+n}=1}^{J}Q_{s_{1}}\ldots Q_{s_{m+n}}\int\ldots\int\mathrm{d}^{3}r_{1}\ldots\mathrm{d}^{3}r_{m+n}\nonumber \\
 & \hphantom{\mu_{m,n}^{\prime}=\sum_{s_{1}=1}^{J}\ldots\sum_{s_{m+n}=1}^{J}}\times L(\vec{r}_{1})\ldots L(\vec{r}_{m+n}){\cal G}_{m,n}^{\prime}(s_{1},\ldots,s_{m+n},\vec{r}_{1},\ldots,\vec{r}_{m+n};t)\label{eq:mup_I}
\end{align}
where
\begin{align}
\lefteqn{{\cal G}_{m,n}^{\prime}(s_{1},\ldots,s_{m+n},\vec{r}_{1},\ldots,\vec{r}_{m+n};t)}\nonumber \\
 & \hphantom{\mathcal{G}_{m,n}^{\prime}}=\left\langle C_{s_{1}}(\vec{r}_{1},0)\ldots C_{s_{m}}(\vec{r}_{m},0)C_{s_{m+1}}(\vec{r}_{m+1},t)\ldots C_{s_{m+n}}(\vec{r}_{m+n},t)\right\rangle \label{eq:calGp_def}
\end{align}

The concentration of particles in state $s$ at position $\vec{r}$
at time $t$ is given by

\begin{equation}
C_{s}(\vec{r},t)=\sum_{j=1}^{M}\delta[s,s_{j}(t)]\delta[\vec{r}-\vec{r}_{j}(t)]\label{eq:C_s}
\end{equation}
where $s_{j}(t)$ and $\vec{r}_{j}(t)$ are the state and position
of the $j$th particle at time $t$, respectively. $\delta(s,s^{\prime})$
and $\delta(\vec{r}-\vec{r}^{\prime})$ denote the Kronecker and the
Dirac delta functions, respectively. $M$ is the total number of molecules
in the large sample volume $V$. We will later take the limit $M,V\to\infty$.

Assuming a stationary (ergodic) system, the expected value of concentration
of particles in state $s$ at position $\vec{r}$ can be obtained
by averaging over $s_{j}(t)$ and $\vec{r}_{j}(t)$ as they vary over
time: 
\begin{align*}
\left\langle C_{s}\right\rangle  & =\sum_{j=1}^{M}\left\langle \delta[s,s_{j}(t)]\delta[\vec{r}-\vec{r}_{j}(t)]\right\rangle _{\vec{r}_{j},s_{j}}
\end{align*}
where
\begin{align}
\left\langle \delta[s,s_{j}(t)]\delta[\vec{r}-\vec{r}_{j}(t)]\right\rangle  & =\sum_{s_{j}=1}^{J}P(s_{j})\delta(s,s_{j})\int_{V}\mathrm{d}^{3}r_{j}P(\vec{r}_{j})\delta(\vec{r}-\vec{r}_{j})\label{eq:expctd_delta_1}\\
 & =P(s)/V\label{eq:expctd_delta_2}
\end{align}
Here, $P(s)$ denotes the probability that a given particle is in
state $s$ and 
\[
P(\vec{r})=\frac{1}{V}
\]
 is a uniform probability density that the particle is found anywhere
in the solution. Denoting the expected number of particles in state
$s$ by 
\[
M_{s}=P(s)M
\]
 we get 
\begin{equation}
\left\langle C_{s}\right\rangle =\frac{M_{s}}{V}\label{eq:expctd_C_s_res}
\end{equation}

Also, assuming ergodicity,
\[
\left\langle C_{s}\right\rangle =\lim_{V\rightarrow\infty}\frac{1}{V}\int_{V}C_{s}(\vec{r},t)\mathrm{d}^{3}r
\]
Taking the expectation of (\ref{eq:I}) and using (\ref{eq:expctd_C_s_res}),
the mean detection count rate is found to be 
\begin{equation}
\left\langle I\right\rangle =\sum_{s=1}^{J}Q_{s}N_{s}\label{eq:expctd_I}
\end{equation}
where we have defined
\begin{equation}
N_{s}=\frac{V_{\mathrm{MDF}}}{V}M_{s}\label{eq:N_s}
\end{equation}
and 
\begin{equation}
V_{\mathrm{MDF}}=\int_{V}L(\vec{r})\mathrm{d}^{3}r\label{eq:V_MDF}
\end{equation}
In the limit $V\to\infty$, $V_{\mathrm{MDF}}$ approaches the volume
of the molecular detection function (observation volume, or probe
region), and $N_{s}$ approaches the average number of molecules in
state $s$ in the observation volume.

Consider a single particle, for example the $j$th one. We define
and evaluate 
\begin{equation}
U^{(1)}(s_{1},s_{2};\vec{r}_{1},\vec{r}_{2};t)=\left\langle \delta(s_{1},s_{j}(0))\delta(\vec{r}_{1}-\vec{r}_{j}(0))\delta(s_{2},s_{j}(t))\delta(\vec{r}_{2}-\vec{r}_{j}(t))\right\rangle \label{eq:U_1_def}
\end{equation}
For brevity, take $s=s_{j}(0)$, $\vec{r}=\vec{r}_{j}(0)$, $s^{\prime}=s_{j}(t)$,
and $\vec{r}^{\prime}=\vec{r}_{j}(t)$. Then

\begin{align*}
\lefteqn{\left\langle \delta(s_{1},s)\delta(\vec{r}_{1}-\vec{r})\delta(s_{2},s^{\prime})\delta(\vec{r}_{2}-\vec{r}^{\prime})\right\rangle }\\
 & \hphantom{\delta(s_{1},s)}=\sum_{s=1}^{J}\sum_{s^{\prime}=1}^{J}\delta(s_{1},s)\delta(s_{2},s^{\prime})P(s,s^{\prime};t)\int_{V}\int_{V}\delta(\vec{r}_{1}-\vec{r})\delta(\vec{r}_{2}-\vec{r}^{\prime})P(\vec{r},\vec{r}^{\prime};t)\mathrm{d}^{3}\vec{r}\mathrm{d}^{3}\vec{r}^{\prime}\\
 & \hphantom{\delta(s_{1},s)}=P(s_{1},s_{2};t)P(\vec{r}_{1},\vec{r}_{2};t)
\end{align*}
where $P(s,s^{\prime};t)P(\vec{r},\vec{r}^{\prime};t)$ denotes the
joint probability that the particle is at state $s$ and position
$\vec{r}$ at time $0$, and at state $s^{\prime}$ and position $\vec{r}^{\prime}$
at time $t$. It can be expressed in terms of conditional probabilities,
\[
\begin{array}{l}
P(s_{1},s_{2};t)=P(s_{2}|s_{1};t)P(s_{1}),\\
P(\vec{r}_{1},\vec{r}_{2};t)=P(\vec{r}_{2}|\vec{r}_{1};t)P(\vec{r}_{1})
\end{array}
\]
where $P(s_{2}|s_{1};t)$ denotes the probability that the particle
is found in state $s_{2}$ at time $t$ given it was in state $s_{1}$
at time $0$, and is {[}obtained by solving linear rate equations,
Appendix~\ref{sec:Two-state-transition-factors}{]}:

\[
P(s_{2}|s_{1};t)=Z_{s_{2},s_{1}}(t)
\]
Similarly, the probability density function of the particle being
at position $\vec{r}_{2}$ at time $t$ given it was at position $\vec{r}_{1}$
at time $0$ is {[}obtained from solving the diffusion equation{]}
\[
P(\vec{r}_{2}|\vec{r}_{1};t)=\frac{\exp\left[-|\vec{r}_{2}-\vec{r}_{1}|^{2}/4Dt\right]}{(4\pi Dt)^{3/2}}
\]
Therefore for a single molecule we obtain
\begin{equation}
U^{(1)}(s_{1},s_{2},\vec{r}_{1},\vec{r}_{2};t)=\frac{P(s_{1})}{V}Z_{s_{2},s_{1}}(t)\frac{\exp\left[-|\vec{r}_{1}-\vec{r}_{2}|^{2}/4Dt\right]}{(4\pi Dt)^{3/2}}\label{eq:U_1_res}
\end{equation}

Next, it will be helpful to define and evaluate
\begin{equation}
U(s_{1},s_{2},\vec{r}_{1},\vec{r}_{2};t)=\sum_{j=1}^{M}\left\langle \delta[s_{1},s_{j}(0)]\delta[\vec{r}_{1}-\vec{r}_{j}(0)]\delta[s_{2},s_{j}(t)]\delta[\vec{r}_{2}-\vec{r}_{j}(t)]\right\rangle \label{eq:U_def}
\end{equation}
Immediately from (\ref{eq:U_1_def}), (\ref{eq:U_1_res}), and (\ref{eq:expctd_C_s_res})
we get: 
\begin{align}
U(s_{1},s_{2},\vec{r}_{1},\vec{r}_{2};t) & =MU^{(1)}(s_{1},s_{2},\vec{r}_{1},\vec{r}_{2};t)\label{eq:U_U1}\\
 & =\left\langle C_{s_{1}}\right\rangle Z_{s_{2},s_{1}}(t)\frac{\exp\left[-|\vec{r}_{1}-\vec{r}_{2}|^{2}/4Dt\right]}{(4\pi Dt)^{3/2}}\label{eq:U_res}
\end{align}
Setting $t=0$ we have:
\[
U(s_{1},s_{2},\vec{r}_{1},\vec{r}_{2};0)=\left\langle C_{s_{1}}\right\rangle \delta(s_{1},s_{2})\delta(\vec{r}_{1}-r_{2})
\]

\subsection{Palmer-Thompson approach}

This method follows the work of Palmer and Thompson \cite{palmer87}
where they directly evaluate ${\cal G}_{m,n}^{\prime}$. We have modified
the notation to incorporate multiple states, and will show that this
method is equivalent to the cumulant-based formulation. 

Using (\ref{eq:C_s}) we have
\begin{align*}
\lefteqn{{\cal G}_{m,n}^{\prime}(s_{1},\ldots,s_{m+n},\vec{r}_{1},\ldots,\vec{r}_{m+n};t)}\\
 & \hphantom{\mathcal{G}_{m,n}^{\prime}}=\sum_{i_{1}=1}^{M}\sum_{i_{2}=1}^{M}\dots\sum_{i_{m+n}=1}^{M}\left\langle \delta[s_{1},s_{i_{1}}(0)]\delta[\vec{r}_{1}-\vec{r}_{i_{1}}(0)]\dots\delta[s_{m},s_{i_{m}}(0)]\delta[\vec{r}_{m}-\vec{r}_{i_{m}}(0)]\right.\\
 & \hphantom{\mathcal{G}_{m,n}^{\prime}=\sum_{i_{1}=1}^{M}\sum_{i_{2}=1}^{M}\dots\sum_{i_{m+n}=1}^{M}}\left.\times\delta[s_{m+1},s_{i_{m+1}}(t)]\delta[\vec{r}_{m+1}-\vec{r}_{i_{m+1}}(t)]\ldots\delta[s_{m+n},s_{i_{m+n}}(t)]\delta[\vec{r}_{m+n}-\vec{r}_{i_{m+n}}(t)]\right\rangle 
\end{align*}

We break the summation according to the following cases:
\begin{itemize}
\item When two or more particle indices are different (at any lag time)
the expectation operator of product of independent random variables
factorizes. As a special case, when all particles at both lag times
are different: 
\begin{multline*}
\mathop{\sum_{i_{1}=1}^{M}\ldots\sum_{i_{m+n}=1}^{M}}_{\text{(no two indices equal)}}\left\langle \delta[s_{1},s_{i_{1}}(0)]\delta[\vec{r}_{1}-\vec{r}_{i_{1}}(0)]\right\rangle \times\ldots\\
\times\left\langle \delta[s_{m},s_{i_{m}}(0)]\delta[\vec{r}_{m}-\vec{r}_{i_{m}}(0)]\right\rangle \times\left\langle \delta[s_{m+1},s_{i_{m+1}}(t)]\delta[\vec{r}_{m+1}-\vec{r}_{i_{m+1}}(t)]\right\rangle \\
\times\ldots\times\left\langle \delta[s_{m+n},s_{i_{m+n}}(t)]\delta[\vec{r}_{m+n}-\vec{r}_{i_{m+n}}(t)]\right\rangle \\
\simeq\left\langle C_{s_{1}}\right\rangle \ldots\left\langle C_{s_{m+n}}\right\rangle 
\end{multline*}
To justify this approximation, notice that according to (\ref{eq:expctd_delta_2})
each term in the sum is of order $V^{-(m+n)}$. The summation on the
left has exactly $M(M-1)\ldots(M-m-n)$ terms, and the right hand
side, written as products of sums and expanded, has exactly $M^{m+n}$
terms. The sum of the extra terms on the right hand side (pertaining
to two or more equal indices) is then of order $1/V$. In the thermodynamic
limit $V,M\rightarrow\infty$ these extra terms become negligible
and the relation becomes exact.
\item When two or more particles indices are the same at the same lag time,
the expectation operator factorizes into products of delta functions
and a single concentration. As a special case, when all particle indices
at lag time $0$ are identical:
\begin{eqnarray*}
\lefteqn{\sum_{j=1}^{M}\left\langle \delta[s_{1},s_{j}(0)]\delta[\vec{r}_{1}-\vec{r}_{j}(0)]\ldots\delta[s_{m},s_{j}(0)]\delta[\vec{r}_{m}-\vec{r}_{j}(0)]\right\rangle }\\
 & = & \delta(s_{1},s_{2})\delta(\vec{r}_{1}-\vec{r}_{2})\ldots\delta(s_{1},s_{m})\delta(\vec{r}_{1}-\vec{r}_{m})\sum_{j=1}^{M}\left\langle \delta[s_{1},s_{j}(0)]\delta[\vec{r}_{1}-\vec{r}_{j}(0)]\right\rangle \\
 & = & \delta(s_{1},s_{2})\delta(\vec{r}_{1}-\vec{r}_{2})\ldots\delta(s_{1},s_{m})\delta(\vec{r}_{1}-\vec{r}_{m})\left\langle C_{s_{1}}\right\rangle 
\end{eqnarray*}
This can be justified by writing the expectation operator in the explicit
form of (\ref{eq:expctd_delta_1}) and using the change of variables
$\vec{y}=\vec{r}_{1}-\vec{r}_{j}$ for the spatial part. The discrete
part is nonzero only when all indices are identical.
\item When two or more particle indices are the same at different lag times,
the expectation operator factorizes into products of delta functions
and a single propagator. As a special case, when all indices at lag
time $0$ and $t$ are identical:
\begin{eqnarray}
\lefteqn{\sum_{j=1}^{M}\left\langle \delta[s_{1},s_{j}(0)]\delta[\vec{r}_{1}-\vec{r}_{j}(0)]\ldots\delta[s_{m},s_{j}(0)]\delta[\vec{r}_{m}-\vec{r}_{j}(0)]\right.}\nonumber \\
 &  & \hphantom{=}\left.\times\delta[s_{m+1},s_{j}(t)]\delta[\vec{r}_{m+1}-\vec{r}_{j}(t)]\ldots\delta[s_{m+n},s_{j}(t)]\delta[\vec{r}_{m+n}-\vec{r}_{j}(t)]\right\rangle \nonumber \\
 &  & =\delta(s_{1},s_{2})\delta(\vec{r}_{1}-\vec{r}_{2})\ldots\delta(s_{1},s_{m})\delta(\vec{r}_{1}-\vec{r}_{m})\nonumber \\
 &  & \hphantom{=\delta}\times\delta(s_{m+1},s_{m+2})\delta(\vec{r}_{m+1}-\vec{r}_{m+2})\ldots\delta(s_{m+1},s_{m+n})\delta(\vec{r}_{m+1}-\vec{r}_{m+n})\nonumber \\
 &  & \hphantom{=\delta}\times U(s_{1},s_{m+1},\vec{r}_{1},\vec{r}_{m+1};t)\label{eq:sumbreak3}
\end{eqnarray}
where $U(s_{1},s_{m+1},\vec{r}_{1},\vec{r}_{m+1};t)$ is defined in
(\ref{eq:U_def}).
\end{itemize}
Let us start with calculating the simpler cases of ${\cal G}_{m,0}^{\prime}$
and ${\cal G}_{0,n}^{\prime}$. We have by definition (\ref{eq:calGp_def})
\begin{align*}
{\cal G}_{1,0}^{\prime}(s_{1},\vec{r}_{1}) & =\left\langle C_{s_{1}}(\vec{r}_{1},0)\right\rangle =\left\langle C_{s_{1}}\right\rangle \\
{\cal G}_{0,1}^{\prime}(s_{1},\vec{r}_{1}) & =\left\langle C_{s_{1}}(\vec{r}_{1},t)\right\rangle =\left\langle C_{s_{1}}\right\rangle 
\end{align*}

For moments of order $(2,0)$, the summation in ${\cal G}_{20}^{\prime}$
runs over two particle indices. Therefore, it can be broken into two
cases where the two indices are either equal or not equal:
\begin{eqnarray*}
\lefteqn{{\cal G}_{2,0}^{\prime}(s_{1},s_{2},\vec{r}_{1},\vec{r}_{2})}\\
 & = & \sum_{j=1}^{M}\sum_{k=1}^{M}\left\langle \delta[s_{1},s_{j}(0)]\delta[\vec{r}_{1}-\vec{r}_{j}(0)]\delta[s_{2},s_{k}(0)]\delta[\vec{r}_{2}-\vec{r}_{k}(0)]\right\rangle \\
 & = & \sum_{j=1}^{M}\left\langle \delta[s_{1},s_{j}(0)]\delta[\vec{r}_{1}-\vec{r}_{j}(0)]\delta[s_{2},s_{j}(0)]\delta[\vec{r}_{2}-\vec{r}_{j}(0)]\right\rangle \\
 &  & +\mathop{\sum_{j=1}^{M}\sum_{k=1}^{M}}_{j\neq k}\left\langle \delta[s_{1},s_{j}(0)]\delta[\vec{r}_{1}-\vec{r}_{j}(0)]\right\rangle \left\langle \delta[s_{2},s_{k}(0)]\delta[\vec{r}_{2}-\vec{r}_{k}(0)]\right\rangle \\
 & = & \left\langle C_{s_{1}}\right\rangle \delta(s_{1},s_{2})\delta(\vec{r}_{1}-\vec{r}_{2})+\left\langle C_{s_{1}}\right\rangle \left\langle C_{s_{2}}\right\rangle 
\end{eqnarray*}
A similar result is obtained for ${\cal G}_{0,2}^{\prime}(s_{1},s_{2},\vec{r}_{1},\vec{r}_{2})$,
in a similar fashion.

For correlation of order $(1,1)$, the summation in ${\cal G}_{11}^{\prime}$
is also over two particle indices only, and it can be broken in a
similar way: 
\begin{eqnarray}
\lefteqn{{\cal G}_{1,1}^{\prime}(s_{1},s_{2},\vec{r}_{1},\vec{r}_{2};t)}\nonumber \\
 & = & \sum_{j=1}^{M}\sum_{k=1}^{M}\left\langle \delta[s_{1},s_{j}(0)]\delta[\vec{r}_{1}-\vec{r}_{j}(0)]\delta[s_{2},s_{k}(t)]\delta[\vec{r}_{2}-\vec{r}_{k}(t)]\right\rangle \nonumber \\
 & = & \sum_{j=1}^{M}\left\langle \delta[s_{1},s_{j}(0)]\delta[\vec{r}_{1}-\vec{r}_{j}(0)]\delta[s_{2},s_{j}(t)]\delta[\vec{r}_{2}-\vec{r}_{j}(t)]\right\rangle \nonumber \\
 &  & +\mathop{\sum_{j=1}^{M}\sum_{k=1}^{M}}_{j\neq k}\left\langle \delta[s_{1},s_{j}(0)]\delta[\vec{r}_{1}-\vec{r}_{j}(0)]\right\rangle \left\langle \delta[s_{2},s_{k}(t)]\delta[\vec{r}_{2}-\vec{r}_{k}(t)]\right\rangle \nonumber \\
 & = & U(s_{1},s_{2},\vec{r}_{1},\vec{r}_{2};t)+\left\langle C_{s_{1}}\right\rangle \left\langle C_{s_{2}}\right\rangle \label{eq:calGp_11_result}
\end{eqnarray}

\begin{figure}
\begin{centering}
\includegraphics[width=1\textwidth]{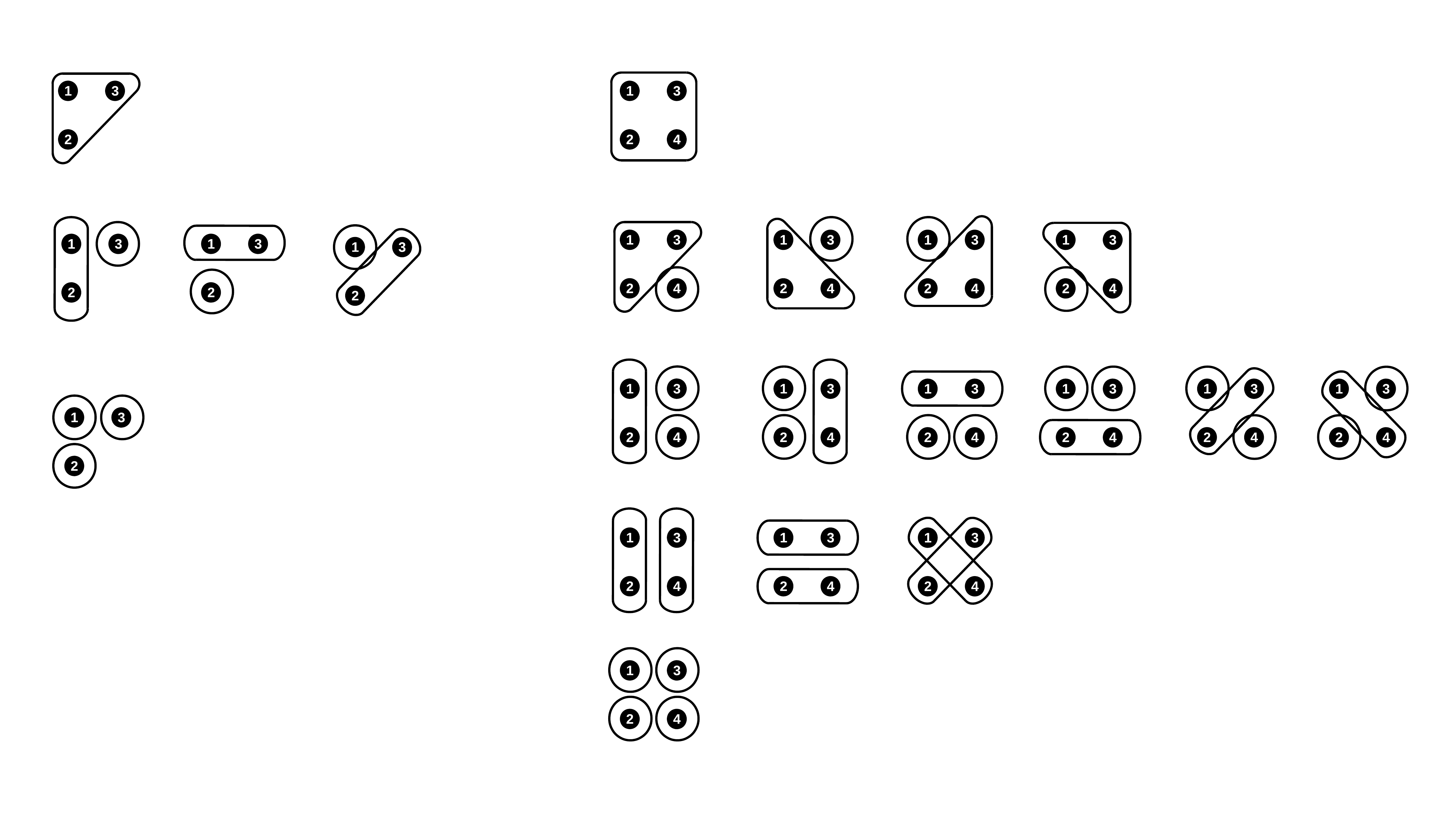}
\par\end{centering}
\caption{The 5 partitions of a set with 3 elements (left) and the 15 partitions
of a set with 4 elements (right).\label{fig:partitions3_4}}
\end{figure}

For order $(2,1)$, there are three particle indices: 
\begin{align}
\lefteqn{{\cal G}_{2,1}^{\prime}(s_{1},s_{2},s_{3},\vec{r}_{1},\vec{r}_{2},\vec{r}_{3};t)}\nonumber \\
 & =\sum_{j=1}^{M}\sum_{k=1}^{M}\sum_{l=1}^{M}\left\langle \delta[s_{1},s_{j}(0)]\delta[\vec{r}_{1}-\vec{r}_{j}(0)]\delta[s_{2},s_{k}(0)]\delta[\vec{r}_{2}-\vec{r}_{k}(0)]\right.\nonumber \\
 & \hphantom{=\sum_{i=1}^{M}\sum_{j=1}^{M}\sum_{k=1}^{M}\left\langle \right.}\left.\times\delta[s_{3},s_{l}(t)]\delta[\vec{r}_{3}-\vec{r}_{l}(t)]\right\rangle \nonumber \\
\intertext{\text{and we break the sum into five categories:}} & =\sum_{j=1}^{M}\left\langle \delta[s_{1},s_{j}(0)]\delta[\vec{r}_{1}-\vec{r}_{j}(0)]\delta[s_{2},s_{j}(0)]\delta[\vec{r}_{2}-\vec{r}_{j}(0)]\right.\nonumber \\
 & \hphantom{=\sum_{j=1}^{M}\left\langle \right.}\left.\times\delta[s_{3},s_{j}(t)]\delta[\vec{r}_{3}-\vec{r}_{j}(t)]\right\rangle \nonumber \\
 & \hphantom{=\,}+\mathop{\sum_{j=1}^{M}\sum_{k=1}^{M}}_{j\neq k}\left\langle \delta[s_{1},s_{j}(0)]\delta[\vec{r}_{1}-\vec{r}_{j}(0)]\delta[s_{2},s_{j}(0)]\delta[\vec{r}_{2}-\vec{r}_{j}(0)]\right\rangle \nonumber \\
 & \hphantom{=+\mathop{\sum_{j=1}^{M}\sum_{k=1}^{M}}\left\langle \right.}\times\left\langle \delta[s_{3},s_{k}(t)]\delta[\vec{r}_{3}-\vec{r}_{k}(t)]\right\rangle \nonumber \\
 & \hphantom{=\,}+\mathop{\sum_{j=1}^{M}\sum_{k=1}^{M}}_{j\neq k}\left\langle \delta[s_{1},s_{j}(0)]\delta[\vec{r}_{1}-\vec{r}_{j}(0)]\delta[s_{3},s_{j}(t)]\delta[\vec{r}_{3}-\vec{r}_{j}(t)]\right\rangle \nonumber \\
 & \hphantom{=+\mathop{\sum_{j=1}^{M}\sum_{k=1}^{M}}\left\langle \right.}\times\left\langle \delta[s_{2},s_{k}(0)]\delta[\vec{r}_{2}-\vec{r}_{k}(0)]\right\rangle \nonumber \\
 & \hphantom{=\,}+\mathop{\sum_{j=1}^{M}\sum_{k=1}^{M}}_{j\neq k}\left\langle \delta[s_{2},s_{j}(0)]\delta[\vec{r}_{2}-\vec{r}_{j}(0)]\delta[s_{3},s_{j}(t)]\delta[\vec{r}_{3}-\vec{r}_{j}(t)]\right\rangle \nonumber \\
 & \hphantom{=+\mathop{\sum_{j=1}^{M}\sum_{k=1}^{M}}\left\langle \right.}\times\left\langle \delta[s_{1},s_{k}(t)]\delta[\vec{r}_{1}-\vec{r}_{k}(t)]\right\rangle \nonumber \\
 & \hphantom{=\,}+\mathop{\sum_{j=1}^{M}\sum_{k=1}^{M}\sum_{l=1}^{M}}_{\text{(no two indices equal)}}\left\langle \delta[s_{1},s_{j}(0)]\delta[\vec{r}_{1}-\vec{r}_{j}(0)]\right\rangle \left\langle \delta[s_{2},s_{k}(0)]\delta[\vec{r}_{2}-\vec{r}_{k}(0)]\right\rangle \nonumber \\
 & \hphantom{=\,+\mathop{\sum_{j=1}^{M}\sum_{k=1}^{M}\sum_{l=1}^{M}}_{\text{(no two indices equal)}}\left\langle \right.}\times\left\langle \delta[s_{3},s_{l}(t)]\delta[\vec{r}_{3}-\vec{r}_{l}(t)]\right\rangle \nonumber \\
 & =\delta(s_{1},s_{2})\delta(\vec{r}_{1}-\vec{r}_{2})U(s_{1},s_{3},\vec{r}_{1},\vec{r}_{3};t)\nonumber \\
 & \hphantom{=\,}+\left\langle C_{s_{3}}\right\rangle \left\langle C_{s_{1}}\right\rangle \delta(s_{1},s_{2})\delta(\vec{r}_{1}-\vec{r}_{2})\nonumber \\
 & \hphantom{=\,}+\left\langle C_{s_{2}}\right\rangle U(s_{1},s_{3},\vec{r}_{1},\vec{r}_{3};t)\nonumber \\
 & \hphantom{=\,}+\left\langle C_{s_{1}}\right\rangle U(s_{2},s_{3},\vec{r}_{2},\vec{r}_{3};t)\nonumber \\
 & \hphantom{=\,}+\left\langle C_{s_{1}}\right\rangle \left\langle C_{s_{2}}\right\rangle \left\langle C_{s_{3}}\right\rangle \label{eq:calGp_21_result}
\end{align}
The categorization of the terms in the summation above corresponds
to partitioning of a set of $3$ items. A partition of a set $A$
is defined as a set of nonempty, pairwise disjoint subsets of $A$
whose union is $A$. There is a one-to-one correspondence between
the $5$ terms in (\ref{eq:calGp_21_result}) and the $5$ partitions
of a set of $3$ items shown in Figure~\ref{fig:partitions3_4},
left.

For order $(2,2)$ we have:
\begin{align*}
\lefteqn{{\cal G}_{2,2}^{\prime}(s_{1},s_{2},s_{3},s_{4},\vec{r}_{1},\vec{r}_{2},\vec{r}_{3},\vec{r}_{4};t)}\\
 & =\sum_{j=1}^{M}\sum_{k=1}^{M}\sum_{l=1}^{M}\sum_{m=1}^{M}\left\langle \delta[s_{1},s_{j}(0)]\delta[\vec{r}_{1}-\vec{r}_{j}(0)]\delta[s_{2},s_{k}(0)]\delta[\vec{r}_{2}-\vec{r}_{k}(0)]\right.\\
 & \hphantom{=\sum_{j=1}^{M}\sum_{k=1}^{M}\sum_{l=1}^{M}\sum_{m=1}^{M}\left\langle \right.}\left.\times\delta[s_{3},s_{l}(t)]\delta[\vec{r}_{3}-\vec{r}_{l}(t)]\delta[s_{4},s_{m}(t)]\delta[\vec{r}_{4}-\vec{r}_{m}(t)]\right\rangle 
\end{align*}
Once again, we break the summation depending on which particle indices
are equal. This will correspond to partitioning a set of $4$ elements
as shown in Figure~\ref{fig:partitions3_4}, right. If all indices
are equal, corresponding to the first row in the figure, we have a
single term
\begin{align*}
 & \sum_{j=1}^{M}\left\langle \delta[s_{1},s_{j}(0)]\delta[\vec{r}_{1}-\vec{r}_{j}(0)]\delta[s_{2},s_{j}(0)]\delta[\vec{r}_{2}-\vec{r}_{j}(0)]\right.\\
 & \hphantom{=\sum_{j=1}^{M}\left\langle \right.}\left.\times\delta[s_{3},s_{j}(t)]\delta[\vec{r}_{3}-\vec{r}_{j}(t)]\delta[s_{4},s_{j}(t)]\delta[\vec{r}_{4}-\vec{r}_{j}(t)]\right\rangle 
\end{align*}
Corresponding to the second row in the figure, we have four terms
of the form
\begin{align*}
 & \mathop{\sum_{j=1}^{M}\sum_{k=1}^{M}}_{j\neq k}\left\langle \delta[s_{1},s_{j}(0)]\delta[\vec{r}_{1}-\vec{r}_{j}(0)]\delta[s_{2},s_{j}(0)]\delta[\vec{r}_{2}-\vec{r}_{j}(0)]\right.\\
 & \hphantom{=\mathop{\sum_{j=1}^{M}\sum_{k=1}^{M}}\left\langle \right.}\left.\times\delta[s_{3},s_{j}(t)]\delta[\vec{r}_{3}-\vec{r}_{j}(t)]\right\rangle \left\langle \delta[s_{4},s_{k}(t)]\delta[\vec{r}_{4}-\vec{r}_{k}(t)]\right\rangle 
\end{align*}
Corresponding to the third row, we have two terms of the form
\begin{align*}
 & \mathop{\sum_{j=1}^{M}\sum_{k=1}^{M}\sum_{l=1}^{M}}_{\text{(no two indices equal)}}\left\langle \delta[s_{1},s_{j}(0)]\delta[\vec{r}_{1}-\vec{r}_{j}(0)]\delta[s_{2},s_{j}(0)]\delta[\vec{r}_{2}-\vec{r}_{j}(0)]\right\rangle \\
 & \hphantom{=\mathop{\sum_{j=1}^{M}\sum_{k=1}^{M}\sum_{l=1}^{M}}_{\text{(no two indices equal)}}\left\langle \right.}\times\left\langle \delta[s_{3},s_{k}(t)]\delta[\vec{r}_{3}-\vec{r}_{k}(t)]\right\rangle \left\langle \delta[s_{4},s_{l}(t)]\delta[\vec{r}_{4}-\vec{r}_{l}(t)]\right\rangle 
\end{align*}
and four terms of the form
\begin{align*}
 & \mathop{\sum_{j=1}^{M}\sum_{k=1}^{M}\sum_{l=1}^{M}}_{\text{(no two indices equal)}}\left\langle \delta[s_{1},s_{j}(0)]\delta[\vec{r}_{1}-\vec{r}_{j}(0)]\delta[s_{3},s_{j}(t)]\delta[\vec{r}_{3}-\vec{r}_{j}(t)]\right\rangle \\
 & \hphantom{=\mathop{\sum_{j=1}^{M}\sum_{k=1}^{M}\sum_{l=1}^{M}}_{\text{(no two indices equal)}}\left\langle \right.}\times\left\langle \delta[s_{2},s_{k}(0)]\delta[\vec{r}_{2}-\vec{r}_{k}(0)]\right\rangle \left\langle \delta[s_{4},s_{l}(t)]\delta[\vec{r}_{4}-\vec{r}_{l}(t)]\right\rangle 
\end{align*}
Corresponding to the fourth row we have a term of the form
\begin{align*}
 & \mathop{\sum_{j=1}^{M}\sum_{k=1}^{M}}_{j\neq k}\left\langle \delta[s_{1},s_{j}(0)]\delta[\vec{r}_{1}-\vec{r}_{j}(0)]\delta[s_{2},s_{j}(0)]\delta[\vec{r}_{2}-\vec{r}_{j}(0)]\right\rangle \\
 & \hphantom{=\mathop{\sum_{j=1}^{M}\sum_{k=1}^{M}}_{j\neq k}\left\langle \right.}\times\left\langle \delta[s_{3},s_{k}(t)]\delta[\vec{r}_{3}-\vec{r}_{k}(t)]\delta[s_{4},s_{k}(t)]\delta[\vec{r}_{4}-\vec{r}_{k}(t)]\right\rangle 
\end{align*}
and two terms of the form
\begin{align*}
 & \mathop{\sum_{j=1}^{M}\sum_{k=1}^{M}}_{j\neq k}\left\langle \delta[s_{1},s_{j}(0)]\delta[\vec{r}_{1}-\vec{r}_{j}(0)]\delta[s_{3},s_{j}(t)]\delta[\vec{r}_{3}-\vec{r}_{j}(t)]\right\rangle \\
 & \hphantom{=\mathop{\sum_{j=1}^{M}\sum_{k=1}^{M}}_{j\neq k}\left\langle \right.}\times\left\langle \delta[s_{2},s_{k}(0)]\delta[\vec{r}_{2}-\vec{r}_{k}(0)]\delta[s_{4},s_{k}(t)]\delta[\vec{r}_{4}-\vec{r}_{k}(t)]\right\rangle 
\end{align*}
Finally, corresponding to the last row we have a single term
\begin{align*}
 & \mathop{\sum_{j=1}^{M}\sum_{k=1}^{M}\sum_{l=1}^{M}\sum_{m=1}^{M}}_{\text{(no two indices equal)}}\left\langle \delta[s_{1},s_{j}(0)]\delta[\vec{r}_{1}-\vec{r}_{j}(0)]\right\rangle \left\langle \delta[s_{2},s_{k}(0)]\delta[\vec{r}_{2}-\vec{r}_{k}(0)]\right\rangle \\
 & \hphantom{=\mathop{\sum_{j=1}^{M}\sum_{k=1}^{M}\sum_{l=1}^{M}\sum_{m=1}^{M}}_{\text{(no two indices equal)}}\left\langle \right.}\times\left\langle \delta[s_{3},s_{l}(t)]\delta[\vec{r}_{3}-\vec{r}_{l}(t)]\right\rangle \left\langle \delta[s_{4},s_{m}(t)]\delta[\vec{r}_{4}-\vec{r}_{m}(t)]\right\rangle 
\end{align*}
Putting all these terms together, we get, in the order presented above,
\begin{align}
\lefteqn{{\cal G}_{2,2}^{\prime}(s_{1},s_{2},s_{3},s_{4},\vec{r}_{1},\vec{r}_{2},\vec{r}_{3},\vec{r}_{4};t)}\nonumber \\
 & =\delta(s_{1},s_{2})\delta(\vec{r}_{1}-\vec{r}_{2})\delta(s_{3},s_{4})\delta(\vec{r}_{3}-\vec{r}_{4})U(s_{1},s_{3},\vec{r}_{1},\vec{r}_{3};t)\nonumber \\
 & \hphantom{=\,}+\left\langle C_{s_{4}}\right\rangle \delta(s_{1},s_{2})\delta(\vec{r}_{1}-\vec{r}_{2})U(s_{1},s_{3},\vec{r}_{1},\vec{r}_{3};t)\nonumber \\
 & \hphantom{=\,}+\left\langle C_{s_{3}}\right\rangle \delta(s_{1},s_{2})\delta(\vec{r}_{1}-\vec{r}_{2})U(s_{1},s_{4},\vec{r}_{1},\vec{r}_{4};t)\nonumber \\
 & \hphantom{=\,}+\left\langle C_{s_{1}}\right\rangle \delta(s_{3},s_{4})\delta(\vec{r}_{3}-\vec{r}_{4})U(s_{2},s_{3},\vec{r}_{2},\vec{r}_{3};t)\nonumber \\
 & \hphantom{=\,}+\left\langle C_{s_{2}}\right\rangle \delta(s_{3},s_{4})\delta(\vec{r}_{3}-\vec{r}_{4})U(s_{1},s_{3},\vec{r}_{1},\vec{r}_{3};t)\nonumber \\
 & \hphantom{=\,}+\left\langle C_{s_{1}}\right\rangle \left\langle C_{s_{3}}\right\rangle \left\langle C_{s_{4}}\right\rangle \delta(s_{1},s_{2})\delta(\vec{r}_{1}-\vec{r}_{2})\nonumber \\
 & \hphantom{=\,}+\left\langle C_{s_{1}}\right\rangle \left\langle C_{s_{2}}\right\rangle \left\langle C_{s_{3}}\right\rangle \delta(s_{3},s_{4})\delta(\vec{r}_{3}-\vec{r}_{4})\nonumber \\
 & \hphantom{=\,}+\left\langle C_{s_{2}}\right\rangle \left\langle C_{s_{4}}\right\rangle U(s_{1},s_{3},\vec{r}_{1},\vec{r}_{3};t)\nonumber \\
 & \hphantom{=\,}+\left\langle C_{s_{1}}\right\rangle \left\langle C_{s_{3}}\right\rangle U(s_{2},s_{4},\vec{r}_{2},\vec{r}_{4};t)\nonumber \\
 & \hphantom{=\,}+\left\langle C_{s_{1}}\right\rangle \left\langle C_{s_{4}}\right\rangle U(s_{2},s_{3},\vec{r}_{2},\vec{r}_{3};t)\nonumber \\
 & \hphantom{=\,}+\left\langle C_{s_{2}}\right\rangle \left\langle C_{s_{3}}\right\rangle U(s_{1},s_{4},\vec{r}_{1},\vec{r}_{4};t)\nonumber \\
 & \hphantom{=\,}+\left\langle C_{s_{1}}\right\rangle \left\langle C_{s_{3}}\right\rangle \delta(s_{1},s_{2})\delta(\vec{r}_{1}-\vec{r}_{2})\delta(s_{3},s_{4})\delta(\vec{r}_{3}-\vec{r}_{4})\nonumber \\
 & \hphantom{=\,}+U(s_{1},s_{3},\vec{r}_{1},\vec{r}_{3};t)U(s_{2},s_{4},\vec{r}_{2},\vec{r}_{4};t)\nonumber \\
 & \hphantom{=\,}+U(s_{1},s_{4},\vec{r}_{1},\vec{r}_{4};t)U(s_{2},s_{3},\vec{r}_{2},\vec{r}_{3};t)\nonumber \\
 & \hphantom{=\,}+\left\langle C_{s_{1}}\right\rangle \left\langle C_{s_{2}}\right\rangle \left\langle C_{s_{3}}\right\rangle \left\langle C_{s_{4}}\right\rangle \label{eq:calGp_22_result}
\end{align}

It is now time to revisit the relations between the moments and the
cumulants of distribution. To better illustrate the point, we start
with a univariate distribution $X$. Following the notation developed
in section~\ref{subsec:(Central)-moments-and}, we have from the
relation between the moment generating function and the cumulant generating
function of $X$:
\begin{align*}
\sum_{r=0}^{\infty}\mu_{m}^{\prime}\frac{t^{m}}{m!} & =\exp\left(\sum_{p=0}^{\infty}\kappa_{p}\frac{t^{p}}{p!}\right)\\
 & =\prod_{p=0}^{\infty}\exp\left(\kappa_{p}\frac{t^{p}}{p!}\right)\\
 & =\prod_{p=0}^{\infty}\sum_{\pi_{p}=0}^{\infty}\frac{1}{\pi_{p}!}{\left(\kappa_{p}\frac{t^{p}}{p!}\right)^{\pi_{p}}}
\end{align*}
Multiplying both sides by $m!$ and picking out the terms in the exponential
expansions which, when multiplied together, give a power of $t^{m}$
we have 
\begin{equation}
\mu_{m}^{\prime}=\sum\left(\frac{\kappa_{p_{1}}}{p_{1}!}\right)^{\pi_{p_{1}}}\left(\frac{\kappa_{p_{2}}}{p_{2}!}\right)^{\pi_{p_{2}}}\ldots\left(\frac{\kappa_{p_{l}}}{p_{l}!}\right)^{\pi_{p_{l}}}\frac{m!}{\pi_{p_{1}}!\pi_{p_{2}}!\ldots\pi_{p_{l}}!}\label{eq:mu_kappa_partition}
\end{equation}
In the summation, each term corresponds to a set of distinct integers
$\{p_{1},p_{2},\ldots,p_{l}\}$ and their associated multiplicities
$\{\pi_{p_{1}},\pi_{p_{2}},\ldots,\pi_{p_{l}}\}$ such that 
\[
p_{1}\pi_{p_{1}}+p_{2}\pi_{p_{2}}+\ldots+p_{l}\pi_{p_{l}}=m
\]
and the summation runs over all such sets. The summation can be limited
to $p_{i}\geq1$ and $\pi_{p_{i}}\geq1$, because $\kappa_{0}=0$,
and zero multiplicities $\pi_{p_{i}}=0$ are not significant. Each
term then represents an ``integer partition'' of the number $m$
into $\sum_{i=1}^{l}\pi_{p_{i}}$ positive integer summands. 

\begin{figure}
\begin{centering}
\includegraphics[height=0.6\textwidth]{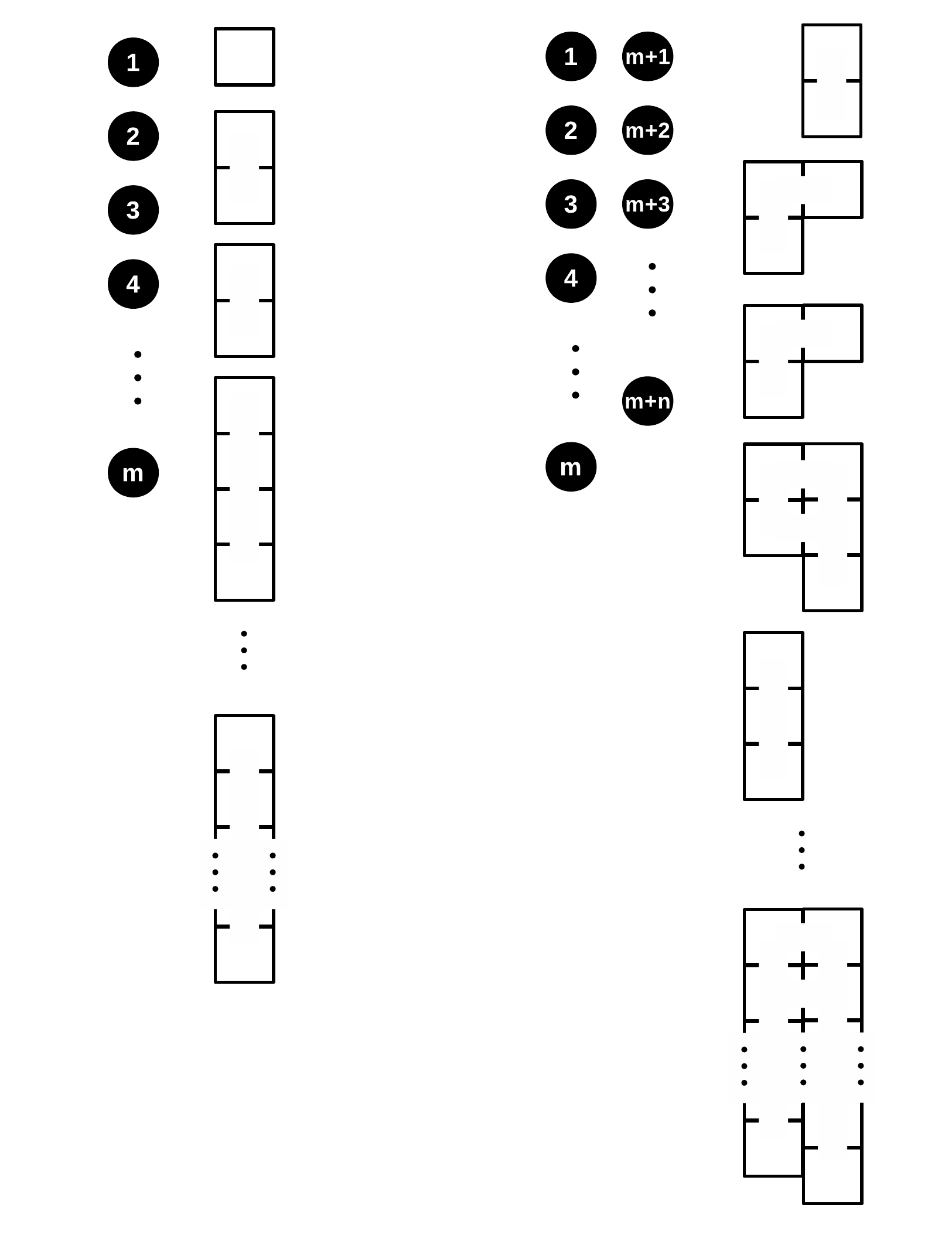}
\par\end{centering}
\caption{Example partition blocks of a set of $m$ elements (left) and those
of a set of $m+n$ elements divided into two subsets (right). \label{fig:partition_general}}
\end{figure}

Now consider a partition of a set of $m$ distinct objects, as shown
in Figure~\ref{fig:partition_general}, left, that includes $\pi_{p_{1}}$
blocks of size $p_{1}$, $\pi_{p_{2}}$ blocks of size $p_{2}$, $\ldots$,
and $\pi_{p_{l}}$ blocks of size $p_{l}$. The number of all possible
partitions with those given blocks and their sizes is exactly equal
to the coefficient of $\kappa_{p_{1}}^{\pi_{p_{1}}}\kappa_{p_{2}}^{\pi_{p_{2}}}\ldots\kappa_{p_{l}}^{\pi_{p_{l}}}$
in (\ref{eq:mu_kappa_partition}): 
\[
\frac{1}{\pi_{p_{1}}!\pi_{p_{2}}!\ldots\pi_{p_{l}}!}\cdot\frac{m!}{(p_{1}!)^{\pi_{p_{1}}}(p_{2}!)^{\pi_{p_{2}}}\ldots(p_{l}!)^{\pi_{p_{l}}}}
\]
This can be understood by first considering all the $m!$ permutations
of the $m$ objects in and among the blocks. The order of placement
of the objects within each block does not matter, therefore the division
by $(p_{1}!)^{\pi_{p_{1}}}(p_{2}!)^{\pi_{p_{2}}}\ldots(p_{l}!)^{\pi_{p_{l}}}$
appears. Finally, the blocks themselves are not ordered, that is,
exchanging all the members of a block with all the members of another
block of the same size does not create a new partition. Therefore,
a division by $\pi_{p_{1}}!\pi_{p_{2}}!\ldots\pi_{p_{l}}!$ is necessary.

As a result, a one-to-one correspondence (bijection) exists between
each factor $\kappa_{p_{1}}^{\pi_{p_{1}}}\kappa_{p_{2}}^{\pi_{p_{2}}}\ldots\kappa_{p_{l}}^{\pi_{p_{l}}}$
that appears in the expansion of $\mu_{m}^{\prime}$ and each partition
of a set of $m$ distinct elements. The mapping is unique in the sense
that identical factors of $\kappa_{p_{1}}^{\pi_{p_{1}}}\kappa_{p_{2}}^{\pi_{p_{2}}}\ldots\kappa_{p_{l}}^{\pi_{p_{l}}}$
are indistinguishable therefore their permutation yields the same
mapping.

There is a parallel situation in the case of a bivariate distribution
$\vec{X}=(X_{1},X_{2})$. Once again, we have from the relation between
the moment generating function and the cumulant generating function
of $\vec{X}$:
\begin{align*}
\sum_{m=0}^{\infty}\sum_{n=0}^{\infty}\mu_{m,n}^{\prime}\frac{t_{1}^{m}t_{2}^{n}}{m!n!} & =\exp\left(\sum_{p=0}^{\infty}\sum_{q=0}^{\infty}\kappa_{p,q}\frac{t_{1}^{p}t_{2}^{q}}{p!q!}\right)\\
 & =\prod_{p=0}^{\infty}\prod_{q=0}^{\infty}\exp\left(\kappa_{p,q}\frac{t_{1}^{p}t_{2}^{q}}{p!q!}\right)\\
 & =\prod_{p=0}^{\infty}\prod_{q=0}^{\infty}\sum_{\pi_{p,q}=0}^{\infty}\frac{1}{\pi_{p,q}!}{\left(\kappa_{p,q}\frac{t_{1}^{p}t_{2}^{q}}{p!q!}\right)^{\pi_{p,q}}}
\end{align*}
Multiplying both sides by $m!n!$ and picking out the terms in the
exponential expansions which, when multiplied together, give a power
of $t_{1}^{m}t_{2}^{n}$ we get
\begin{equation}
\mu_{m,n}^{\prime}=\sum\left(\frac{\kappa_{p_{1},q_{1}}}{p_{1}!q_{1}!}\right)^{\pi_{p_{1},q_{1}}}\left(\frac{\kappa_{p_{2},q_{2}}}{p_{2}!q_{2}!}\right)^{\pi_{p_{2},q_{2}}}\ldots\left(\frac{\kappa_{p_{l},q_{l}}}{p_{l}!q_{l}!}\right)^{\pi_{p_{l},q_{l}}}\frac{m!n!}{\pi_{p_{1},q_{1}}!\pi_{p_{2},q_{2}}!\ldots\pi_{p_{l},q_{l}}!}\label{eq:mu_kappa_partition-bivar}
\end{equation}
Each term in the summation corresponds to a set of distinct ordered\emph{
}pairs $\{(p_{1},q_{1}),(p_{2},q_{2}),\ldots,(p_{l},q_{l})\}$ with
multiplicities $\{\pi_{p_{1},q_{1}},\pi_{p_{2},q_{2}},\ldots,\pi_{p_{l},q_{l}}\}$
subject to the conditions
\begin{align}
p_{1}\pi_{p_{1},q_{1}}+p_{2}\pi_{p_{2},q_{2}}+\ldots+p_{l}\pi_{p_{l},q_{l}} & =m\label{eq:p_pi_cond}
\end{align}
and
\begin{equation}
q_{1}\pi_{p_{1},q_{1}}+q_{2}\pi_{p_{2},q_{2}}+\ldots+q_{l}\pi_{p_{l},q_{l}}=n\label{eq:q_pi_cond}
\end{equation}
and the summation runs over all such sets. The summation can be limited
to $(p_{i}+q_{i})\geq1$ (while either $p_{i}$ or $q_{i}$ can be
zero), and nonzero multiplicities $\pi_{p_{i},q_{i}}\geq1$. This
will correspond to integer partitioning the ordered pair of positive
integers $(m,n)$ into $\sum_{i=1}^{l}\pi_{p_{i},q_{i}}$ ordered
pairs of positive integers. By ordered, we mean $(p_{i},q_{i})$ and
$(q_{i},p_{i})$ are different pairs.

Now consider a partition of a set of $m+n$ distinct elements, as
shown in Figure~\ref{fig:partition_general}, right. The set, denoted
by $S$, is divided into two complementary subsets $L$ and $R$ of
sizes $m$ and $n$ respectively. Each block of an arbitrary partition
of $S$ consists of $p_{i}$ elements in $L$ and $q_{i}$ elements
in $R$, and there are $\pi_{p_{i},q_{i}}$ such blocks. Notice that
the a block of size $(p_{i},q_{i})$ is considered to have a different
size than a block of size $(q_{i},p_{i})$. We claim that the total
number of partitions with a given set of blocks and their sizes, that
is a given integer partition of $(m,n)$, is equal to the coefficient
of $\kappa_{p_{1},q_{1}}^{\pi_{p_{1},q_{1}}}\kappa_{p_{2},q_{2}}^{\pi_{p_{2},q_{2}}}\ldots\kappa_{p_{2},q_{2}}^{\pi_{p_{2},q_{2}}}$
in (\ref{eq:mu_kappa_partition-bivar}):
\begin{equation}
\frac{1}{\pi_{p_{1},q_{1}}!\pi_{p_{2},q_{2}}!\ldots\pi_{p_{l},q_{l}}!}\cdot\frac{m!}{(p_{1}!)^{\pi_{p_{1},q_{1}}}(p_{2}!)^{\pi_{p_{2},q_{2}}}\ldots(p_{l}!)^{\pi_{p_{l},q_{l}}}}\cdot\frac{n!}{(q_{1}!)^{\pi_{p_{1},q_{1}}}(q_{2}!)^{\pi_{p_{2},q_{2}}}\ldots(q_{l}!)^{\pi_{p_{l},q_{l}}}}\label{eq:coeff_kappa}
\end{equation}
To show this, we argue that only by permuting the members of $L$
within $L$ and the members of $R$ within $R$, all the set partitions
of $S$ relevant to the given integer partition of $(m,n)$ are generated.
Consider exchanging a member of $L$ with a member of $R$. If the
exchange happens within the same block, no new partition is created.
Therefore it is not counted. If the exchange happens between two distinct
blocks, the sizes of the blocks have changed by definition and the
resultant set partition of $S$ is counted under a different integer
partition of $(m,n)$. Coefficient (\ref{eq:coeff_kappa}) then gives
the number of all the set partitions relevant to a given number partition,
by arguments similar to the univariate case. Each and every partition
of $S$ is constructed by a partition of $(m,n)$ together with a
permutation of the members of $L$ and $R$ within their home subsets. 

As a result, there is a one-to-one correspondence between the partitions
of a set of $m+n$ distinct elements and the products of bivariate
cumulants appearing in the expansion of $\mu_{m,n}^{\prime}$. For
each partition there is a product term in the expansion, and for each
block of size $(p_{i},q_{i})$ in the partition there is a $\kappa_{p_{i},q_{i}}$
in the product term. 

On the other hand, we saw that the partitions of a set of $m+n$ distinct
elements corresponds uniquely to the terms appearing in the expansion
of ${\cal G}_{m,n}^{\prime}(s_{1},\ldots,s_{m+n},\vec{r}_{1},\ldots,\vec{r}_{m+n};t)$.
Each term in the expansion corresponds to a partition of the set.
In each term, groups of equal particle indices, corresponding to the
blocks of the partition, break the summation into products of some
factors; each factor being of the form:
\begin{align*}
{\cal F}_{1,0}(s_{1},\vec{r}_{1}) & ={\cal F}_{0,1}(s_{1},\vec{r}_{1})=\left\langle C_{s_{1}}\right\rangle \\
{\cal F}_{2,0}(s_{1},s_{2},\vec{r}_{1},\vec{r}_{2}) & ={\cal F}_{0,2}(s_{1},s_{2},\vec{r}_{1},\vec{r}_{2})=\left\langle C_{s_{1}}\right\rangle \delta(s_{1},s_{2})\delta(\vec{r}_{1}-\vec{r}_{2})
\end{align*}
\begin{align*}
{\cal F}_{1,1}(s_{1},s_{2},\vec{r}_{1},\vec{r}_{2};t) & =U(s_{1},s_{2},\vec{r}_{1},\vec{r}_{2};t)\\
{\cal F}_{2,1}(s_{1},s_{2},s_{3},\vec{r}_{1},\vec{r}_{2},\vec{r}_{3};t) & =\delta(s_{1},s_{2})\delta(\vec{r}_{1}-\vec{r}_{2})U(s_{1},s_{3},\vec{r}_{1},\vec{r}_{3};t)\\
{\cal F}_{1,2}(s_{1},s_{2},s_{3},\vec{r}_{1},\vec{r}_{2},\vec{r}_{3};t) & =\delta(s_{2},s_{3})\delta(\vec{r}_{2}-\vec{r}_{3})U(s_{1},s_{2},\vec{r}_{1},\vec{r}_{2};t)\\
{\cal F}_{2,2}(s_{1},s_{2},s_{3},s_{4},\vec{r}_{1},\vec{r}_{2},\vec{r}_{3},\vec{r}_{4};t) & =\delta(s_{1},s_{2})\delta(\vec{r}_{1}-\vec{r}_{2})\\
 & \phantom{=\,}\times\delta(s_{3},s_{4})\delta(\vec{r}_{3}-\vec{r}_{4})U(s_{1},s_{3},\vec{r}_{1},\vec{r}_{3};t)
\end{align*}
and in general, for $m,n\geq1$,
\begin{align}
\lefteqn{{\cal F}_{m,n}(s_{1},\ldots,s_{m+n},\vec{r}_{1},\ldots,\vec{r}_{m+n};t)}\nonumber \\
 & =\delta(s_{1},s_{2})\delta(\vec{r}_{1}-\vec{r}_{2})\ldots\delta(s_{1},s_{m})\delta(\vec{r}_{1}-\vec{r}_{m})\nonumber \\
 & \hphantom{=\,}\times\delta(s_{m+1},s_{m+2})\delta(\vec{r}_{m+1}-\vec{r}_{m+2})\ldots\delta(s_{m+1},s_{m+n})\delta(\vec{r}_{m+1}-\vec{r}_{m+n})\nonumber \\
 & \hphantom{=\:}\times U(s_{1},s_{m+1},\vec{r}_{1},\vec{r}_{m+1};t)\label{eq:calF_def}
\end{align}
and, for $m,n,n^{\prime}\geq1$,
\begin{align*}
{\cal F}_{m,0}(s_{1},\ldots,s_{m},\vec{r}_{1},\ldots,\vec{r}_{m}) & ={\cal F}_{0,m}(s_{1},\ldots,s_{m},\vec{r}_{1},\ldots,\vec{r}_{m})\\
 & ={\cal F}_{n,n^{\prime}}(s_{1},\ldots,s_{m},\vec{r}_{1},\ldots,\vec{r}_{m};0)\\
 & \tag{\ensuremath{n+n^{\prime}=m}}
\end{align*}
For example, we write the first few ${\cal G}_{m,n}^{\prime}$ in
terms of ${\cal F}_{p,q}$:
\begin{align*}
{\cal G}_{1,0}^{\prime}(s_{1},\vec{r}_{1}) & ={\cal F}_{1,0}(s_{1},\vec{r}_{1})\\
{\cal G}_{2,0}^{\prime}(s_{1},s_{2},\vec{r}_{1},\vec{r}_{2}) & ={\cal F}_{2,0}(s_{1},s_{2},\vec{r}_{1},\vec{r}_{2})+{\cal F}_{1,0}(s_{1},\vec{r}_{1}){\cal F}_{1,0}(s_{2},\vec{r}_{2})
\end{align*}
and 
\begin{equation}
{\cal G}_{1,1}^{\prime}(s_{1},s_{2},\vec{r}_{1},\vec{r}_{2};t)={\cal F}_{1,1}(s_{1},s_{2},\vec{r}_{1},\vec{r}_{2};t)+{\cal F}_{1,0}(s_{1},\vec{r}_{1}){\cal F}_{0,1}(s_{2},\vec{r}_{2})\label{eq:calGp_11_calF}
\end{equation}
\begin{align}
{\cal G}_{2,1}^{\prime}(s_{1},s_{2},s_{3},\vec{r}_{1},\vec{r}_{2},\vec{r}_{3};t) & ={\cal F}_{2,1}(s_{1},s_{2},s_{3},\vec{r}_{1},\vec{r}_{2},\vec{r}_{3};t)\nonumber \\
 & \hphantom{=\,}+{\cal F}_{2,0}(s_{1},s_{2},\vec{r}_{1},\vec{r}_{2}){\cal F}_{0,1}(s_{3},\vec{r}_{3})\nonumber \\
 & \hphantom{=\,}+{\cal F}_{1,0}(s_{2},\vec{r}_{2}){\cal F}_{1,1}(s_{1},s_{3},\vec{r}_{1},\vec{r}_{3};t)\nonumber \\
 & \hphantom{=\,}+{\cal F}_{1,0}(s_{1},\vec{r}_{1}){\cal F}_{1,1}(s_{2},s_{3},\vec{r}_{2},\vec{r}_{3};t)\nonumber \\
 & \hphantom{=\,}+{\cal F}_{1,0}(s_{1},\vec{r}_{1}){\cal F}_{1,0}(s_{2},\vec{r}_{2}){\cal F}_{0,1}(s_{3},\vec{r}_{3})\label{eq:calGp_21_calF}
\end{align}
\begin{align}
\lefteqn{{\cal G}_{2,2}^{\prime}(s_{1},s_{2},s_{3},s_{4},\vec{r}_{1},\vec{r}_{2},\vec{r}_{3},\vec{r}_{4};t)}\nonumber \\
 & ={\cal F}_{2,2}(s_{1},s_{2},s_{3},s_{4},\vec{r}_{1},\vec{r}_{2},\vec{r}_{3},\vec{r}_{4};t)\nonumber \\
 & \hphantom{=\,}+{\cal F}_{0,1}(s_{4},\vec{r}_{4}){\cal F}_{2,1}(s_{1},s_{2},s_{3},\vec{r}_{1},\vec{r}_{2},\vec{r}_{3};t)\nonumber \\
 & \hphantom{=\,}+{\cal F}_{0,1}(s_{3},\vec{r}_{3}){\cal F}_{2,1}(s_{1},s_{2},s_{4},\vec{r}_{1},\vec{r}_{2},\vec{r}_{4};t)\nonumber \\
 & \hphantom{=\,}+{\cal F}_{1,0}(s_{1},\vec{r}_{1}){\cal F}_{1,2}(s_{2},s_{3},s_{4},\vec{r}_{2},\vec{r}_{3},\vec{r}_{4};t)\nonumber \\
 & \hphantom{=\,}+{\cal F}_{1,0}(s_{2},\vec{r}_{2}){\cal F}_{1,2}(s_{1},s_{3},s_{4},\vec{r}_{1},\vec{r}_{3},\vec{r}_{4};t)\nonumber \\
 & \hphantom{=\,}+{\cal F}_{0,1}(s_{3},\vec{r}_{3}){\cal F}_{0,1}(s_{4},\vec{r}_{4}){\cal F}_{2,0}(s_{1},s_{2},\vec{r}_{1},\vec{r}_{2})\nonumber \\
 & \hphantom{=\,}+{\cal F}_{1,0}(s_{1},\vec{r}_{1}){\cal F}_{1,0}(s_{2},\vec{r}_{2}){\cal F}_{0,2}(s_{3},s_{4},\vec{r}_{3},\vec{r}_{4})\nonumber \\
 & \hphantom{=\,}+{\cal F}_{1,0}(s_{2},\vec{r}_{2}){\cal F}_{0,1}(s_{4},\vec{r}_{4}){\cal F}_{1,1}(s_{1},s_{3},\vec{r}_{1},\vec{r}_{3};t)\nonumber \\
 & \hphantom{=\,}+{\cal F}_{1,0}(s_{1},\vec{r}_{1}){\cal F}_{0,1}(s_{3},\vec{r}_{3}){\cal F}_{1,1}(s_{2},s_{4},\vec{r}_{2},\vec{r}_{4};t)\nonumber \\
 & \hphantom{=\,}+{\cal F}_{1,0}(s_{1},\vec{r}_{1}){\cal F}_{0,1}(s_{4},\vec{r}_{4}){\cal F}_{1,1}(s_{2},s_{3},\vec{r}_{2},\vec{r}_{3};t)\nonumber \\
 & \hphantom{=\,}+{\cal F}_{1,0}(s_{2},\vec{r}_{2}){\cal F}_{0,1}(s_{3},\vec{r}_{3}){\cal F}_{1,1}(s_{1},s_{4},\vec{r}_{1},\vec{r}_{4};t)\nonumber \\
 & \hphantom{=\,}+{\cal F}_{2,0}(s_{1},s_{2},\vec{r}_{1},\vec{r}_{2}){\cal F}_{0,2}(s_{3},s_{4},\vec{r}_{3},\vec{r}_{4})\nonumber \\
 & \hphantom{=\,}+{\cal F}_{1,1}(s_{1},s_{3},\vec{r}_{1},\vec{r}_{3};t){\cal F}_{1,1}(s_{2},s_{4},\vec{r}_{2},\vec{r}_{4};t)\nonumber \\
 & \hphantom{=\,}+{\cal F}_{1,1}(s_{1},s_{4},\vec{r}_{1},\vec{r}_{4};t){\cal F}_{1,1}(s_{2},s_{3},\vec{r}_{2},\vec{r}_{3};t)\nonumber \\
 & \hphantom{=\,}+{\cal F}_{1,0}(s_{1},\vec{r}_{1}){\cal F}_{1,0}(s_{2},\vec{r}_{2}){\cal F}_{0,1}(s_{3},\vec{r}_{3}){\cal F}_{0,1}(s_{4},\vec{r}_{4})\label{eq:calGp_22_calF_rel}
\end{align}

The functions ${\cal G}_{m,n}^{\prime}$ and ${\cal F}_{m,n}$ are
not yet in the form of bivariate moments and cumulants, respectively.
They are in fact $(m+n)$-variate moments and cumulants of order $(1,1,\ldots,1)$
of the local concentration variables on which they depend. However,
consider summation and integration of both sides of the following
form; and define 
\begin{multline*}
G_{m,n}^{\prime}(t)=\sum_{s_{1}=1}^{J}\ldots\sum_{s_{m+n}=1}^{J}Q_{s_{1}}\ldots Q_{s_{m+n}}\int\ldots\int\mathrm{d}^{3}r_{1}\ldots\mathrm{d}^{3}r_{m+n}\\
\times L(\vec{r}_{1})\ldots L(\vec{r}_{m+n}){\cal G}_{m,n}^{\prime}(s_{1},\ldots,s_{m+n},\vec{r}_{1},\ldots,\vec{r}_{m+n};t)
\end{multline*}
\begin{multline}
F_{m,n}(t)=\sum_{s_{1}=1}^{J}\ldots\sum_{s_{m+n}=1}^{J}Q_{s_{1}}\ldots Q_{s_{m+n}}\int\ldots\int\mathrm{d}^{3}r_{1}\ldots\mathrm{d}^{3}r_{m+n}\\
\times L(\vec{r}_{1})\ldots L(\vec{r}_{m+n}){\cal F}_{m,n}(s_{1},\ldots,s_{m+n},\vec{r}_{1},\ldots,\vec{r}_{m+n};t)\label{eq:F_def}
\end{multline}
Since each ${\cal F}_{p,q}$ in each term of the expansion of ${\cal G}_{m,n}^{\prime}$
corresponds to a separate block of variables, the integral of their
product factorizes into the product of integrals, $F_{p_{1},q_{1}}^{\pi_{p_{1},q_{1}}}F_{p_{2},q_{2}}^{\pi_{p_{2},q_{2}}}\ldots F_{p_{2},q_{2}}^{\pi_{p_{2},q_{2}}}$.
Each product term corresponds to an integer partition of the number
$m+n$, and the number of occurrences of such terms is equal to the
number of partitions consisting of those block sizes, of a set of
$m+n$ distinct objects. All such partitions are constructed by permuting
the two groups of $m$ and $n$ particle indices among themselves,
and all such permutations yield the same integration result. Therefore,
the relations between $G_{m,n}^{\prime}$ and $F_{m,n}$ are exactly
those between moments and cumulants of a bivariate distribution. Finally,
noting that $G_{m,n}^{\prime}$ is indeed the $(m,n)$th moment of
the bivariate distribution $\vec{I}=\left(I(0),I(t)\right)$ (see~(\ref{eq:mup_I})),
\[
G_{m,n}^{\prime}(t)=\left\langle I^{m}(0)I^{n}(t)\right\rangle =\mu_{m,n}^{\prime}\left[I(0),I(t)\right]
\]
we conclude that $F_{m,n}$ are indeed the cumulants of $\vec{I}$:
\begin{equation}
F_{m,n}(t)=\kappa_{m,n}[I(0),I(t)]\label{eq:F_kappaI}
\end{equation}

We can simply write the first few moments of $\vec{I}$ in terms of
its cumulants using (\ref{eq:calGp_22_calF_rel}):
\begin{equation}
\begin{array}{ll}
G_{1,0}^{\prime} & =F_{1,0}\\
G_{2,0}^{\prime} & =F_{2,0}+F_{1,0}^{2}\\
G_{1,1}^{\prime}(t) & =F_{1,1}(t)+F_{1,0}F_{0,1}\\
G_{2,1}^{\prime}(t) & =F_{2,1}(t)+F_{2,0}F_{0,1}+2F_{1,0}F_{1,1}(t)+F_{1,0}^{2}F_{0,1}\\
G_{2,2}^{\prime}(t) & =F_{2,2}(t)+2F_{0,1}F_{2,1}(t)+2F_{1,0}F_{1,2}(t)+F_{0,1}^{2}F_{2,0}+F_{1,0}^{2}F_{0,2}\\
 & \multicolumn{1}{r}{+4F_{1,0}F_{0,1}F_{1,1}(t)+F_{2,0}F_{0,2}+2F_{1,1}^{2}(t)+F_{1,0}^{2}F_{0,1}^{2}}
\end{array}\label{eq:Gp_F_rel}
\end{equation}

The advantage of cumulant functions, $F_{m,n}$, is that the integrals
yield simple analytical forms, and the advantage of $G_{m,n}^{\prime}$
functions is that they are simply related to observable intensities.
Since the collection of all nonzero cumulants also gives a complete
description of the distribution, we can primarily define higher order
correlations based on cumulants. To experimentally compute cumulants
based on the observed moments of intensity, the inverse relations
can be found from (\ref{eq:Gp_F_rel}):
\begin{equation}
\begin{array}{ll}
F_{1,0} & =G_{1,0}^{\prime}\\
F_{2,0} & =G_{2,0}^{\prime}-G_{1,0}^{\prime2}\\
F_{1,1}(t) & =G_{1,1}^{\prime}(t)-G_{1,0}^{\prime}G_{0,1}^{\prime}\\
F_{2,1}(t) & =G_{2,1}^{\prime}(t)-G_{2,0}^{\prime}G_{0,1}^{\prime}-2G_{1,0}^{\prime}G_{1,1}^{\prime}(t)+2G_{1,0}^{\prime2}G_{0,1}^{\prime}\\
F_{2,2}(t) & =G_{2,2}^{\prime}(t)-2G_{0,1}^{\prime}G_{2,1}^{\prime}(t)-2G_{1,0}^{\prime}G_{1,2}^{\prime}(t)+2G_{0,1}^{\prime2}G_{2,0}^{\prime}+2G_{1,0}^{\prime2}G_{0,2}^{\prime}\\
 & \multicolumn{1}{r}{+8G_{1,0}^{\prime}G_{0,1}^{\prime}G_{1,1}^{\prime}(t)-G_{2,0}^{\prime}G_{0,2}^{\prime}-2G_{1,1}^{\prime2}(t)-6G_{1,0}^{\prime2}G_{0,1}^{\prime2}}
\end{array}\label{eq:F_Gp_rel}
\end{equation}
Notice that e.g. $F_{12}$ can be found by exchanging the subscripts
in the relation for $F_{21}$, and similarly for $G_{12}^{\prime}$
and $G_{21}^{\prime}$. As a general rule, the coefficients in (\ref{eq:F_Gp_rel})
are obtained from the coefficients in (\ref{eq:Gp_F_rel}) multiplied
by $(-1)^{\rho-1}(\rho-1)!$ where $\rho$ is the total number of
blocks in the partition (To see this, compare (\ref{eq:mu_kappa_partition})
and (\ref{eq:kappa_mu_partition})).

\subsubsection*{Conversion to central moments}

The relations involving central moments rather than moments are significantly
simpler to write and handle. We define 
\begin{align*}
\lefteqn{{\cal G}_{m,n}(s_{1},\ldots,s_{m+n},\vec{r}_{1},\ldots,\vec{r}_{m+n};t)}\\
 & =\left\langle \delta C_{s_{1}}(\vec{r}_{1},0)\ldots\delta C_{s_{m}}(\vec{r}_{m},0)\delta C_{s_{m+1}}(\vec{r}_{m+1},t)\ldots\delta C_{s_{m+n}}(\vec{r}_{m+n},t)\right\rangle 
\end{align*}
where
\[
\delta C_{s}(\vec{r},t)=C_{s}(\vec{r},t)-\left\langle C_{s}\right\rangle 
\]

We first explain the method suggested by Palmer and Thompson, then
explain an easier method of shifting the origin. They write

\begin{align*}
{\cal G}_{1,1}^{\prime}(s_{1},s_{2},\vec{r}_{1},\vec{r}_{2};t) & =\left\langle C_{s_{1}}(\vec{r}_{1},0)C_{s_{2}}(\vec{r}_{2},t)\right\rangle \\
 & =\left\langle [\delta C_{s_{1}}(\vec{r}_{1},0)+\left\langle C_{s_{1}}\right\rangle ][\delta C_{s_{2}}(\vec{r}_{2},t)+\left\langle C_{s_{2}}\right\rangle ]\right\rangle \\
 & =\left\langle \delta C_{s_{1}}(\vec{r}_{1},0)\delta C_{s_{2}}(\vec{r}_{2},t)\right\rangle +\left\langle C_{s_{1}}\right\rangle \left\langle C_{s_{2}}\right\rangle \\
 & ={\cal G}_{1,1}(s_{1},s_{2},\vec{r}_{1},\vec{r}_{2};t)+{\cal F}_{1,0}(s_{1},\vec{r}_{1}){\cal F}_{0,1}(s_{2},\vec{r}_{2})
\end{align*}
Comparing with (\ref{eq:calGp_11_calF}) we obtain 
\[
{\cal G}_{1,1}(s_{1},s_{2},\vec{r}_{1},\vec{r}_{2};t)={\cal F}_{1,1}(s_{1},s_{2},\vec{r}_{1},\vec{r}_{2};t)
\]
Similarly,
\[
{\cal G}_{2,0}(s_{1},s_{2},\vec{r}_{1},\vec{r}_{2})={\cal F}_{2,0}(s_{1},s_{2},\vec{r}_{1},\vec{r}_{2})
\]
For order $(2,1)$ we have 
\begin{align*}
{\cal G}_{2,1}^{\prime}(s_{1},s_{2},s_{3},\vec{r}_{1},\vec{r}_{2},\vec{r}_{3};t) & =\left\langle C_{s_{1}}(\vec{r}_{1},0)C_{s_{2}}(\vec{r}_{2},0)C_{s_{3}}(\vec{r}_{3},t)\right\rangle \\
 & =\left\langle \delta C_{s_{1}}(\vec{r}_{1},0)\delta C_{s_{2}}(\vec{r}_{2},0)\delta C_{s_{3}}(\vec{r}_{3},t)\right\rangle \\
 & \hphantom{=\,}+\left\langle C_{s_{3}}\right\rangle \left\langle \delta C_{s_{1}}(\vec{r}_{1},0)\delta C_{s_{2}}(\vec{r}_{2},0)\right\rangle \\
 & \hphantom{=\,}+\left\langle C_{s_{2}}\right\rangle \left\langle \delta C_{s_{1}}(\vec{r}_{1},0)\delta C_{s_{3}}(\vec{r}_{3},t)\right\rangle \\
 & \hphantom{=\,}+\left\langle C_{s_{1}}\right\rangle \left\langle \delta C_{s_{2}}(\vec{r}_{2},0)\delta C_{s_{3}}(\vec{r}_{3},t)\right\rangle \\
 & \hphantom{=\,}+\left\langle C_{s_{1}}\right\rangle \left\langle C_{s_{2}}\right\rangle \left\langle C_{s_{3}}\right\rangle \\
 & ={\cal G}_{2,1}(s_{1},s_{2},s_{3},\vec{r}_{1},\vec{r}_{2},\vec{r}_{3};t)\\
 & \hphantom{=\,}+{\cal F}_{0,1}(s_{3},\vec{r}_{3}){\cal G}_{2,0}(s_{1},s_{2},\vec{r}_{1},\vec{r}_{2})\\
 & \hphantom{=\,}+{\cal F}_{1,0}(s_{2},\vec{r}_{2}){\cal G}_{1,1}(s_{1},s_{3},\vec{r}_{1},\vec{r}_{3};t)\\
 & \hphantom{=\,}+{\cal F}_{1,0}(s_{1},\vec{r}_{1}){\cal G}_{1,1}(s_{2},s_{3},\vec{r}_{2},\vec{r}_{3};t)\\
 & \hphantom{=\,}+{\cal F}_{1,0}(s_{1},\vec{r}_{1}){\cal F}_{1,0}(s_{2},\vec{r}_{2}){\cal F}_{0,1}(s_{3},\vec{r}_{3})
\end{align*}
Then substituting the already calculated ${\cal G}_{2,0}$ and ${\cal G}_{1,1}$
with ${\cal F}_{2,0}$ and ${\cal F}_{1,1}$ respectively and comparing
with (\ref{eq:calGp_21_calF}) we obtain
\[
{\cal G}_{2,1}(s_{1},s_{2},s_{3},\vec{r}_{1},\vec{r}_{2},\vec{r}_{3};t)={\cal F}_{2,1}(s_{1},s_{2},s_{3},\vec{r}_{1},\vec{r}_{2},\vec{r}_{3};t)
\]
and the procedure can continue to all higher orders.

However, one can see that this procedure is equivalent to shifting
the origin such that 
\begin{align*}
\left\langle C_{s}\right\rangle  & ={\cal F}_{1,0}(s,\vec{r})={\cal F}_{0,1}(s,\vec{r})\\
 & ={\cal G}_{1,0}^{\prime}(s,\vec{r})={\cal G}_{0,1}^{\prime}(s,\vec{r})\\
 & =0
\end{align*}
for all $s$, also 
\begin{align*}
\left\langle I(t)\right\rangle  & =F_{1,0}=F_{0,1}\\
 & =G_{1,0}^{\prime}=G_{0,1}^{\prime}\\
 & =0
\end{align*}
Therefore, the relations for central moments instead of moments are
obtained simply by dropping the terms that contain any $\left\langle C_{s}\right\rangle $
(any $s$) in (\ref{eq:calGp_11_result})--(\ref{eq:calGp_22_result}),
the terms that contain any ${\cal F}_{1,0}$ or ${\cal F}_{0,1}$
in (\ref{eq:calGp_11_calF})--(\ref{eq:calGp_22_calF_rel}), the
terms that contain any $F_{1,0}$ or $F_{0,1}$ in (\ref{eq:Gp_F_rel}),
and the terms that contain any $G_{1,0}^{\prime}$ or $G_{0,1}^{\prime}$
in (\ref{eq:F_Gp_rel}), and removing the primes on any remaining
${\cal G}$ and $G$ letters. The latter two sets of relations then
become the known relations between the central moments and the cumulants
of the bivariate distribution $\vec{I}=\left(I(0),I(t)\right)$:
\[
\begin{array}{ll}
G_{1,0} & =F_{1,0}\\
G_{2,0} & =F_{2,0}\\
G_{1,1}(t) & =F_{1,1}(t)\\
G_{2,1}(t) & =F_{2,1}(t)\\
G_{2,2}(t) & =F_{2,2}(t)+2F_{1,1}^{2}(t)+F_{2,0}F_{0,2}
\end{array}
\]
and
\[
F_{2,2}(t)=G_{2,2}(t)-2G_{1,1}^{2}(t)-G_{2,0}G_{0,2}
\]

\subsection{Melnykov-Hall approach}

This derivation exploits the additive property of cumulants. It was
first presented by Melnykov and Hall \cite{melnykov09} to derive
higher order correlations, following the work of Müller \cite{muller04}
in the context of Fluorescence Cumulant Analysis. We review the derivation
in the context of multi-state reacting and diffusing particles.

For a single particle we have (similar to (\ref{eq:sumbreak3}) but
without summation)
\begin{align}
\lefteqn{{\cal G}_{m,n}^{\prime(1)}(s_{1},\ldots,s_{m+n},\vec{r}_{1},\ldots,\vec{r}_{m+n};t)}\nonumber \\
 & \hphantom{\mathcal{G}_{m,n}^{\prime(1)}}=\delta(s_{1},s_{2})\delta(\vec{r}_{1}-\vec{r}_{2})\ldots\delta(s_{1},s_{m})\delta(\vec{r}_{1}-\vec{r}_{m})\nonumber \\
 & \hphantom{\mathcal{G}_{m,n}^{\prime(1)}=\,}\times\delta(s_{m+1},s_{m+2})\delta(\vec{r}_{m+1}-\vec{r}_{m+2})\ldots\delta(s_{m+1},s_{m+n})\delta(\vec{r}_{m+1}-\vec{r}_{m+n})\nonumber \\
 & \hphantom{\mathcal{G}_{m,n}^{\prime(1)}=\,}\times U^{(1)}(s_{1},s_{m+1},\vec{r}_{1},\vec{r}_{m+1};t)\label{eq:calGp_1_res}
\end{align}
where $U^{(1)}$ is given by (\ref{eq:U_1_res}). Substituting this
into (\ref{eq:mup_I}) we obtain, after summation and integration
of delta functions, the $(m,n)$th moment of the intensity vector
$\vec{I}^{(1)}=\left[I^{(1)}(0),I^{(1)}(t)\right]$ for a single particle:
\begin{align}
\mu_{m,n}^{\prime}[\vec{I}^{(1)}] & =\gamma_{m+n}X_{m,n}^{(1)}(t)Y_{m,n}(t)\label{eq:mom_I_single}
\end{align}
where we have defined:
\begin{equation}
\gamma_{k}=\frac{\int_{V}L^{k}(\vec{r})\mathrm{d}^{3}r}{\int_{V}L(\vec{r})\mathrm{d}^{3}r}\label{eq:gamma_def}
\end{equation}
the single-molecule reaction factor:
\[
X_{m,n}^{(1)}(t)=\sum_{s=1}^{J}\sum_{s^{\prime}=1}^{J}\frac{V_{\mathrm{MDF}}}{V}P(s)Q_{s}^{m}Q_{s^{\prime}}^{n}Z_{s^{\prime},s}(t)
\]
and the spatial factor:
\begin{equation}
Y_{m,n}(t)=\frac{1}{\gamma_{m+n}V_{\mathrm{MDF}}}\int_{V}\int_{V}L^{m}(\vec{r})L^{n}(\vec{r}^{\prime})\frac{\exp\left[-|\vec{r}-\vec{r}^{\prime}|^{2}/4Dt\right]}{(4\pi Dt)^{3/2}}\mathrm{d}^{3}r\mathrm{d}^{3}r^{\prime}\label{eq:Y_def}
\end{equation}
The volume of the molecular detection function, $V_{\mathrm{MDF}}$,
is defined in (\ref{eq:V_MDF}). 

Now consider the relation between the cumulant generating function
and the moment generating function of a bivariate distribution (see
(\ref{eq:mc-3-14-1})):
\begin{align*}
\sum_{m=0}^{\infty}\sum_{n=0}^{\infty}\kappa_{m,n}\frac{t_{1}^{m}t_{2}^{n}}{m!n!} & =\ln\left(\sum_{p=0}^{\infty}\sum_{q=0}^{\infty}\mu_{p,q}^{\prime}\frac{t_{1}^{p}t_{2}^{q}}{p!q!}\right)\\
 & =\sum_{\rho=1}^{\infty}\frac{(-1)^{\rho+1}}{\rho}\left(\sum_{p=0}^{\infty}\sum_{q=0}^{\infty}\mu_{p,q}^{\prime}\frac{t_{1}^{p}t_{2}^{q}}{p!q!}-1\right)^{\rho}\\
 & =\sum_{\rho=1}^{\infty}\frac{(-1)^{\rho+1}}{\rho}\left(\mathop{\sum_{p=0}^{\infty}\sum_{q=0}^{\infty}}_{p+q\geq1}\mu_{p,q}^{\prime}\frac{t_{1}^{p}t_{2}^{q}}{p!q!}\right)^{\rho}
\end{align*}
where we have used $\mu_{0,0}^{\prime}=1$. Through multinomial expansion
of the last expression and equating the powers of $t_{1}^{m}t_{2}^{n}$
on the two sides, we obtain:
\begin{equation}
\kappa_{m,n}=\sum\left(\frac{\mu_{p_{1},q_{1}}^{\prime}}{p_{1}!q_{1}!}\right)^{\pi_{p_{1},q_{1}}}\left(\frac{\mu_{p_{2},q_{2}}^{\prime}}{p_{2}!q_{2}!}\right)^{\pi_{p_{2},q_{2}}}\ldots\left(\frac{\mu_{p_{l},q_{l}}^{\prime}}{p_{l}!q_{l}!}\right)^{\pi_{p_{l},q_{l}}}\frac{(-1)^{\rho-1}(\rho-1)!m!n!}{\pi_{p_{1},q_{1}}!\pi_{p_{2},q_{2}}!\ldots\pi_{p_{l},q_{l}}!}\label{eq:kappa_mu_partition}
\end{equation}
where the summation is subject to the conditions (\ref{eq:p_pi_cond})
and (\ref{eq:q_pi_cond}), corresponding to integer partitioning of
the ordered pair $(m,n)$ into blocks of size $(p_{i},q_{i})$ and
multiplicity $\pi_{p_{i},q_{i}}$, and $\rho$ is the number of blocks
in each partition:
\[
\rho=\sum_{i=1}^{l}\pi_{p_{i},q_{i}}
\]
This shows that for the intensity vector $\vec{I}^{(1)}=\left[I^{(1)}(0),I^{(1)}(t)\right]$
arising from one particle, we have 
\[
\kappa_{m,n}[\vec{I}^{(1)}]=\mu_{m,n}^{\prime}[\vec{I}^{(1)}]+f(\mu_{p,q}^{\prime}[\vec{I}^{(1)}])
\]
where $f(\mu_{p,q}^{\prime})$ denotes a linear combination of the
\emph{products} of the moments of lower order. Now we use the additive
property of cumulants, (\ref{eq:cum_of_sum}), to write the cumulant
of the sum of $M$ independent intensity vectors, $\vec{I}=\sum_{j=1}^{M}\vec{I}_{j}^{(1)}$,
arising from $M$ independent molecules in the sample volume $V$:
\begin{align*}
\kappa_{m,n}(\vec{I}) & =M\kappa_{m,n}[\vec{I}^{(1)}]\\
 & =M\mu_{m,n}^{\prime}[\vec{I}^{(1)}]+Mf(\mu_{p,q}^{\prime}[\vec{I}^{(1)}])
\end{align*}
In the thermodynamic limit $M,V\to\infty$, this becomes
\begin{equation}
\kappa_{m,n}(\vec{I})=M\mu_{m,n}^{\prime}[\vec{I}^{(1)}]\label{eq:kappa_mup_1}
\end{equation}
which, upon substitution from (\ref{eq:mom_I_single}), yields
\begin{equation}
\kappa_{m,n}(\vec{I})=\gamma_{m+n}X_{m,n}(t)Y_{m,n}(t)\label{eq:kappa_res-I}
\end{equation}
and, using (\ref{eq:mup_W}),
\begin{equation}
\kappa_{m,n}(\vec{W})\approx T^{m+n}\gamma_{m+n}X_{m,n}(t)Y_{m,n}(t)\label{eq:kappa_res_W}
\end{equation}
where we have defined
\begin{equation}
X_{m,n}(t)=\sum_{s=1}^{J}\sum_{s^{\prime}=1}^{J}N_{s}Q_{s}^{m}Q_{s^{\prime}}^{n}Z_{s^{\prime},s}(t)\label{eq:X_def}
\end{equation}
and 
\[
N_{s}=\frac{V_{\mathrm{MDF}}}{V}P(s)M
\]
 (also defined in (\ref{eq:N_s})) is the expected number of molecules
in state $s$ in the observation volume.

The connection of this approach to that inspired by Palmer and Thompson
can be seen by comparing the key relation (\ref{eq:kappa_mup_1})
with 
\[
{\cal F}_{m,n}=M{\cal G}_{m,n}^{\prime(1)}
\]
which follows from (\ref{eq:calF_def}), (\ref{eq:calGp_1_res}),
and (\ref{eq:U_U1}).

\subsection[Normalized higher order correlations]{Normalized higher order correlations\protect\titlefootnote[*]{The content of this section is available in references \cite{abdollahnia16a,abdollahnia16b}
and is brought here for the sake of continuity.}}

Given the simpler analytical form of cumulants to study a system of
diffusing molecules in solution, we define normalized higher order
correlations, with $\vec{W}=\left[W(0),W(t)\right]$, as
\[
g_{m,n}(t)=\frac{\kappa_{m,n}(\vec{W})}{\kappa_{m,0}(\vec{W})\kappa_{0,n}(\vec{W})}
\]
which, in a multi-detector experiment (and/or sub-binning approach)
with $\vec{n}$ defined in (\ref{eq:vec_n_multi}), becomes
\begin{equation}
g_{m,n}(t)=\frac{\kappa_{\vec{1}_{m+n}}(\vec{n})}{\kappa_{\vec{1}_{m},\vec{0}_{n}}(\vec{n})\kappa_{\vec{0}_{m},\vec{1}_{n}}(\vec{n})}\label{eq:g_multi-detector}
\end{equation}
and in a single-detector experiment, with $\vec{n}_{\mathrm{1d}}=\left[n(0),n(t)\right]$,
becomes, as obtained by Melnykov and Hall\cite{melnykov09}, 
\begin{equation}
g_{m,n}(t)=\frac{\kappa_{[m,n]}(\vec{n}_{\mathrm{1d}})}{\kappa_{[m,0]}(\vec{n}_{\mathrm{1d}})\kappa_{[0,n]}(\vec{n}_{\mathrm{1d}})}\label{eq:g_single_detector}
\end{equation}
The two forms follow directly from~(\ref{eq:kappa_n1d_kappa_W})
and~(\ref{eq:kappa_n_kappa_W}):
\begin{equation}
\kappa_{\vec{1}_{m+n}}(\vec{n})=\kappa_{[m,n]}(\vec{n}_{\mathrm{1d}})=\kappa_{m,n}(\vec{W})\label{eq:kappa_n_kappa_n1d_kappa_W}
\end{equation}

Using (\ref{eq:kappa_res_W}) in the limit $T\to0$, we obtain the
simple analytical form
\begin{equation}
g_{m,n}(t)=\gamma_{m,n}R_{m,n}(t)Y_{m,n}(t)\label{eq:g_def}
\end{equation}
for the normalized correlation functions, where 
\[
\gamma_{m,n}=\frac{\gamma_{m+n}}{\gamma_{m}\gamma_{n}}
\]
$\gamma_{k}$ is defined in (\ref{eq:gamma_def}),

\begin{align*}
R_{m,n}(t) & =\frac{X_{m,n}(t)}{X_{m,0}X_{0,n}}\\
 & =\frac{\sum_{s=1}^{J}\sum_{s^{\prime}=1}^{J}N_{s}Q_{s}^{m}Q_{s^{\prime}}^{n}Z_{s^{\prime},s}(t)}{\left(\sum_{s=1}^{J}N_{s}Q_{s}^{m}\right)\left(\sum_{s^{\prime}=1}^{J}N_{s^{\prime}}Q_{s^{\prime}}^{n}\right)}
\end{align*}
and $Y_{m,n}(t)$ is defined in (\ref{eq:Y_def}). 

In practice, one of the parameters can be found from the mean channel
count rate, (\ref{eq:expctd_I}): 
\[
\left\langle I\right\rangle =\sum_{s=1}^{J}Q_{s}N_{s}
\]
Without loss of generality, it can be taken to be the first brightness
level $Q_{1}$. Thus a set of $\{N_{1},\ldots,N_{J}\}$ and $\{Q_{2},\ldots,Q_{J}\}$
remain to be determined, as well as the reaction relaxation times
which result from $Z_{s^{\prime},s}(t)$. 

Alternatively, we can define 
\[
N=\sum_{s=1}^{J}N_{s}
\]
as the total number of molecules in the probe region regardless of
their state. Then we have 
\[
R_{m,n}(t)=\frac{1}{N}\frac{\sum_{s=1}^{J}\sum_{s^{\prime}=1}^{J}K_{s}q_{s}^{m}q_{s^{\prime}}^{n}Z_{s^{\prime},s}(t)}{\left(\sum_{s=1}^{J}x_{s}q_{s}^{m}\right)\left(\sum_{s^{\prime}=1}^{J}x_{s^{\prime}}q_{s^{\prime}}^{n}\right)}
\]
where
\[
K_{s}=\frac{N_{s}}{N_{1}}
\]
and 
\[
q_{s}=\frac{Q_{s}}{Q_{1}}
\]
are the concentration (equilibrium constant) and brightness of state
$s$ relative to state 1. Obviously, $q_{1}=1$ and $K_{1}=1$. Thus
the number of independent parameters has not changed: $\{N,K_{2},\ldots,K_{J}\}$
and $\{q_{2},\ldots,q_{J}\}$ (again, plus the relaxation times).
In practice, the rate constants constructing $\mathbf{Z}(t)$ may
be more desirable to find. In that case, the system is inversely solved
to find the rate constants using the relations that link $K_{i}$
and relaxation times to the rate constants, and possibly the detailed
balance relations.

Here, the usefulness of the factorized form of (\ref{eq:g_def}) for
obtaining information on multi-state reactions becomes evident. The
factors $\gamma_{m,n}$ and $Y_{m,n}(t)$ depend only on the illumination
profile and the diffusion constant. Therefore, if the reaction parameters,
including rates, relative concentrations, and the relative brightness
values are of interest only, then higher order correlations $g_{m,n}^{(\mathrm{ref})}(t)$
from a ``reference'' sample with identical diffusional properties
can be used to extract the relative reaction function
\begin{align*}
R_{m,n}^{(\mathrm{rel})}(t) & =\frac{g_{m,n}(t)}{g_{m,n}^{(\mathrm{ref})}(t)}\\
 & =\frac{R_{m,n}(t)}{R_{m,n}^{(\mathrm{ref})}(t)}
\end{align*}
This eliminates the need to characterize the illumination profile
and calibrate the beam shape and diffusion parameters, greatly simplifying
the technique and making it more accurate. In practice, the reference
sample can consist of non-reacting molecules, or reacting molecules
labeled such as to remains in a single brightness state. In this case,
we simply have $R_{m,n}^{(\mathrm{ref})}(t)=1/N^{(\mathrm{ref})}$
and we get 
\begin{equation}
R_{m,n}^{(\mathrm{rel})}(t)=\frac{1}{N^{(\mathrm{rel})}}\frac{\sum_{s=1}^{J}\sum_{s^{\prime}=1}^{J}x_{s}q_{s}^{m}q_{s^{\prime}}^{n}Z_{s^{\prime},s}(t)}{\left(\sum_{s=1}^{J}x_{s}q_{s}^{m}\right)\left(\sum_{s^{\prime}=1}^{J}x_{s^{\prime}}q_{s^{\prime}}^{n}\right)}\label{eq:R_rel}
\end{equation}
where 
\begin{align*}
N^{(\mathrm{rel})} & =\frac{N}{N^{(\mathrm{ref})}}\\
 & =\frac{\left\langle C\right\rangle }{\left\langle C\right\rangle ^{(\mathrm{ref})}}
\end{align*}
The ratio of the concentration of the sample of interest (``test''
sample) to that of the reference sample, $N^{(\mathrm{rel})}$, can
be obtained either as a fitting parameter in higher-order FCS, or,
if possible, through independent techniques such as UV-Vis to reduce
the number of higher order correlations required. The values of the
absolute parameters $N$, $N^{(\mathrm{ref})}$, $N_{s}$ and $Q_{s}$
are usually of no general interest since they depend on the experimental
setup. However, they can be determined using $N^{(\mathrm{ref})}$,
measured for example by second-order FCS, and the mean detector count,
$\left\langle I\right\rangle $.

\section{Variance of correlations\label{sec:Variance-of-correlations}}

In a multi-detector (and/or sub-binning) approach we have 
\[
g_{m,n}(t)=\frac{\kappa_{\vec{1}_{m+n}}(\vec{n})}{\kappa_{\vec{1}_{m},\vec{0}_{n}}(\vec{n})\kappa_{\vec{0}_{m},\vec{1}_{n}}(\vec{n})}
\]
in a population, and 
\[
\tilde{g}_{m,n}(t)=\frac{k_{\vec{1}_{m+n}}(\vec{n})}{k_{\vec{1}_{m},\vec{0}_{n}}(\vec{n})k_{\vec{0}_{m},\vec{1}_{n}}(\vec{n})}
\]
in a sample. 

Using~(\ref{eq:K-9-1}), the sampling variance of $\tilde{g}_{m,n}$
is related to sampling moments of $k$-statistics: 
\begin{multline}
\mathrm{var}(\tilde{g}_{m,n})=\tilde{g}_{m,n}^{2}\left\{ \frac{\mathrm{var}(k_{\vec{1}_{m+n}})}{\kappa_{\vec{1}_{m+n}}^{2}}+\frac{\mathrm{var}(k_{\vec{1}_{m},\vec{0}_{n}})}{\kappa_{\vec{1}_{m},\vec{0}_{n}}^{2}}+\frac{\mathrm{var}(k_{\vec{0}_{m},\vec{1}_{n}})}{\kappa_{\vec{0}_{m},\vec{1}_{n}}^{2}}-2\frac{\mathrm{cov}(k_{\vec{1}_{m+n}},k_{\vec{1}_{m},\vec{0}_{n}})}{\kappa_{\vec{1}_{m+n}}\kappa_{\vec{1}_{m},\vec{0}_{n}}}\right.\\
\left.-2\frac{\mathrm{cov}(k_{\vec{1}_{m+n}},k_{\vec{0}_{m},\vec{1}_{n}})}{\kappa_{\vec{1}_{m+n}}\kappa_{\vec{0}_{m},\vec{1}_{n}}}+2\frac{\mathrm{cov}(k_{\vec{1}_{m},\vec{0}_{n}},k_{\vec{0}_{m},\vec{1}_{n}})}{\kappa_{\vec{1}_{m},\vec{0}_{n}}\kappa_{\vec{0}_{m},\vec{1}_{n}}}\right\} \label{eq:var_gmn_tilde}
\end{multline}
Unless specified otherwise, the cumulants and $k$-statistics are
of random vector $\vec{n}$ (the multi-channel photon count) defined
as 
\[
\vec{n}=\left[n_{1}(0),n_{2}(0),\ldots,n_{p}(0),n_{1}(t),n_{2}(t),\ldots,n_{q}(t)\right]
\]

The signal-to-noise ratio is defined as
\[
\mathrm{SNR}_{m,n}=\sqrt{\frac{\tilde{g}_{m,n}^{2}}{\var(\tilde{g}_{m,n})}}
\]

\subsection{Order (1,1)}

To calculate the sampling variance of 
\[
\tilde{g}_{11}=\frac{k_{11}}{k_{10}k_{01}}
\]
equation~(\ref{eq:var_gmn_tilde}) becomes 
\begin{multline}
\mathrm{var}(\tilde{g}_{11})=\tilde{g}_{11}^{2}\left\{ \frac{\mathrm{var}(k_{11})}{\kappa_{11}^{2}}+\frac{\mathrm{var}(k_{10})}{\kappa_{10}^{2}}+\frac{\mathrm{var}(k_{01})}{\kappa_{01}^{2}}-2\frac{\mathrm{cov}(k_{11},k_{10})}{\kappa_{11}\kappa_{10}}\right.\\
\left.-2\frac{\mathrm{cov}(k_{11},k_{01})}{\kappa_{11}\kappa_{01}}+2\frac{\mathrm{cov}(k_{10},k_{01})}{\kappa_{10}\kappa_{01}}\right\} \label{eq:K-21-1-1}
\end{multline}

We need to find 
\begin{align*}
 & \mathrm{var}(k_{11})=\kappa\left(\begin{array}{cc}
1 & 1\\
1 & 1
\end{array}\right)\\
 & \mathrm{var}(k_{10})=\kappa\left(\begin{array}{cc}
1 & 1\\
0 & 0
\end{array}\right)\\
 & \mathrm{var}(k_{01})=\kappa\left(\begin{array}{cc}
0 & 0\\
1 & 1
\end{array}\right)\\
 & \mathrm{cov}(k_{11},k_{10})=\kappa\left(\begin{array}{cc}
1 & 1\\
1 & 0
\end{array}\right)\\
 & \mathrm{cov}(k_{11},k_{01})=\kappa\left(\begin{array}{cc}
1 & 0\\
1 & 1
\end{array}\right)\\
 & \mathrm{cov}(k_{10},k_{01})=\kappa\left(\begin{array}{cc}
1 & 0\\
0 & 1
\end{array}\right)
\end{align*}

By definition
\[
k_{1}=m_{1}^{\prime}
\]
and as shown in~(\ref{eq:K-4-1})
\[
\mathrm{var}(m_{r}^{\prime})=\frac{1}{n}(\mu_{2r}^{\prime}-\mu_{r}^{\prime2})
\]
Thus
\[
\mathrm{var}(k_{1})=\mathrm{var}(m_{1}^{\prime})=\frac{\mu_{2}}{n}=\frac{\kappa_{2}}{n}
\]
Now we can add a zero subscript to both sides:
\begin{equation}
\begin{aligned}\mathrm{var}(k_{10}) & =\frac{\kappa_{20}}{n}\\
\mathrm{var}(k_{01}) & =\frac{\kappa_{02}}{n}
\end{aligned}
\label{eq:K-19-1}
\end{equation}
Here we will use the formulae written in ``tensor notation'' by
Kaplan\cite{kaplan52} (also listed in~\cite{kendall94}). In tensor
notation, 
\[
\kappa(i,j)=\kappa\left(\begin{array}{cc}
1 & 0\\
0 & 1
\end{array}\right)
\]
\[
\kappa(i,kl)=\kappa\left(\begin{array}{cc}
1 & 0\\
0 & 1\\
0 & 1
\end{array}\right)
\]
and so forth.

As a simple formula in this notation (see~(\ref{eq:kappa_ij})),
\[
\kappa(i,j)=\frac{\kappa_{ij}}{n}
\]
 which means
\begin{equation}
\kappa\left(\begin{array}{cc}
1 & 0\\
0 & 1
\end{array}\right)=\frac{\kappa_{11}}{n}\label{eq:K-19-2}
\end{equation}
As another useful formula (see~(\ref{eq:kappa_ikl})):
\[
\kappa(i,kl)=\frac{\kappa_{ikl}}{n}
\]
meaning
\[
\kappa\left(\begin{array}{cc}
1 & 0\\
0 & 1\\
0 & 1
\end{array}\right)=\frac{\kappa_{111}}{n}
\]
Merging (adding and replacing) the first two rows (i.e. the first
two variates are identical) we get
\begin{equation}
\begin{aligned}\kappa\left(\begin{array}{cc}
1 & 1\\
1 & 0
\end{array}\right) & =\kappa\left(\begin{array}{cc}
1 & 1\\
0 & 1
\end{array}\right)=\frac{\kappa_{21}}{n}\end{aligned}
\label{eq:K-19-3}
\end{equation}
Multivariate cumulants are symmetric functions, thus we can exchange
the matrix columns.

In a similar way we have:
\[
\kappa(ij,k)=\frac{\kappa_{ijk}}{n}
\]
which means
\begin{equation}
\kappa\left(\begin{array}{cc}
1 & 0\\
1 & 0\\
0 & 1
\end{array}\right)=\frac{\kappa_{111}}{n}\label{eq:K-20-1-0}
\end{equation}
Merging the last two rows
\begin{equation}
\begin{aligned}\kappa\left(\begin{array}{cc}
1 & 0\\
1 & 1
\end{array}\right) & =\kappa\left(\begin{array}{cc}
0 & 1\\
1 & 1
\end{array}\right)=\frac{\kappa_{12}}{n}\end{aligned}
\label{eq:K-20-1}
\end{equation}
We could also obtain this by exchanging the subscripts in~(\ref{eq:K-19-3}),
however, the intermediate relation will be useful~(\ref{eq:K-20-1-0})
later.

Kaplan's Equation (3) reads:
\[
\kappa(ab,ij)=\frac{1}{n}\kappa_{abij}+\frac{1}{n-1}\sum^{2}\kappa_{ai}\kappa_{bj}
\]
In the summation, $a$ and $b$ indicate whichever of the first two
subscripts that are nonzero. $i$ and $j$ indicate whichever of the
last two subscripts that are nonzero. All other subscripts in each
$\kappa$ must be zero. $a$, $b$, $i$ and $j$ must be non-identical.
The summation superscript $2$ indicates that there are two distinct
terms of such kind. Thus we have:
\begin{equation}
\kappa\left(\begin{array}{cc}
1 & 0\\
1 & 0\\
0 & 1\\
0 & 1
\end{array}\right)=\frac{1}{n}\kappa_{1111}+\frac{1}{n-1}(\kappa_{1010}\kappa_{0101}+\kappa_{0110}\kappa_{1001})\label{eq:K-21-0}
\end{equation}
Merging row $1$ with row $3$, and row $2$ with row $4$, we get
\begin{equation}
\kappa\left(\begin{array}{cc}
1 & 1\\
1 & 1
\end{array}\right)=\frac{1}{n}\kappa_{22}+\frac{1}{n-1}(\kappa_{20}\kappa_{02}+\kappa_{11}^{2})\label{eq:K-21-1}
\end{equation}
The right hand side is in terms of cumulants of $\vec{n}$ (the multi-channel
photon count) which have not been calculated, while the cumulants
of $\vec{W}=\left[W(0),W(t)\right]$ (the underlying integrated intensity)
have been calculated in~(\ref{eq:kappa_res_W}): 
\begin{equation}
\kappa_{m,n}(\vec{W})=T^{m+n}\gamma_{m+n}X_{m,n}(t)Y_{m,n}(t)\label{eq:K-21-2}
\end{equation}
To simplify matters, we only consider the shot-noise-dominant limit
$t\to0$ in which $Y_{m,n}(0)=1$ and 
\[
X_{m,n}(0)=\sum_{s=1}^{J}N_{s}Q_{s}^{m+n}
\]
and to further simplify, we only consider a single species, thus:
\[
X_{m,n}(0)=NQ^{m+n}
\]
Therefore~(\ref{eq:K-21-2}) becomes
\begin{equation}
\kappa_{m,n}(\vec{W})=\gamma_{m+n}N\lambda^{m+n}\label{eq:K-21-3}
\end{equation}
where $\lambda:=QT$ is the average number of photons per bin. Here,
we assume the shot-noise-dominant regime, that is
\[
\lambda\ll1
\]

On the other hand, factorial cumulants of $\vec{n}$ are equal to
cumulants of $\vec{W}$, by~(\ref{eq:fc_n_c_W}):
\begin{equation}
\kappa_{[\vec{r}]}(\vec{n})=\kappa_{m,n}(\vec{W})\label{eq:K-21-3-2}
\end{equation}
where $\vec{r}$ can have elements greater than $1$, and $(m,n)$
is constructed by merging the elements of $\vec{r}$ that are at identical
lag times (i.e. identical $W(t)$). Therefore, to calculate the cumulants
of $\vec{n}$, as in~(\ref{eq:K-21-1}), we express them in terms
of factorial cumulants of $\vec{n}$. The conversion relations are
formally similar to those between moments and factorial moments, and
are given in~(\ref{eq:c_fc_list}). Using those, and~(\ref{eq:K-21-1})
in the limit $n\gg1$, we write 
\begin{align*}
\frac{\var(k_{11})}{\kappa_{11}^{2}} & =\frac{1}{n}(1+\frac{\kappa_{20}\kappa_{02}}{\kappa_{11}^{2}}+\frac{\kappa_{22}}{\kappa_{11}^{2}})\\
 & =\frac{1}{n}(1+\frac{(\kappa_{[20]}+\kappa_{[10]})(\kappa_{[02]}+\kappa_{[01]})}{\kappa_{[11]}^{2}}+\frac{\kappa_{[22]}+\kappa_{[21]}+\kappa_{[12]}+\kappa_{[11]}}{\kappa_{[11]}^{2}})\\
\intertext{\text{then using\,\eqref{eq:K-21-3} and\,\eqref{eq:K-21-3-2}}} & =\frac{1}{n}(2+2\frac{\gamma_{2}\gamma_{1}}{\gamma_{2}^{2}\lambda}+\frac{\gamma_{1}^{2}}{\gamma_{2}^{2}\lambda^{2}}+\frac{\gamma_{4}}{\gamma_{2}^{2}N}+2\frac{\gamma_{3}}{\gamma_{2}^{2}N\lambda}+\frac{1}{\gamma_{2}N\lambda^{2}})\\
 & =\frac{1}{n}(\frac{N+\gamma_{2}}{\gamma_{2}^{2}N\lambda^{2}})[1+\mathcal{O}(\lambda)]
\end{align*}
having used $\gamma_{1}=1$ .

Following a similar approach, one can show that the order difference
between the denominator and the numerator does not exceed $1$ in
any other term in~(\ref{eq:K-21-1-1}) when written in the form of
factorial cumulants, thus they can all be neglected, and the SNR becomes
\begin{equation}
\mathrm{SNR}_{11}\approx\sqrt{n}\gamma_{2}QT\sqrt{\frac{N}{N+\gamma_{2}}}\label{eq:SNR11}
\end{equation}

\subsection{Order (2,1)}

For the sampling variance of 
\[
\tilde{g}_{21}=\frac{k_{111}}{k_{110}k_{001}}
\]
we need to calculate
\begin{align*}
 & \mathrm{var}(k_{111})=\kappa\left(\begin{array}{cc}
1 & 1\\
1 & 1\\
1 & 1
\end{array}\right)\\
 & \mathrm{var}(k_{110})=\kappa\left(\begin{array}{cc}
1 & 1\\
1 & 1\\
0 & 0
\end{array}\right)\\
 & \mathrm{var}(k_{001})=\kappa\left(\begin{array}{cc}
0 & 0\\
0 & 0\\
1 & 1
\end{array}\right)\\
 & \mathrm{cov}(k_{111},k_{110})=\kappa\left(\begin{array}{cc}
1 & 1\\
1 & 1\\
1 & 0
\end{array}\right)\\
 & \mathrm{cov}(k_{111},k_{001})=\kappa\left(\begin{array}{cc}
1 & 0\\
1 & 0\\
1 & 1
\end{array}\right)\\
 & \mathrm{cov}(k_{110},k_{001})=\kappa\left(\begin{array}{cc}
1 & 0\\
1 & 0\\
0 & 1
\end{array}\right)
\end{align*}

Extending the univariate case by adding a neutral $0$ subscript,
as in~(\ref{eq:K-19-1}),
\begin{align}
\var(k_{001}) & =\frac{\kappa_{002}}{n}\nonumber \\
 & =\frac{\kappa_{[001]}}{n}+\mathcal{O}(\lambda^{2})\label{eq:K-24-1}
\end{align}
and extending the bivariate case~(\ref{eq:K-21-1}) 
\begin{align}
\var(\kappa_{110}) & =\frac{1}{n}\kappa_{220}+\frac{1}{n-1}(\kappa_{200}\kappa_{020}+\kappa_{110}^{2})\nonumber \\
 & =\frac{1}{n}\kappa_{[110]}+\frac{1}{n-1}\kappa_{[100]}\kappa_{[010]}+\mathcal{O}(\lambda^{3})\label{eq:K-24-2}
\end{align}
where we have kept the lowest-order term(s) in conversion to factorial
cumulants for later use.

Using the rule of Appendix~\ref{sec:tensor-rule} again we have:
\[
\kappa(abc,i)=\frac{\kappa_{abci}}{n}
\]
meaning
\[
\kappa\left(\begin{array}{cc}
1 & 0\\
1 & 0\\
1 & 0\\
0 & 1
\end{array}\right)=\frac{\kappa_{1111}}{n}
\]
Merging the last two rows:
\begin{align}
\kappa\left(\begin{array}{cc}
1 & 0\\
1 & 0\\
1 & 1
\end{array}\right) & =\frac{\kappa_{112}}{n}\nonumber \\
 & =\frac{1}{n}\kappa_{[111]}+\mathcal{O}(\lambda^{4})\label{eq:K-24-3}
\end{align}
And, $\mathrm{cov}(k_{110},k_{001})$ is already calculated in~(\ref{eq:K-20-1-0}).

Kaplan's Equation~(4) reads: 
\[
\kappa(ab,ijk)=\frac{1}{n}\kappa_{abijk}+\frac{1}{n-1}\sum^{6}\kappa_{ai}\kappa_{bjk}
\]
which is a shorthand notation for 
\[
\kappa\left(\begin{array}{cc}
1 & 0\\
1 & 0\\
0 & 1\\
0 & 1\\
0 & 1
\end{array}\right)=\frac{\kappa_{11111}}{n}+\frac{1}{n-1}\sum\left(\begin{array}{c}
\kappa_{10100}\kappa_{01011}\\
\kappa_{10010}\kappa_{01101}\\
\kappa_{10001}\kappa_{01110}\\
\kappa_{01100}\kappa_{10011}\\
\kappa_{01010}\kappa_{10101}\\
\kappa_{01001}\kappa_{10110}
\end{array}\right)
\]
where the matrix in front of the summation lists the summed terms.
To obtain $\mathrm{cov}(k_{110},k_{111})$, we merge the rows according
to a ``subscript identicality pattern'' of the form $\{1,2,1,2,3\}$,
which means the first and the third rows (subscripts) are merged together
and used as the new first row, the second and the fourth rows are
merged together to form the new second row, and the fifth row forms
the new third row as is. This yields 
\begin{align}
\kappa\left(\begin{array}{cc}
1 & 1\\
1 & 1\\
0 & 1
\end{array}\right) & =\frac{\kappa_{221}}{n}+\frac{1}{n-1}\sum\left(\begin{array}{c}
\kappa_{200}\kappa_{021}\\
\kappa_{110}\kappa_{111}\\
\kappa_{101}\kappa_{120}\\
\kappa_{110}\kappa_{111}\\
\kappa_{020}\kappa_{201}\\
\kappa_{011}\kappa_{210}
\end{array}\right)\label{eq:K-25-2-0}\\
 & =\frac{\kappa_{[111]}}{n}+\frac{1}{n-1}(\kappa_{[100]}\kappa_{[011]}+\kappa_{[010]}\kappa_{[101]})+\mathcal{O}(\lambda^{4})\label{eq:K-25-2}
\end{align}
Upon conversion to factorial cumulants, a $\kappa$ with a subscript
of, say, $021$ breaks down to $[021]$ plus $[011]$, and we care
about the lowest-order terms in $\lambda$ only.

Using Kaplan's Equation (7):
\begin{multline*}
\kappa(abc,ijk)=\frac{\kappa_{abcijk}}{n}+\frac{1}{n-1}(\sum^{9}\kappa_{ai}\kappa_{bcjk}+\sum^{9}\kappa_{abi}\kappa_{cjk})\\
+\frac{n}{(n-1)(n-2)}\sum^{6}\kappa_{ai}\kappa_{bj}\kappa_{ck}
\end{multline*}
which becomes
\begin{align*}
\kappa\left(\begin{array}{cc}
1 & 0\\
1 & 0\\
1 & 0\\
0 & 1\\
0 & 1\\
0 & 1
\end{array}\right) & =\frac{\kappa_{111111}}{n}+\frac{1}{n-1}\sum^{9}\left(\begin{array}{c}
\kappa_{100100}\kappa_{011011}\\
\vdots\\
\kappa_{001001}\kappa_{110110}
\end{array}\right)\\
 & \hphantom{=\frac{\kappa_{111111}}{n}}+\frac{1}{n-1}\sum^{9}\left(\begin{array}{c}
\kappa_{110100}\kappa_{001011}\\
\vdots\\
\kappa_{011001}\kappa_{100110}
\end{array}\right)\\
 & \hphantom{=\frac{\kappa_{111111}}{n}}+\frac{n}{(n-1)(n-2)}\sum^{6}\left(\begin{array}{c}
\kappa_{100100}\kappa_{010010}\kappa_{001001}\\
\vdots\\
\kappa_{100001}\kappa_{010010}\kappa_{001100}
\end{array}\right)
\end{align*}
The procedure can be computerized to produce all terms of a particular
form. The program is included in Supporting Information. 

Using the identicality pattern $\{123123\}$, then converting to factorial
cumulants and keeping the lowest-order terms: 
\begin{align}
\kappa\left(\begin{array}{cc}
1 & 1\\
1 & 1\\
1 & 1
\end{array}\right) & =\frac{\kappa_{222}}{n}+\frac{1}{n-1}\sum^{9}\left(\begin{array}{c}
\kappa_{200}\kappa_{022}\\
\vdots\\
\kappa_{002}\kappa_{220}
\end{array}\right)+\frac{1}{n-1}\sum^{9}\left(\begin{array}{c}
\kappa_{210}\kappa_{012}\\
\vdots\\
\kappa_{012}\kappa_{210}
\end{array}\right)\nonumber \\
 & \hphantom{=\frac{\kappa_{222}}{n}}+\frac{n}{(n-1)(n-2)}\sum^{6}\left(\begin{array}{c}
\kappa_{200}\kappa_{020}\kappa_{002}\\
\vdots\\
\kappa_{101}\kappa_{020}\kappa_{101}
\end{array}\right)\label{eq:K-26-0}\\
 & =\frac{\kappa_{[111]}}{n}+\frac{1}{n-1}(\kappa_{[100]}\kappa_{[011]}+\kappa_{[010]}\kappa_{[101]}+\kappa_{[001]}\kappa_{[110]})\nonumber \\
 & \hphantom{=\frac{\kappa_{[111]}}{n}}+\frac{n}{(n-1)(n-2)}\kappa_{[100]}\kappa_{[010]}\kappa_{[001]}+\mathcal{O}(\lambda^{4})\label{eq:K-26-1}
\end{align}
We are now ready to evaluate the following terms: \\
From~(\ref{eq:K-24-1}):
\[
\frac{\var(k_{001})}{\kappa_{001}^{2}}=\frac{\mathcal{O}(\lambda)}{\mathcal{O}(\lambda^{2})}=\mathcal{O}(\lambda^{-1})
\]
From~(\ref{eq:K-24-2}):
\[
\frac{\var(k_{110})}{\kappa_{110}^{2}}=\frac{\mathcal{O}(\lambda^{2})}{\mathcal{O}(\lambda^{4})}=\mathcal{O}(\lambda^{-2})
\]
From~(\ref{eq:K-26-1}):
\[
\frac{\var(k_{111})}{\kappa_{111}^{2}}=\frac{\mathcal{O}(\lambda^{3})}{\mathcal{O}(\lambda^{6})}=\mathcal{O}(\lambda^{-3})
\]
From~(\ref{eq:K-24-3}):
\[
\frac{\mathrm{cov}(k_{111},k_{001})}{\kappa_{111}\kappa_{001}}=\frac{\mathcal{O}(\lambda^{3})}{\mathcal{O}(\lambda^{4})}=\mathcal{O}(\lambda^{-1})
\]
From~(\ref{eq:K-25-2}):
\[
\frac{\mathrm{cov}(k_{111},k_{110})}{\kappa_{111}\kappa_{110}}=\frac{\mathcal{O}(\lambda^{3})}{\mathcal{O}(\lambda^{5})}=\mathcal{O}(\lambda^{-2})
\]
From~(\ref{eq:K-20-1-0}):
\[
\frac{\mathrm{cov}(k_{110},k_{001})}{\kappa_{110}\kappa_{001}}=\frac{\mathcal{O}(\lambda^{3})}{\mathcal{O}(\lambda^{3})}=\mathcal{O}(\lambda^{0})
\]
Therefore the largest oder in $\frac{1}{\lambda}$ is the term $\frac{\var(k_{111})}{\kappa_{111}^{2}}$
which dominates the relative uncertainty. This was expected, as the
$k$-statistics of higher order have larger relative uncertainties
in the shot-noise-dominated regime. Thus, using~(\ref{eq:K-26-1})
\begin{align*}
\frac{\var(k_{111})}{\kappa_{111}^{2}} & =\frac{\frac{1}{n}(\gamma_{3}N\lambda^{3}+3\gamma_{1}\gamma_{2}N^{2}\lambda^{3}+\gamma_{1}^{3}N^{3}\lambda^{3})+\mathcal{O}(\lambda^{4})}{(\gamma_{3}N\lambda^{3})^{2}}\\
 & \approx\frac{1}{n\gamma_{3}^{2}\lambda^{3}}(\frac{\gamma_{3}+3\gamma_{2}N+N^{2}}{N})
\end{align*}
Finally,
\begin{equation}
\mathrm{SNR}_{21}\approx\sqrt{n}\gamma_{3}(QT)^{3/2}\sqrt{\frac{N}{\gamma_{3}+3\gamma_{2}N+N^{2}}}\label{eq:SNR21}
\end{equation}
Comparing with the single-detector result reported by Melnykov and
Hall \cite{melnykov09},
\[
\mathrm{SNR}_{21}^{\mathrm{(1d)}}\approx\sqrt{n}\gamma_{3}(QT)^{3/2}\sqrt{\frac{N}{2\gamma_{3}+6\gamma_{2}N+2N^{2}}}
\]
we see that the presence of two independent channels contributes positively
by a factor of $\sqrt{2}$ . This is intuitive by considering the
fact that if $x,x_{1},x_{2},\ldots,x_{p}$ are independent random
variables with identical distribution, we have 
\[
\frac{\delta(x^{p})}{x^{p}}=p\frac{\delta x}{x}
\]
\[
\frac{\delta(x_{1}x_{2}\ldots x_{p})}{x_{1}x_{2}\ldots x_{p}}=\sqrt{p}\frac{\delta x}{x}
\]
However, it should also be noted that splitting the beam among $m$
detectors is equivalent to reducing the brightness $Q$ by a factor
of $m$. The SNR has a power of $(QT)^{1/2}$ for each higher order.
Also, sub-binning reduces the effective bin size, $T$. Combining
multiple ways to selection sub-bins partially makes up for the loss.
Therefore, comparing the overall single-detector and multi-detector
SNR requires more detailed analysis for each order. An optimally large
bin size usually ensures sufficient SNR in each method.

Before we finish this section, we point out that the single-detector
$\mathrm{SNR}_{21}^{\mathrm{(1d)}}$ can be obtained from 
\[
\frac{\var(k_{[21]})}{\kappa_{[21]}^{2}}
\]
 where the random variables are $\vec{n}_{\mathrm{1d}}=\left[n(0),n(t)\right]$.
The denominator is the same as in multi-detector case. To calculate
the numerator, we use 
\[
k_{[21]}=k_{21}-k_{11}
\]
which yields
\[
\var(k_{[21]})=\var(k_{21})+\var(k_{11})-2\mathrm{cov}(k_{21},k_{11})
\]
We can obtain $\var(k_{21})$ by merging the first two subscripts
in the result for $\var(k_{111})$ (Equation~(\ref{eq:K-26-0})).
Similarly, $\mathrm{cov}(k_{21},k_{11})$ is obtained by merging the
last two subscripts in $\mathrm{cov}(k_{111},k_{110})$ (Equation~(\ref{eq:K-25-2-0}))
and swapping the subscripts. The result then has to be converted back
to factorial cumulants and the lowest-order terms kept to obtain $\mathrm{SNR}_{21}^{\mathrm{(1d)}}$.

\subsection{Order (2,2)\label{subsec:Order-(2,2)}}

Now for the sampling variance of 
\[
\tilde{g}_{22}=\frac{\kappa_{1111}}{\kappa_{1100}\kappa_{0011}}
\]
we need to calculate 
\begin{align*}
 & \mathrm{var}(k_{1111})\\
 & \mathrm{var}(k_{1100})\\
 & \mathrm{var}(k_{0011})\\
 & \mathrm{cov}(k_{1111},k_{1100})\\
 & \mathrm{cov}(k_{1111},k_{0011})\\
 & \mathrm{cov}(k_{1100},k_{0011})
\end{align*}
Simply by rewriting~(\ref{eq:K-24-2}) with a neutral 4th subscript,
we have
\[
\begin{aligned}\var(k_{1100}) & =\frac{1}{n}\kappa_{2200}+\frac{1}{n-1}(\kappa_{2000}\kappa_{0200}+\kappa_{1100}^{2})\\
 & =\frac{1}{n}\kappa_{[1100]}+\frac{1}{n-1}\kappa_{[1000]}\kappa_{[0100]}+\mathcal{O}(\lambda^{3})
\end{aligned}
\]
and $\var(k_{0011})$ is obtained by swapping the first two and last
two indices. 

Also, $\var(k_{1100},k_{0011})$ is already given in~(\ref{eq:K-21-0}),
which has the same factorial form.

We invoke Kaplan's Equation (5)
\[
\kappa(ab,ijkl)=\frac{\kappa_{abijkl}}{n}+\frac{1}{n-1}(\sum^{8}\kappa_{ai}\kappa_{bjkl}+\sum^{6}\kappa_{aij}\kappa_{bkl})
\]
which yields
\begin{align*}
\mathrm{cov}(k_{110000},k_{001111}) & =\frac{\kappa_{111111}}{n}+\frac{1}{n-1}\sum^{8}\left(\begin{array}{c}
\kappa_{101000}\kappa_{010111}\\
\vdots\\
\kappa_{010001}\kappa_{101110}
\end{array}\right)\\
 & \hphantom{=\frac{\kappa_{111111}}{n}}+\frac{1}{n-1}\sum^{6}\left(\begin{array}{c}
\kappa_{101100}\kappa_{010011}\\
\vdots\\
\kappa_{100011}\kappa_{011100}
\end{array}\right)
\end{align*}
With subscript identicality pattern $\{1,2,1,2,3,4\}$ we get
\begin{align*}
\mathrm{cov}(k_{1100},k_{1111}) & =\frac{\kappa_{2211}}{n}+\frac{1}{n-1}\{\sum^{8}\left(\begin{array}{c}
\kappa_{2000}\kappa_{0211}\\
\vdots\\
\kappa_{0101}\kappa_{2110}
\end{array}\right)+\sum^{6}\left(\begin{array}{c}
\kappa_{2100}\kappa_{0111}\\
\vdots\\
\kappa_{1011}\kappa_{1200}
\end{array}\right)\}\\
 & =\frac{\kappa_{[1111]}}{n}+\frac{1}{n-1}(\kappa_{[1000]}\kappa_{[0111]}+\kappa_{[0100]}\kappa_{[1011]})\\
 & \hphantom{=\,\,\frac{\kappa_{[1111]}}{n}}+\frac{1}{n-1}(\kappa_{[1010]}\kappa_{[0101]}+\kappa_{[1001]}\kappa_{[0110]})+\mathcal{O}(\lambda^{5})
\end{align*}
Similarly, $\mathrm{cov}(k_{1111},k_{0011})$ is of order $\mathcal{O}(\lambda^{4})$.

To calculate $\var(k_{1111})$ we use Kaplan's Equation~(10), (with
a correction in the number of terms of $\kappa_{abi}\kappa_{cdjkl}$):
\[
\kappa(abcd,ijkl)=\frac{\kappa_{abcdijkl}}{n}+\frac{\sum^{16}\kappa_{ai}\kappa_{bcdjkl}}{n-1}+\frac{\sum^{24}\kappa_{abi}\kappa_{cdjkl}}{n-1}+\ldots
\]
This directly yields
\begin{align*}
\mathrm{cov}(k_{11110000},k_{00001111}) & =\frac{\kappa_{11111111}}{n}\\
 & \hphantom{=}+\frac{1}{n-1}\sum^{16}\left(\begin{array}{c}
\kappa_{10001000}\kappa_{01110111}\\
\vdots\\
\kappa_{00010001}\kappa_{11101110}
\end{array}\right)\\
 & \hphantom{=}+\frac{1}{n-1}\sum^{24}\left(\begin{array}{c}
\kappa_{11001000}\kappa_{00110111}\\
\vdots\\
\kappa_{00110001}\kappa_{11001110}
\end{array}\right)\\
 & \hphantom{=}+\frac{n}{(n-1)(n-2)}\sum^{72}\left(\begin{array}{c}
\kappa_{10001000}\kappa_{01000100}\kappa_{00110011}\\
\vdots\\
\kappa_{00100001}\kappa_{00010010}\kappa_{11001100}
\end{array}\right)\\
 & \hphantom{=}+\frac{1}{n-1}\sum^{16}\left(\begin{array}{c}
\kappa_{10001110}\kappa_{01110001}\\
\vdots\\
\kappa_{00010111}\kappa_{11101000}
\end{array}\right)\\
 & \hphantom{=}+\frac{1}{n-1}\sum^{18}\left(\begin{array}{c}
\kappa_{11001100}\kappa_{00110011}\\
\vdots\\
\kappa_{10010011}\kappa_{01101100}
\end{array}\right)\\
 & \hphantom{=}+\frac{n}{(n-1)(n-2)}\sum^{144}\left(\begin{array}{c}
\kappa_{1001000}\kappa_{01100100}\kappa_{00010011}\\
\vdots\\
\kappa_{00010001}\kappa_{01100010}\kappa_{10001100}
\end{array}\right)\\
 & \hphantom{=}+\frac{n(n+1)}{(n-1)(n-2)(n-3)}\sum^{24}\left(\begin{array}{c}
\kappa_{10001000}\kappa_{01000100}\kappa_{00100010}\kappa_{00010001}\\
\vdots\\
\kappa_{10000001}\kappa_{01000010}\kappa_{00100100}\kappa_{00011000}
\end{array}\right)
\end{align*}
With the identicality pattern $\{1,2,3,4,1,2,3,4\}$,
\begin{align*}
\var(k_{1111}) & =\frac{\kappa_{2222}}{n}+\frac{1}{n-1}\sum^{16}\left(\begin{array}{c}
\kappa_{2000}\kappa_{0222}\\
\vdots\\
\kappa_{0002}\kappa_{2220}
\end{array}\right)\\
 & \hphantom{=\,\,\frac{\kappa_{2222}}{n}}+\frac{1}{n-1}\sum^{24}\left(\begin{array}{c}
\kappa_{2100}\kappa_{0122}\\
\vdots\\
\kappa_{0012}\kappa_{2210}
\end{array}\right)\\
 & \hphantom{=\,\,\frac{\kappa_{2222}}{n}}+\frac{n}{(n-1)(n-2)}\sum^{72}\left(\begin{array}{c}
\kappa_{2000}\kappa_{0200}\kappa_{0022}\\
\vdots\\
\kappa_{0011}\kappa_{0011}\kappa_{2200}
\end{array}\right)\\
 & \hphantom{=\,\,}+\frac{1}{n-1}\sum^{16}\left(\begin{array}{c}
\kappa_{2110}\kappa_{0112}\\
\vdots\\
\kappa_{0112}\kappa_{2110}
\end{array}\right)+\frac{1}{n-1}\sum^{18}\left(\begin{array}{c}
\kappa_{2200}\kappa_{0022}\\
\vdots\\
\kappa_{1012}\kappa_{1210}
\end{array}\right)\\
 & \hphantom{=\,\,\frac{\kappa_{2222}}{n}}+\frac{n}{(n-1)(n-2)}\sum^{144}\left(\begin{array}{c}
\kappa_{2000}\kappa_{0210}\kappa_{0012}\\
\vdots\\
\kappa_{0002}\kappa_{0120}\kappa_{2100}
\end{array}\right)\\
 & \hphantom{=\,\,}+\frac{n(n+1)}{(n-1)(n-2)(n-3)}\sum^{24}\left(\begin{array}{c}
\kappa_{2000}\kappa_{0200}\kappa_{0020}\kappa_{0002}\\
\vdots\\
\kappa_{1001}\kappa_{0110}\kappa_{0110}\kappa_{1001}
\end{array}\right)
\end{align*}
Converting to factorial cumulants and keeping the lowest-order terms
\begin{alignat*}{1}
\var(k_{1111}) & =\frac{\kappa_{[1111]}}{n}+\frac{1}{n-1}(\kappa_{[1000]}\kappa_{[0111]}+\kappa_{[0100]}\kappa_{[1011]}+\kappa_{[0010]}\kappa_{[1101]}\\
 & \hphantom{=}+\kappa_{[0001]}\kappa_{[1110]}+\mathcal{O}(\lambda^{5}))+\frac{1}{n-1}\mathcal{O}(\lambda^{5})\\
 & \hphantom{=}+\frac{n}{(n-1)(n-2)}(\kappa_{[1000]}\kappa_{[0100]}\kappa_{[0011]}+\kappa_{[1000]}\kappa_{[0010]}\kappa_{[0101]}\\
 & \hphantom{=}+\kappa_{[1000]}\kappa_{[0001]}\kappa_{[0110]}+\kappa_{[0100]}\kappa_{[0010]}\kappa_{[1001]}+\kappa_{[0100]}\kappa_{[0001]}\kappa_{[1010]}\\
 & \hphantom{=}+\kappa_{[0010]}\kappa_{[0001]}\kappa_{[1100]}+\mathcal{O}(\lambda^{5}))+\frac{1}{n-1}\mathcal{O}(\lambda^{6})\\
 & \hphantom{=}+\frac{1}{n-1}(\kappa_{[1100]}\kappa_{[0011]}+\kappa_{[1010]}\kappa_{[0101]}+\kappa_{[1001]}\kappa_{[0110]}+\mathcal{O}(\lambda^{5}))\\
 & \hphantom{=}+\frac{n}{(n-1)(n-2)}\mathcal{O}(\lambda^{5})+\\
 & \hphantom{=}+\frac{n(n+1)}{(n-1)(n-2)(n-3)}(\kappa_{[1000]}\kappa_{[0100]}\kappa_{[0010]}\kappa_{[0001]}+\mathcal{O}(\lambda^{5}))
\end{alignat*}
Assuming $n\gg1$ and using~(\ref{eq:K-21-3}),
\begin{equation}
\begin{aligned}\var(k_{1111}) & \approx\frac{1}{n}(\gamma_{4}N\lambda^{4}+4\gamma_{1}\gamma_{3}N^{2}\lambda^{4}+6\gamma_{1}^{2}\gamma_{2}N^{3}\lambda^{4}\\
 & \hphantom{\approx\,\,\frac{1}{n}(\gamma_{4}N\lambda^{4}}+3\gamma_{2}^{2}N^{2}\lambda^{4}+\gamma_{1}^{4}N^{4}\lambda^{4})+\mathcal{O}(\lambda^{5})
\end{aligned}
\label{eq:K-33-1}
\end{equation}

Now we can verify,
\[
\frac{\var(k_{1100})}{\kappa_{1100}^{2}}=\frac{\var(k_{0011})}{\kappa_{0011}^{2}}=\frac{\mathcal{O}(\lambda^{2})}{\mathcal{O}(\lambda^{4})}=\mathcal{O}(\lambda^{-2})
\]
\[
\frac{\var(k_{1111})}{\kappa_{1111}^{2}}=\frac{\mathcal{O}(\lambda^{4})}{\mathcal{O}(\lambda^{8})}=\mathcal{O}(\lambda^{-4})
\]
\[
\frac{\mathrm{cov}(k_{1100},k_{0011})}{\kappa_{1100}\kappa_{0011}}=\frac{\mathcal{O}(\lambda^{4})}{\mathcal{O}(\lambda^{4})}=\mathcal{O}(\lambda^{0})
\]
\[
\frac{\mathrm{cov}(k_{1100},k_{1111})}{\kappa_{1100}\kappa_{1111}}=\frac{\mathrm{cov}(k_{1111},k_{0011})}{\kappa_{1111}\kappa_{0011}}=\frac{\mathcal{O}(\lambda^{4})}{\mathcal{O}(\lambda^{6})}=\mathcal{O}(\lambda^{-2})
\]
Therefore $\frac{\var(k_{1111})}{\kappa_{1111}^{2}}$ has the lowest
order in $\lambda^{-1}$ as expected. 

Using~(\ref{eq:K-33-1}),
\[
\begin{aligned}\frac{\var(k_{1111})}{\kappa_{1111}^{2}} & =\frac{\var(k_{1111})}{(\gamma_{4}N\lambda^{4})^{2}}\\
 & \approx\frac{1}{n\gamma_{4}^{2}\lambda^{4}}(\frac{\gamma_{4}+4\gamma_{3}N+6\gamma_{2}N^{2}+3\gamma_{2}^{2}N+N^{3}}{N})
\end{aligned}
\]
Therefore, 
\begin{equation}
\mathrm{SNR}_{22}\approx\sqrt{n}\gamma_{4}(QT)^{2}\sqrt{\frac{N}{\gamma_{4}+(4\gamma_{3}+3\gamma_{2}^{2})N+6\gamma_{2}N^{2}+N^{3}}}\label{eq:SNR22}
\end{equation}
(multi-detector case).

The SNR results, (\ref{eq:SNR11}), (\ref{eq:SNR21}), and (\ref{eq:SNR22}),
are consistent with
\[
\begin{cases}
\mathrm{SNR_{m,n}}\approx\sqrt{n}\gamma_{m+n}^{1/2}(QT)^{(m+n)/2}\sqrt{N} & \text{for }N\ll1\\
\mathrm{SNR_{m,n}}\approx\sqrt{n}\gamma_{m+n}(QT)^{(m+n)/2}N^{1-(m+n)/2} & \text{for }N\gg1
\end{cases}
\]
reported by Melnykov and Hall (they have not reported $\mathrm{SNR}_{22}^{\mathrm{(1d)}}$
specifically, and we have not calculated it yet either; it is, as
one might expect, a tedious task). ``In the high-concentration limit,
a fluorescence signal becomes Gaussian and therefore correlations
other than $g_{11}$ tend to 0. When $m=n=1$ SNR is independent of
the number of molecules in the high concentration limit and is proportional
to N in the low concentration limit.'' In our current report and
experiments, the highest order studied is $(2,2)$ which has our limiting
uncertainty. Therefore, assuming a Gaussian illumination profile which
yields
\[
\gamma_{m}=m^{-3/2}
\]
we estimate the optimal value of $N$ that maximizes~(\ref{eq:SNR22})
to be $N\approx0.22$, with half maxima at $(0.016,\,1.9)$, corresponding
to correlation amplitudes of 
\[
g_{11}(0)=\frac{\gamma_{2}}{N}=\frac{1}{N_{\mathrm{FCS}}}\approx1.6\,\,(0.19,\,22)
\]
or $N_{\mathrm{FCS}}=N/\gamma_{2}\approx0.62\,(0.046,\,5.4)$ ($N_{\mathrm{FCS}}$
is a more common definition of the number of molecules in the probe
region in conventional FCS.) The numbers in parentheses indicate the
$N$ values at half maxima of $\mathrm{SNR}_{22}$~. 

To examine the validity of the assumption $QT\ll1$ at large lag times,
consider the lag time $t\approx1\,\mathrm{ms}$ and $T=0.1t=100\,\mu s$,
with $N\approx1$ and channel count rate $R\approx10\,k\mathrm{Hz}$.
We then have $Q=R/N\approx10\,k\mathrm{Hz}$ then $QT\approx1$. But
this is about the limit in our experiments. As long as $QT\gg1$ is
not true, and $N\approx1$, the truncated formulas above give the
right order of magnitude.

In our program implementation, we estimate $QT$ using $QT=\kappa_{10}/N=\kappa_{01}/N$,
with $\kappa_{10}$ (ideally, $=\kappa_{01}$) being the mean photon
count per bin, and $N$ calculated using the inverse of $g_{11}(0)$:
\[
N=\frac{\gamma_{2}}{g_{11}(0)}=\frac{\gamma_{2}\kappa_{10}\kappa_{01}}{\kappa_{11}(0)}=\frac{\gamma_{2}\kappa_{10}\kappa_{01}}{\kappa_{20}}
\]
This result assumes not reaction. For reactions, further analysis
is needed.

\appendix

\section{Variance and covariance of a function }

The following derivation is described in~\cite{kendall94}. Suppose
for $x_{1},x_{2},\ldots,x_{k}$ we have 
\[
\begin{array}{l}
\mathrm{E}(x_{i})=\theta_{i}\\
\mathrm{var}(x_{i})=\mathcal{O}(n^{-r})\\
\mathrm{cov}(x_{i},x_{j})=\mathcal{O}(n^{-r})
\end{array}
\]
(usually $r=1$) then $x_{i}\to\theta_{i}$ as $n\to\infty$. For
example, $x_{i}$ can be statistics.

Consider 
\[
f(x):=f(x_{1},x_{2},\ldots,x_{k})
\]
\[
f_{i}^{\prime}(\theta):=\left.\frac{\partial f(x)}{\partial x_{i}}\right|_{\theta_{1},\theta_{2},\ldots,\theta_{k}}
\]

Through Taylor expansion we have
\[
f(x)=f(\theta)+\sum_{i=1}^{k}f_{i}^{\prime}(\theta)(x_{i}-\theta_{i})+\mathcal{O}(n^{-r})
\]
Since $\mathrm{E}(x_{i})=\theta_{i}$ we have
\[
\mathrm{E}[f(x)]=f(\theta)+\mathcal{O}(n^{-r})
\]
Now
\begin{align}
\mathrm{var}[f(x)] & =\mathrm{E}\left[\left\{ f(x)-\mathrm{E}[f(x)]\right\} ^{2}\right]\nonumber \\
 & =\mathrm{E}\left[\left\{ \sum_{i=1}^{k}f_{i}^{\prime}(\theta)(x_{i}-\theta_{i})\right\} ^{2}\right]+\mathcal{O}(n^{-r})\nonumber \\
\intertext{\text{assuming not all \ensuremath{f_{i}^{\prime}(\theta)=0} }} & =\sum_{i=1}^{k}\left\{ f_{i}^{\prime}(\theta)\right\} ^{2}\mathrm{E}\left[(x_{i}-\theta_{i})^{2}\right]\nonumber \\
 & \hphantom{=\sum_{i=1}^{k}\left\{ f_{i}^{\prime}(\theta)\right\} ^{2}}+\mathop{\sum_{i=1}^{k}\sum_{j=1}^{k}}_{i\neq j}f_{i}^{\prime}(\theta)f_{j}^{\prime}(\theta)\mathrm{E}\left[(x_{i}-\theta_{i})(x_{j}-\theta_{j})\right]+\mathcal{O}(n^{-r})\nonumber \\
 & =\sum_{i=1}^{k}\left\{ f_{i}^{\prime}(\theta)\right\} ^{2}\mathrm{var}(x_{i})+\mathop{\sum_{i=1}^{k}\sum_{j=1}^{k}}_{i\neq j}f_{i}^{\prime}(\theta)f_{j}^{\prime}(\theta)\mathrm{cov}(x_{i},x_{j})+\mathcal{O}(n^{-r})\label{eq:K-9-1}
\end{align}
Permutation is allowed in the second summation. 

In particular, if $f$ is a linear function of random variables: $f(x_{1},\ldots,x_{k})=\sum a_{i}x_{i}$
then 
\[
\mathrm{var}[f(x)]=\sum a_{i}^{2}\mathrm{var}(x_{i})+\sum_{i\neq j}a_{i}a_{j}\mathrm{cov}(x_{i},x_{j})\mbox{\tag{exact}}
\]

Similarly, one can show
\[
\mathrm{cov}[f(x),g(x)]=\sum_{i=1}^{k}f_{i}^{\prime}(\theta)g_{i}^{\prime}(\theta)\mathrm{var}(x_{i})+\mathop{\sum_{i=1}^{k}\sum_{j=1}^{k}}_{i\neq j}f_{i}^{\prime}(\theta)g_{j}^{\prime}(\theta)\mathrm{cov}(x_{i},x_{j})+\mathcal{O}(n^{-r})
\]

\section{Some sampling moments}

There are three types of moments:
\begin{itemize}
\item Population moments (moments of the population), such as $\mu_{r}^{\prime}$,
$\mu_{r}$, etc.
\item Sample moments (moments of the sample), such as $m_{r}^{\prime}$,
$m_{r}$, etc.
\item Sampling moments (moments of the sampling distribution), such as $\mbox{\ensuremath{\mathrm{E}}}(m_{r}^{\prime})$,
$\mbox{\ensuremath{\mathrm{var}}}(m_{r}^{\prime})$, $\mbox{\ensuremath{\mathrm{E}}}(m_{r})$,
$\mbox{\ensuremath{\mathrm{var}}}(m_{r})$, etc.
\end{itemize}
Similarly we have population cumulants, sample cumulants, and sampling
cumulants.

Specifically, the $r$th moment is $\mu_{r}^{\prime}$, the $r$th
moment statistic is 
\[
m_{r}^{\prime}=\frac{1}{n}\sum_{j=1}^{n}x_{j}^{r}
\]
and the $r$th central moment statistic (moment about the sample mean)
is 
\begin{equation}
m_{r}=\frac{1}{n}\sum_{j=1}^{n}(x_{j}-m_{1}^{\prime})^{r}\label{eq:K-4-0}
\end{equation}
Here, we aim to find to the first few sampling moments. The material
is adopted from~\cite{kendall94}.

The expectation of $m_{r}^{\prime}$ is 

\[
\begin{aligned}\mathrm{E}(m_{r}^{\prime}) & =\frac{1}{n}\sum\mathrm{E}(x^{r})\\
 & =\mathrm{E}(x^{r})\\
 & =\mu_{r}^{\prime}
\end{aligned}
\]

The sampling variance of $m_{r}^{\prime}$ is
\begin{align}
\mathrm{var}(m_{r}^{\prime}) & =\mathrm{E}\left[\left\{ m_{r}^{\prime}-\mathrm{E}(m_{r}^{\prime})\right\} ^{2}\right]\nonumber \\
 & =\mathrm{E}\left[\left\{ \frac{1}{n}\sum x^{r}-\mu_{r}^{\prime}\right\} ^{2}\right]\nonumber \\
 & =\frac{1}{n^{2}}\mathrm{E}\left[\left\{ \sum x^{r}\right\} ^{2}\right]-\mu_{r}^{\prime2}\nonumber \\
 & =\frac{1}{n^{2}}\mathrm{E}\left[\sum x^{2r}+\sum_{j\neq k}x_{j}^{r}x_{k}^{r}\right]-\mu_{r}^{\prime2}\nonumber \\
\intertext{\text{The second summation has \ensuremath{n(n-1)} terms with \ensuremath{x_{j}} and \ensuremath{x_{k}} independent}} & =\frac{1}{n^{2}}\left\{ n\mu_{2r}^{\prime}+n(n-1)\mu_{r}^{\prime2}\right\} -\mu_{r}^{\prime2}\nonumber \\
 & =\frac{1}{n}(\mu_{2r}^{\prime}-\mu_{r}^{\prime2})\label{eq:K-4-1}
\end{align}
which is an exact result.

The expectation of $m_{2}$ is
\[
\begin{aligned}\mathrm{E(m_{2})} & =\mathrm{E}\left[\frac{\sum x^{2}}{n}-\left(\frac{\sum x}{n}\right)^{2}\right]\\
 & =\mathrm{E}\left[\frac{\sum x^{2}}{n}-\frac{1}{n^{2}}\left(\sum x^{2}+\sum_{i\neq j}x_{i}x_{j}\right)\right]
\end{aligned}
\]
Since $x_{i}$ and $x_{j}$ are independent, $\mathrm{E}(x_{i}x_{j})=\mu_{1}^{\prime2}$
and
\[
\mathrm{E}\left(\sum_{i\neq j}x_{i}x_{j}\right)=n(n-1)\mu_{1}^{\prime2}
\]
we get
\[
\begin{aligned}\mathrm{E(m_{2})} & =(\frac{1}{n}-\frac{1}{n^{2}})n\mu_{2}^{\prime}-\frac{1}{n^{2}}n(n-1)\mu_{1}^{\prime2}\\
 & =\frac{n-1}{n}(\mu_{2}^{\prime}-\mu_{1}^{\prime2})\\
 & =\frac{n-1}{n}\mu_{2}
\end{aligned}
\]
 Thus, asymptotically,
\[
\mathrm{E(m_{2})}\doteq\mu_{2}
\]

To calculate the sampling variance of $m_{2}$, we shall generalize
this approach for higher moments. Recall e.g. 
\[
\sum x_{i}^{2}x_{j}x_{l}^{3}x_{k}\tag{all suffixes different}
\]
has $n(n-1)(n-2)(n-3)$ terms. To simplify the notation, we define
the following variants of symmetric functions. A symmetric function
$t(x_{1},\ldots,x_{n})$ remains unchanged if we permute the $x$s. 
\begin{itemize}
\item Augmented symmetric functions: (all subscripts different)
\begin{equation}
[p_{1}^{\pi_{1}}p_{2}^{\pi_{2}}\ldots p_{s}^{\pi_{s}}]=\sum\underbrace{x_{i}^{p_{1}}x_{j}^{p_{1}}\ldots}_{\pi_{1}}\underbrace{x_{q}^{p_{2}}x_{r}^{p_{2}}\ldots}_{\pi_{2}}\underset{\ldots}{\ldots}\underbrace{x_{u}^{p_{s}}x_{v}^{p_{s}}\ldots}_{\pi_{s}}\label{eq:K-16-1}
\end{equation}
E.g.
\[
\sum x_{i}^{2}x_{j}x_{l}^{3}x_{k}=[1^{2}23]
\]
\[
\sum x_{i}^{2}x_{j}^{2}x_{k}^{2}=[2^{3}]
\]
\item Monomial symmetric functions:
\[
(p_{1}^{\pi_{1}}p_{2}^{\pi_{2}}\ldots p_{s}^{\pi_{s}})=\frac{[p_{1}^{\pi_{1}}p_{2}^{\pi_{2}}\ldots p_{s}^{\pi_{s}}]}{\pi_{1}!\pi_{2}!\ldots\pi_{s}!}
\]
\item Unitary functions (no two indices equal):
\[
a_{r}=(1^{r})=\frac{\sum x_{i}x_{j}\ldots x_{l}}{r!}
\]
\item One-part functions or power sums:
\[
s_{r}=(r)=\sum x^{r}=[r]
\]
\end{itemize}
From~(\ref{eq:K-16-1}) 
\begin{equation}
\mathrm{E}[p_{1}^{\pi_{1}}p_{2}^{\pi_{2}}\ldots p_{s}^{\pi_{s}}]=n(n-1)\ldots(n-\rho+1)\mu_{p_{1}}^{\prime\pi_{1}}\mu_{p_{2}}^{\prime\pi_{2}}\ldots\mu_{p_{s}}^{\prime\pi_{s}}\label{eq:K-16-2}
\end{equation}
where $\rho=\sum_{i=1}^{s}\pi_{i}$ and $p:=\sum_{i=1}^{s}p_{i}\pi_{i}$
is the ``weight'' of the symmetric function.

For example, the sample variance~(\ref{eq:K-4-0}) for $r=2$ becomes
\[
\begin{aligned}m_{2} & =\frac{\sum x^{2}}{n}-\left(\frac{\sum x}{n}\right)^{2}\\
 & =\frac{(2)}{n}-\frac{(1)^{2}}{n^{2}}
\end{aligned}
\]
We have, 
\[
\begin{array}{lcl}
(2)=[2] &  & \text{since }1!=1\\
(1)^{2}=[2]+[1^{2}] &  & \text{since }\left(\sum x\right)^{2}=\sum x^{2}+\sum_{i\neq j}x_{i}x_{j}
\end{array}
\]
(or from table). Hence, 
\begin{align}
m_{2} & =\frac{[2]}{n}-\frac{[2]+[1^{2}]}{n^{2}}\nonumber \\
 & =\frac{n-1}{n^{2}}[2]-\frac{1}{n^{2}}[1^{2}]\label{eq:K-17-1}
\end{align}
Now using~(\ref{eq:K-16-2})
\begin{align}
\mathrm{E}(m_{2}) & =\frac{n-1}{n^{2}}n\mu_{2}^{\prime}-\frac{1}{n^{2}}n(n-1)\mu_{1}^{\prime2}\nonumber \\
 & =\frac{n-1}{n}(\mu_{2}^{\prime}-\mu_{1}^{\prime2})\nonumber \\
 & =\frac{n-1}{n}\mu_{2}\label{eq:K-17-2}
\end{align}
For a shortcut to calculate such expectation values, notice that statistic
$m_{2}$ is independent of the origin. Therefore, we can take the
population mean to be zero: $\mu_{1}^{\prime}=0$. Then ignore any
$[\:]$ containing a unit and~(\ref{eq:K-17-1}) immediately gives~(\ref{eq:K-17-2}).

Similarly, 
\begin{align*}
m_{2}^{2} & =\left[\frac{(2)}{n}-\frac{(1)^{2}}{n^{2}}\right]^{2}\\
 & =\frac{(2)^{2}}{n^{2}}-\frac{2(2)(1)^{2}}{n^{3}}+\frac{(1)^{4}}{n^{4}}
\end{align*}
From table or directly
\[
\begin{array}{lcl}
(2)^{2}=[4]+[2^{2}] &  & \text{since }\left(\sum x^{2}\right)^{2}=\sum x^{4}+\sum_{i\neq j}x_{i}^{2}x_{j}^{2}\\
(2)(1)^{2}=[4]+2[31]+[2^{2}]+[21^{2}]
\end{array}
\]
since
\[
\left(\sum x^{2}\right)\left(\sum x\right)^{2}=\sum x^{4}+2\sum x_{i}^{3}x_{j}+\sum x_{i}^{2}x_{j}^{2}+\sum x_{i}^{2}x_{j}x_{k}
\]
and 
\[
(1)^{4}=[4]+4[31]+3[2^{2}]+6[21^{2}]+[1^{4}]
\]
because
\[
\left(\sum x\right)^{4}={4 \choose 4}\sum x^{4}+{4 \choose 3}\sum x_{i}^{3}x_{j}+\frac{1}{2}{4 \choose 2}\sum x_{i}^{2}x_{j}^{2}+{4 \choose 2}\sum x_{i}^{2}x_{j}x_{k}+\sum x_{i}x_{j}x_{k}x_{l}
\]
Ignoring $[\,]$'s containing a unit:
\begin{align*}
\mathrm{E}(m_{2}^{2}) & =\mathrm{E}\left[\frac{[4]+[2^{2}]}{n^{2}}-2\frac{[4]+[2^{2}]}{n^{3}}+\frac{[4]+3[2^{2}]}{n^{4}}\right]\\
 & =\mathrm{E}\left[\frac{(n^{2}-2n+1)}{n^{4}}[4]+\frac{(n^{2}-2n+3)}{n^{4}}[2^{2}]\right]\\
\intertext{\text{and using\,\eqref{eq:K-16-2}}} & =\frac{(n-1)^{2}}{n^{3}}\mu_{4}+\frac{(n-1)(n^{2}-2n+3)}{n^{3}}\mu_{2}^{2}
\end{align*}
 Using this and~(\ref{eq:K-17-2}) we get
\begin{equation}
\begin{aligned}\mathrm{var}(m_{2}) & =\mathrm{E}(m_{2}^{2})-\left\{ \mathrm{E}(m_{2})\right\} ^{2}\\
 & =\frac{(n-1)^{2}}{n^{3}}\mu_{4}-\frac{(n-1)(n-3)}{n^{3}}\mu_{2}^{2}\\
 & \doteq\frac{1}{n}(\mu_{4}-\mu_{2}^{2})
\end{aligned}
\label{eq:K-18-1}
\end{equation}
where $\doteq$ indicates asymptotic equality at large $n$.

Also, if in~(\ref{eq:K-18-1}) we put $\kappa_{4}=\mu_{4}-3\mu_{2}^{2}$
and $\kappa_{2}=\mu_{2}$ and $k_{2}=m_{2}$

\[
\mathrm{var}(k_{2})=\left(\frac{n-1}{n}\right)^{2}\left(\frac{\kappa_{4}}{n}+\frac{2}{n-1}\kappa_{2}^{2}\right)
\]

\section{A tensor notation rule\label{sec:tensor-rule}}

In Kaplan's formulae in tensor notation for cumulants of $k$-statistics\cite{kaplan52},
to insert a single subscript, we affix the subscript in every possible
position and divide by $n$. For example, given
\[
\kappa(ab,ijk)=\frac{\kappa_{abijk}}{n}+\frac{\sum^{6}\kappa_{ai}\kappa_{bjk}}{n-1}
\]
we can construct
\[
\kappa(ab,ijk,p)=\frac{\kappa_{abijkp}}{n^{2}}+\frac{\sum^{6}\kappa_{aip}\kappa_{bjk}+\sum^{6}\kappa_{ai}\kappa_{bjkp}}{n(n-1)}
\]
Other examples:
\begin{equation}
\kappa(i,j)=\frac{\kappa_{ij}}{n}\label{eq:kappa_ij}
\end{equation}
\begin{equation}
\kappa(i,kl)=\frac{\kappa_{ikl}}{n}\label{eq:kappa_ikl}
\end{equation}
\[
\kappa(i,j,kl)=\frac{\kappa_{ijkl}}{n^{2}}
\]

\section{Two-state transition factors\label{sec:Two-state-transition-factors}}

Consider a fluorescent particle alternating between two states:%
\[
\ce{\text{state 1} <=>[k_{\mathrm{f}}][k_{\mathrm{b}}] \text{state 2} }
\]
where ``state 1'' is usually the brighter (unfolded) state, ``state
2'' is usually the darker (folded) state, and $k_{\mathrm{f}}$ and
$k_{\mathrm{b}}$ are the forward and backward (reverse) rates respectively.
Denote the probability that the particle is found in state $1$ at
time time $t$ with $P_{1}(t)$, and similarly for state $2$. The
following equations describe the reaction: 

\begin{align*}
\frac{\mathrm{d}P_{1}(t)}{\mathrm{d}t} & =-k_{\mathrm{f}}P_{1}(t)+k_{\mathrm{b}}P_{2}(t)\\
P_{2}(t) & =1-P_{1}(t)
\end{align*}

The solutions are
\begin{equation}
\begin{aligned}P_{1}(t) & =\frac{1+[(1+k)P_{1}(0)-1]e^{-t/t_{\mathrm{R}}}}{1+k}\\
P_{2}(t) & =\frac{k-[k-(1+k)P_{2}(0)]e^{-t/t_{\mathrm{R}}}}{1+k}
\end{aligned}
\label{eq:rate_solve}
\end{equation}
where $P_{1}(0)$ and $P_{2}(0)$ are the initial probabilities at
time $0$. We have defined
\[
k=\frac{k_{\mathrm{f}}}{k_{\mathrm{b}}}=\frac{N_{2}}{N_{1}}
\]
with $N_{i}$ being the number of molecules in state $i$ in the ensemble,
and
\[
t_{\mathrm{R}}=(k_{\mathrm{f}}+k_{\mathrm{b}})^{-1}
\]
as the overall reaction time constant. Also, defining 
\[
P(i)=P_{i}(\infty)=\frac{N_{i}}{N_{1}+N_{2}}
\]
as the probability of finding the particle in state $i$ independent
of initial conditions, we have
\[
k=\frac{P(2)}{P(1)}
\]

The transition factor $Z_{s_{2},s_{1}}(t)$, denoting the probability
that the particle is found in state $s_{2}$ at time $t$ given it
was in state $s_{1}$ at time $0$, can be found by setting the initial
probabilities in~(\ref{eq:rate_solve}) equal to $1$ or $0$:
\begin{align*}
Z_{1,1}(t) & =\frac{1+ke^{-t/t_{\mathrm{R}}}}{1+k}\\
Z_{1,2}(t) & =\frac{1-e^{-t/t_{\mathrm{R}}}}{1+k}\\
Z_{2,1}(t) & =\frac{k(1-e^{-t/t_{\mathrm{R}}})}{1+k}\\
Z_{2,2}(t) & =\frac{k+e^{-t/t_{\mathrm{R}}}}{1+k}
\end{align*}

\bibliographystyle{plain}
\phantomsection\addcontentsline{toc}{section}{\refname}\bibliography{hofcs}

\end{document}